# BASIC MECHANICS OF PLANET-SATELLITE INTERACTION WITH SPECIAL REFERENCE TO EARTH-MOON SYSTEM.

**INTRODUCTION.**

In 18th Century, German Philosopher Kant had suggested the theory of retardation of Earth's spin based on the ancient records of Solar Eclipses [Stephenson 1986, Stephenson 2003]. Similar kind of studies have been carried out by Kevin Pang at Jet propulsion Laboratory at Pasadena [Morrison 1978, Jong & Soldt 1989]. He happened to step upon certain ancient records regarding Solar Eclipses. A total Solar Eclipse had been observed in the town of Anyang, In Eastern China, on June 5, 1302 B.C. during the reign of Wu Ding. Had Earth maintained the present rate of spin, the Eclipse should have been observed in middle of Europe. This implies that in 1302 B.C. i.e. 3,291 years ago Earth's spin period was shorter by 0.047 seconds. This leads to a slowdown rate of 1.428 seconds per 100,000 years.

In 1879 George Howard Darwin carried out a complete theoretical analysis of Earth-Moon System and put forward a sound hypothesis for explaining the slow down of Earth's spin on its axis.

*Goerge Howard Darwin's Hypothesis:*

According to George Howard Darwin (son of Charles Darwin, father of Evolution Theory) [ Darwin 1879, Darwin 1880], Moon is the daughter of the Earth. At the time of birth of the Solar System, when Earth was accreted from circumsolar disc of gas and dust, Earth was spinning at the rate of 6 hours per day. At the time of birth, Earth had no natural satellites. Hence Solar tides were the only tidal force. Tides came and ebbed at 3 hours interval. The Solar tides created a forced vibration of Earth's bulk at the frequency of 1/3 cycles hr or 0.33 cycles/hr. George calculated the natural frequency of vibration which came out to be the same as the forced frequency of vibration. This caused Resonance leading to catastrophic vibrations of the Earth ultimately tearing asunder a part of the Earth and leading to the birth of Moon from this torn part.

George postulated that Moon was torn out of the Pacific Ocean basin and hurled into an outward spiral trajectory.

Once Earth-Moon System is born, because of proximity, lunar tides became the dominating tidal force and it drove the System off-resonance.

George postulated that Moon was orbiting at an angular velocity $\Omega$ which was slower than the spin angular velocity of the Earth. Hence slow orbiting Moon tries to hold back Earth's spin. This leads to transfer of angular momentum from fast spinning Earth to slow orbiting Moon. This results in outward spiral orbit of Moon.

The lunar tidal torque slows down the fast spinning Earth. Simultaneously Moon gets pushed outward until the two bodies become geo-synchronous. George calculated that the final geo-synchronous orbit $a_G$ will be 1.5 times the present Lunar Orbital Radius of $a_p = 3.844 \times 10^8 m$ i.e. $a_G = 5.766 \times 10^8 m$. George also calculated the geo-synchronism period to be 47 days. That is Solar Day will be 47 days long and there will be no Moon Rise or Moon Set.

George further postulated that after geo-synchronism is achieved, Solar tidal drag will become dominant. Sun will try to bring Earth in synchronous orbit as Moon is in synchronous orbit around Earth.

Moon's synchronism [Hartmann 1978] with respect to Earth means Moon's orbital velocity and Moon's spin velocity are equal leading to Moon keeping the same face towards Earth all the time. As a result we see only one side of Moon. The other side always remains in dark. In 1959 Russian Luna Probes for the first time photographed the dark side of Moon.

Further retardation of Earth's spin velocity with respect to Moon's orbital velocity will lead to reversal of transfer of angular momentum. Now lunar tidal interaction will try to make Earth spin faster and angular momentum will be transferred from Moon to Earth. As a result Moon will be drawn into a collapsing or inward spiral orbit ultimately leading to head on collision of Moon into Earth [Moore ?].

Much earlier than this doomsday, our Sun would have consumed its fusion fuel and left the main sequence to become Red Giant and then collapse into a White dwarf. As our Sun bloats into Red Giant 5 billion years hence, it would devour the terrestrial Planets including Earth-Moon System [Garlick 2002].

Hence Earth would never survive to meet the fateful and devastating head on collision with Moon.

As we will see in this Thesis that Darwin's hypothesis is only partly correct.

*Modern findings about Moon:*

Since George Darwin's proposed Hypothesis, Human-Kind has made spectacular progress in every field of Science and Technology including the field of Astronomy and Astrophysics. The launching of Astronomical Satellites has specially helped us in making precise measurements of different orbital and globe parameters of various Celestial Objects of interest[Kaula 1969].

Ranger Probes have pinned down the value of mG of our Moon at 4902.78±0.05 (km)$^3$/(sec)$^2$. The ratio of masses of Moon and Earth have been improved 30 fold. This has been achieved by tracking the motion of Earth around the barycenter of Earth-Moon System. Through satellites the moment of inertia has been deduced to be $8.02 \times 10^{37}$kg-m$^2$. The mean radius was deduced to be 1738km by the occultation of stars by the edge of Moon. But by Ranger Impacts it has been improved to be (1734.8±0.3)km. Density of Moon was deduced to be 3.361±0.002gm/cm$^3$.

The measurements made by Astronomical Satellites in 1963 gives a more precise set of data about Earth-Moon System which is tabulated in Table (1). This is given at the end of Appendices.

Conventionally Principle Moment of Inertia around the spin axis of Earth is $C = 2/5 M (R_E)^2 = 97.06 \times 10^{36} Kg - m^2$.

By Astronomical Satellites the value comes out to be $C = 80.26 \times 10^{36} Kg - m^2$. This reduction comes out because Earth is not a homogenous spherical mass but a stratified oblate, ellipsoid with highest density of mass at the core and lightest density of mass at the crust [Appendix (V)].

Secondly the rock samples brought from our Moon during Apollo 11 to Apollo 17 Mission and during Luna 16 and Luna 20 Mission conclusively prove that Earth and Moon had been formed from the disc of accretion about 4.53 Gya [Cameron 2002, Kerr 2002] but they were never a single body. The Age of Moon has been found incorrect by recent findings [Toubol, Kliene et al 2007, Stevenson 2008]. The new age is around 4.467Gya.

There were three competing hypotheses about the origin of our Moon [Appendix (III)] :
(i) Fission Model put forward by George Howard Darwin as early as 1879;
(ii) Capture Model;
(iii) Double Planet Theory or Co-formation Theory.

None of these theories were consistent with the facts that emerged after extensive study of the Lunar Samples, after interpreting the data received from the network of seismometers set up on the surface of our Moon and after spectroscopic studies of Moon's crust. Finally in 1984 at the International Conference held at Kona, Hawaii, **GIANT IMPACT THEORY** was accepted as the most consistent theory regarding Moon's Origin [Hartmann 1978].

According to this theory a Supernova Explosion occurred in the neighborhood of our Solar System. The shock waves from this explosion caused a neighbouring Giant Interstellar Cloud of gas and dust to go into spin mode. The rapid spinning of the Giant Cloud flattened it into a pancake called the disc of accretion or the primordial Solar Nebula.

The nucleus of this Solar Nebula gravitationally collapsed into a ball of Nuclear Fusion Furance called our Sun [Maddox 1994]. This gave birth to the protoplanetary disk of gas and dust.

According to the Current Model of planetary formation[Halliday & Wood 2007, Chambers 2004, Raymond 2006, Ogihara 2007, Lissauer & Stevenson 2007], the disc of accretion would form around any forming star. This disc of accretion is not unique to our Solar System. In our case disc was 50 AU where 1 AU(astronomical units) is the distance between Earth and our Sun. It is postulated that at 4.568Gya the Solar Nebula was formed. By 4.567Gya there were dust particles embedded in hydrogen gas. Through gentle collisions these dust particles aggregated into larger particles of centimeter or more size. These larger particles either through collision-agglomeration or through gravitational instability coalesced together to form kilometer sized planetesimals. Meteorites are the residues of these planetesimals . This process was completed in one million year by 4.567Gya. A dense swarm of planetesimals in near circular, low-inclination orbits is gravitationally unstable on a short time scale. In less than one more million year much larger bodies of 10km size are formed through collision and accretion. These larger bodies are planetary embryos and are Moon- to Mars- sized. From this point onward the current model is unsure of the series of events which led to the evolution of our present Solar System.

Inspite of the ambiguity there are a few points where there is a wide consensus:

(i) Formation of Gas Giant Jupiter preceded the formation of all planets in our Solar System;

(ii) For next 5 to 30 million years there was enough hydrogen to give birth to the Gas Giants. After 5 to 30 million years the gas- residual dust disc should get dissipated by photoevaporation and Robertson-Poynting drag;

(iii) This means by next 5 to 30 million years Jupiter, Saturn, Neptune and Uranus formation should be complete;

(iv) Further Jupiter and Saturn migrated from inside to their present orbit which caused 1:2 Mean-Motion-Resonance crossing at about 300My after the birth of Solar Nebula. The after effects of 1:2 MMR crossing was felt 200My later in form of Late Heavy Bombardment Era;

(v) Jupiter's migration had a significant role in the subsequent formation of terrestrial planets;

(vi) Terrestrial Planets including Earth were not formed through gradual accretion by the planetary embryos but through a series of infrequent and highly traumatic impacts separated by periods of cooling and healing;

(vii) The last such event in case of Planet Earth was the lunar forming Giant Impact;

(viii) The latest isotopic evidence place this event at 4.467Gya;

(ix) It is also now accepted that 1:2 MMR crossing was responsible for Late Heavy Bombardment Era occurring at 4.2Gya ;

All these consensual points form a part of the Model of planetary formation and evolution proposed in this dissertation[This paper is the first chapter of D.Sc dissertation].

Pluto is a captured body which was not formed by normal process of accretion or by a normal planetary formation process.

According to the present studies of meteorites, spinning Solar Nebula was born 4.567 Gya and Earth formation was complete by 4.467 Gya in a time span of 100 million years [Toubol, Kliene et al 2007, Stevenson 2008].

About this time Earth experienced a glancing angle collision from a Mars-sized planetary embryo. Earth acquired an angle of Obliquity which presently is $\Phi = 23.5^o$ and a rapid spin of 6 hours per revolution. The Impact generated circumterrestrial disc of debris accreted within Roche's zone to form our Moon [Canup & Esposito 1996]. Giant Impact involved a Mars size mass planetary embryo(10% of Earth's mass) and 90% complete Earth. Most of the impact energy was dissipated as heat but the total angular momentum has been conserved till this day. The core of the projectile merged with the core of the proto Earth. Some of the Earth's atmosphere got blown off. The molten mantle froze before it could differentiate. After the impact a circumterrestrial disk of debris was generated. There was a rapid exchange of material between the molten Earth and vaporized disk. This leads to similarity in Oxygen isotope in Earth and Moon. Tungsten and possibly Silicon isotope also point to this kind of exchange and achieving an equilibrium in the aftermath of Impact.

*It is the finding of this thesis that in any two body system there are two Geosynchronous orbits $a_{G1}$ and $a_{G2}$. If the accreting body or the captured body, as is the case for Martian Satellite Phobos, lies within $a_{G1}$ then the Satellite is gravitationally launched on an inward spiral path towards its sure doom. If the accreting zone lies beyond $a_{G1}$ as is the case with our Moon or the captured body lies beyond $a_{G1}$ as is the case for the Martian Satellite Deimos, then the said satellite experiences an impulsive,gravitational sling shot effect because of Conservation of Energy Principle[Cook 2005, Dukla et al 2004, Epstein 2005,Jones2005]* . The inner Geo-Synchronous orbit is also known as Cassini State 1[Gladman, Quin et al 1996]

*As the differential between the orbital angular velocity of the Satellite ($\Omega$) and the spin angular velocity of the Planet ($\omega$) increases, Planet-Satellite becomes a dissipative system and conservation of energy is destroyed. According to the finding of this paper, the peak of the impulsive sling shot torque occurs at $a_1$ and where the maxima of the radial acceleration of the Satellite also occurs.*

*At $a_2$, the gravitational runaway phase is damped out and then onward the Satellite coasts on its own on an outward spiral trajectory toward the outer Geosynchronous Orbital radius $a_{G2}$. At $a_2$ the recession velocity of the outward spiraling Satellite is a maxima and is defined as Gravitational Resonance Point [Rubicam 1975]. At Gravitational Resonance Point, Orbital Period of the Satellite is twice the Spin Period of the Planet. Then onward recession velocity is continuously retarded. Our Moon had a recession velocity maxima of 17 cm/yr at $a_2$ . But the present recession velocity is 3.8 cm/yr as established by Laser Lunar Ranging[Alley et al 1965,Faller et al 1969, Dickey et al 1994]. The eventual destination is $a_{G2}$.*

*In case of Plutonian Satellite Charon, the outer Geo-synchronism is already achieved and Plutonian Spin ω= Charonian Orbital angular velocity Ω*

The profile of the radial acceleration and the profile of the radial velocity are plotted in Fig (1) and Fig (2)

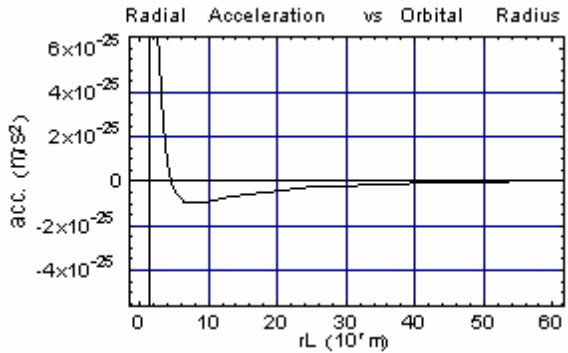

**Figure 1 Radial Acceleration Profile of Moon (Within $a_{G1}$ the Moon is accelerated inward. Beyond $a_{G1}$ the Moon is rapidly accelerated outward under the influence of an impulsive gravitational torque due to rapid transfer of spin rotational energy. The maxima of the outward radial acceleration occurs at $a_1$. (This is the peak of the impulsive sling shot torque.)**

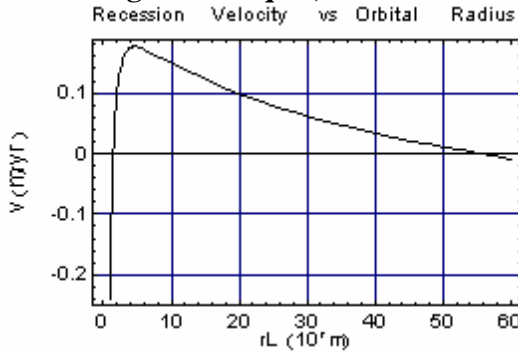

**Figure 2 Radial Velocity Profile of Moon. (Beyond $a_{G1}$, Moon is rapidly accelerated to a maximum radial velocity, $V_{max}$, at $a_2$ where Sling-Shot Effect terminates and radial acceleration is zero. Then onward Moon coasts on it own towards the outer Geo-Synchronous Orbit $a_{G2}$)**

According to the Giant Impact Theory [Hartmann 1978, Canup & Esposioto 1996], the metallic core of the impactor got stuck in Earth's interior and the rocky mantle of the Impactor blew off as rocky debris. Because of impact heating the circumterrestrial debris formed a magma disk in orbit around Earth. The impact heating of the debris drove away all the moisture and volatiles trapped in the rocky debris. This bone dry, volatile-less debris accreted into Moon at about 4.467 Gya. The outer part of the magma disk coalesces to form Moon and inner part falls back onto Earth.

Latest findings indicate[Toubol, Kliene et al 2007 , Stevenson 2008] that $^{182}W/\ ^{184}W$ ratios are identical in the lunar and terrestrial mantle. This implies that either Moon has been mainly derived from Earth's mantle or that tungsten isotopes equilibrated in the aftermath of Giant Impact. A similar proposal has been put forward to account for identical Oxygen Isotopes ratio in two cases. This is the time when Earth's metallic core also precipitated because of the fact that Earth turned into magmatic ball because of impact heating. This permitted magmatic differentiation and stratification by allowing the metallic core to percolate to the centre. As Moon got fully formed in the Roche's zone at 15,000 Km from the centre of the Earth, full scale tidal interaction started between Earth and Moon.

Because of tidal drag and tidal retardation, Earth's spin slowed from 6 hours per solar day to the present 24 hours per solar day. Because of transfer of angular momentum from Earth to Moon, latter's orbital moment of inertia has been increasing and Moon has been receding from the initial 15,000 km to the present 3,84,400 km.

At the point of birth of our Moon, the tidal flexing by the Earth pulverized the rocky debris within Roche's zone into Saturn-like annular rings surrounding the Earth.

During the Late Heavy Bombardment Era when due to 1:2 MMR crossing, there were heavy meteoritic and asteroidic showers on all the planets. This Era lasted from 4.0Gya to 3.8Gya. During this phase annular ring around Earth must have been swept clean. Table (2) gives the comparative study of the Earth and the Moon based on the recent studies. Table (3) gives the new understanding of Earth-Moon system.

The dates arrived at theoretically is consistent with the observed ages of the Moon's samples.

**Table (2) Comparative study of Earth & Moon [Hartmann 1978]**

|  | Earth | Moon |
|---|---|---|
| Ratio of Oxygen isotopes | $R_1$ | $R_1$ |
| Moisture | Blue Planet | Bone Dry |
| Refractory |  | Abundance of refractory such as Alumina, Calcium, Thorium and Rare Earth Elements. Rare Earth Elements are |

| | | 50% higher on Moon as compared to that on Earth. |
|---|---|---|
| Volatiles | Mantle is volatile rich | Much less quantity of volatile materials such as Sodium, Potassium, Bismuth and Thallium. |
| Core | Iron and Nickel (16% by weight of total) | Lighter metals (1% by weight of the total) |
| Average density | 5.5 gms per $(cm)^3$ | 3.3 gms per $(cm)^3$ |
| MgO/FeO | $R_2$ | $R_2$ + 10% |
| $^{182}$W/ $^{184}$W ratios in the mantle. | $R_3$ | $R_3$ |
| Age | 4.467Gyr when accretion was completed. | 4.456±0.040Gyr of the lunar crust |

$^{182}$W/ $^{184}$W ratio of any planetary mantle reflects the timescale of accretion and core formation, the degree of tungsten depletion and the extent of re-equilibration of tungsten isotopes during core formation [Jacobsen 2005, Kliene, Mezgar et al 2004, Nimmo & Agnor 2006]

**Table (3) Apollo's influence on Lunar Science [Hartmann 1978, Stevenson 2008]**

| Topic | Pre-Apollo | Current |
|---|---|---|
| Origin | (1) Fission Model (2) Capture Model (3) Binary Planet Model | Giant Impact Model |
| Time of Giant Impact | | 62(+90 / -10)Myr after the instant when first solids or first planetesimals formed in our Solar System formed |
| Crater | Most impact and some volcanic | Almost all impact and dynamics of ejected debris determined |
| Presence of volatiles | unknown | Bone dry and volatile less but water brought by impacting meteorites. Recently evidence of water obtained on the poles. |
| Rock ages | unknown | Highland rocks are older than 4.1Gy. |

| | | Anarthosites are 4.4Gy old. |
|---|---|---|
| Magma ocean | Not conceived | Initially anarthosites crystallized from magma ocean. Other Highland rocks formed after that |
| Composition of Mare Basin | unknown | Wide variety of Basalts |
| Composition of Highlands | unknown | Wide variety of rock types but all containing more Alumina than that in Mare Basin |
| Mantle | unknown | Olivine & Pyroxine |
| Core | unknown | Lighter elements |

The Modern Day findings and the present study indicate that Solar Nebula was formed 4.567Gya[Toubol, Kliene et al 2007]. By 4.467Gya Jovian Planets and Earth had fully formed. The other terrestrial planets were in the process of formation. At about 4.467Gya, by glancing angle impact from a Mars-sized planetismals, Moon was born out of the impact generated circumterrestrial debris beyond Roche's limit as well as beyond $a_{G1}$, the first Geosynchronous Orbit of E-M System. Therefore no rock on Moon can be older than 4.46Gy which infact is the case.

Table[4] gives the birth of Solar Nebula, the time of formation of Earth and the time of formation of Moon [adapted from Toubol, Kliene et al 2007 and Brandon 2007]

**Table[4] The Timeline of Planetary Formation.**

| Triggering | Birth of Solar Nebula | Dissipation of gas & dust disk | Last Giant Impact | Late Heavy Bombardment | Life |
|---|---|---|---|---|---|
| 4.568Gya | 4.567Gya | 4.558Gya | 4.468Gya | 4.0Gya | 3.568Gya |
| | 0 | 9My | 99My | 567My | 999My |
| Triggering | A supernova explosion in our neighborhood generates shock waves which sets a passing-by interstellar cloud of gas and dust into spin mode. This spinning primordial cloud flattens into a pancake like disc of cloud and dust. | | | | |
| Birth of Solar Nebula | The dust particles may be colliding and sticking giving rise to pebble sized solids. These pebbles further coalesce to form km-size planetesimals. This formally marks the birth of Solar Nebula and the start of Planetary Formation. Planetesimals collide and accrete to form planetary embryos. | | | | |
| Dissipation of | Particles less than micron size and gases are pushed out by photon | | | | |

| gas and dust disk. | pressures. This is known as Photo-evaporation. Particles of micron size and more are acted upon by Robertson-Poynting drag which constrains these particles to spiral inward and eventually fall into the host body. This leads to gradual dispersal and dissipation of gas-dust disk. By this process all the gas and dust will be removed in 30My. This means that within this narrow time slot the Gas Giants should have completed their formation. Hence in first 30My Jupiter and Saturn should complete their formation. Planetary embryos get enveloped by Hydrogen gas through gravitational accretional runaway mechanism terminated by the paucity of material because of a gaping void. When the void gets filled up then the next sequential gravitational accretional runaway process initiated. This process is repeated until all the gases are exhausted. In this way in 30My Jupiter, Saturn, Neptune and Uranus formation is completed. |
|---|---|
| Last Giant Impact | The terrestrial planets are not formed by runaway gravitational accretional mechanisms because such large amounts of material is not present to sustain such a process. Instead a series of infrequent and titanic impacts caused the formation of the present sized Earth, Venus, Mars and Mercury. The Giant Impact was the last such event, atleast in the context of Earth, which formally marked the completion of formation of Earth. |
| Late Heavy Bombardment Era. | About 567My after the birth of Solar Nebula, 1:2 MMR crossing occurred by the spirally expanding orbits of Jupiter and Saturn. This caused Neptune to be flung into Kuiper Belt. The disturbance of Kuiper Belt caused a large amount of comets and asteroids to be flung into the inner part of the Solar System. This caused all the planets to experience a Late Heavy Bombardment Era. The foot prints of this era is well preserved in the petrological record of Moon. |
| Life | After 1 Gy the first organic life based on anaerobic fermentation was initiated. |

Impact heating, accretion heating, core formation and sinking of the metallic core caused immense heating resulting in a sea of magma covering whole of the Moon. Anorthosites composed of white Feldspar were first to crystallize out of Magma Ocean. These are found to be 4.4Gy old [Hartmann 1978].

The earliest history of the Moon and Earth are given in Table[I.5] [Nimmo & Agnor 2006]

**Table[5] The earliest history of the Moon and Earth.**

| 4.567Gya | 4.53Gya | | 4.467Gya | | | 4Gya |
|---|---|---|---|---|---|---|

| 0 | 30My | 60My | 100My | 150My | 200My | 567My |
|---|---|---|---|---|---|---|
| **EARTH** | | | | | | |
| Older View | Earth formation 90% | | | | | |
| New View | | Earth Formation 90% | | | | |
| **MOON** | | | | | | |
| Older View | Giant Impact | LMO starts | | | | Late Heavy Bombardment ERA |
| New View | | Giant starts | Imp.,LMO | | LMO Final? | |
| | | | Earliest crusts | | | |

**Table[6] Stages in Planetary Formation.**[Chambers 2004].

| Planetesimal Hypothesis |
|---|
| Young Sun is surrounded by a disk of gas and fine dust. |
| Rapid Collisions and sticking cause the formation of mountain sized boulders called planetesimals. |
| Collisions and gravitational interactions cause the planetesimals to combine to produce a few tens of Moon-to-Mars-size planetary embryos in roughly 0.1-1 million years. These Planetary embryos form the cores of all subsequent planets formation. |
| In first 30My the Jovian planets namely Jupiter, Saturn, Neptune and Uranus are formed sequentially by runaway gravitational accretion of gas. |
| By the end of 30My all the gas and dust are dissipated by photoevaporation and Robertson-Poynting drag. Once the gas-dust disk is dissipated runaway gravitational accretion cannot be sustained. |
| After 30 My in the next 100My some of the planetary embryos grow by infrequent impacts until the embryos are exhausted. |
| One of the late collisions led to the completion of Earth Formation and birth of Moon. |
| It is not clear why there is an Asteroid Belt between Mars and Jupiter Orbit? |

**Table[7] The stages following the Giant Impact which led to the complete formation of our MOON.[Toubol, Kliene 2007]**

| Around 4.467Gya, while Earth Magma Ocean was cooling and silicate differentiation was taking place, a Mars sized planetary embryo(30% of Earth's mass) made a glancing angle collision with 90% formed Earth. This was the last of the impacts experienced in the formation of Earth. |
|---|
| The impact generated debris is converted into a magma disk orbiting around the |

> Earth while blobs of metal from the impacting body settles down through the Earth's mantle to join the target's core. The magma disk and new bulk Earth were a combination of proto-Earth and the impactor. The impact generated heat produced global magma ocean on Earth. Iron-rich metal from proto-Earth and the impactor merged to form the present Earth's core. Whereas silicate and oxide materials formed the mantle and crust of the Earth.

> The outermost part of the Magma Disk(molten and vaporized ejecta) coalesces to form the Moon as the result of radioactive cooling while the rest of magma disk falls back to Earth. There was a slower rate of Lunar Magma Ocean solidification and it may have continued till 200-250My after the birth of Solar Nebula.

From Laser Lunar Ranging experiments [Alley et al 1965, Faller et al 1969, Dickey et al 1994] the velocity of recession of Moon has been determined to be 3.8 cm/yr $= 12.711 \times 10^{-10}$ m per second.

*On 1<sup>st</sup> August , 10:15 to 12:50UT, with Lick Observatory 120-in(304cm) Telescope and Laser operating at 6943Angstrom, return signals from optical retro-reflector array placed on Moon by Apollo 11 Astronauts was detected. Power Transmitted was 748Joules per pulse and return signal was more than one photo-electron [Faller et al 1969].*

From these precise data we proceed on to a more accurate formulation of Earth-Moon interaction.

*A review of the work done till now:*

The technique first adopted by George Howard Darwin (Darwin 1879,1880), by Genstenkorn (1955), by Macdonald (1964) and Kaula (1964) was that of setting up of dynamical equations and integrating back in time. Rubincam (1975) showed that Moon does have an equatorial origin therefore it could have formed by fission or by accretion. Touma et al (1998) has worked out the eviction resonance and secular inclination-eccentricity resonance that excites 10 degree inclination in the remote past. Williams et al (1998) have worked out the obliquity-oblateness feedback process and tried to rationalize the present day obliquity of Earth at 23.5° and lunar orbital plane inclination at 5°. Ward et.al.(2000) have shown that substantial lunar orbital inclination of about 15° in the remote past could be caused as a natural result of its formation from an impact generated disk.

Touma et.al. (1994) do the Hamiltonian reformulation of the multiply averaged secular theory of Goldreich (1966) and use this to examine various tidal models. They make less severe approximations by including second order effects. Touma uses symplectic integration scheme for studying rotational and orbital motion of extended bodies in the planetary n-body problem.

Cameron, the progenitor of Giant Impact theory, was the first person to do simulation of the impact process. Cameron(1997) along with Benz, Slattery(1986,1987)and Melosh (1989) have published their simulation results in ICARUS. Canup and Esposito published their simulation results also in ICARUS(1996). Simulation work was also done by Kipp and Melosh (1986,1989). A new technique known as Smooth Particle Hydrodynamics (SPH) has been developed and utilized for impact simulation by Canup et.al.(2001). The most recent simulation yields iron-poor Moon as well as the present mass and angular momentum of the Earth- Moon system (Canup et al 2001). This class of impacts involves a smaller and thus more likely impacting object than previously considered viable, and suggests that Moon formed near the very end of Earth's accretion. The latest works on the Dynamical History of E-M System have been done by George E.Williams (2000), by Sigfrido Leschiutta et al (2001) and by G.A. Krasinsky (2002). But to date all the works suffer the pitfall of too short or too long an evolutionary span of time from the inception to the present orbital radius.

*The present work gives the correct evolutionary time span of 4.53Gyrs, the present Age of Moon, by optimizing the structure factor in the velocity of recession expression.*

The new Age of Moon is 4.467Gy. Adjusting the system parameters will give this Age.

There is no empiricism involved in arriving at the best fit value of the various parameters of the Planet-Satellite system. There is a general underlying theory which explains any two body system in the Universe-it may be Planet-Satellite System, Planet-Sun System or Binary Star System [Sharma & Ishwar 2004].

1 THE GENERAL UNDERLYING THEORY OF ANY TWO BODY SYSTEM
[Detailed analysis of Two Body System is described in Appendix(XII) as well as in Appendix(XIV) except for the form of tidal torque. This difference of Tidal Torque Form doesnot effect the results]

The Authors have done the Non-Keplerian analysis of Earth-Moon System in the present work. It is found that Earth-Moon do not precisely obey Kepler's Law :

$$G(M + m) = \Omega^2 a^3 \quad (1)$$

Where $a$ = semi major axis of the orbiting body,
$\Omega$ = orbital angular velocity.

Moon at present is experiencing a residual deceleration.

In Planet-Satellite Systems (or in any two body system) we always have a LOM/LOD (length of month/loength of day) equation where LOM is the sidereal orbital period of the satellite or the orbiting body and LOD is the sidereal spin period of the planet or the central body.

In case there is Binary Star System we can proceed with more massive body as the central body and the less massive as the orbiting body. The sidereal orbital period of the orbiting body will be LOM and sidereal spin period of the central body will be LOD whereas in fact both bodies will be experiencing both spin and orbital motion.

The LOM/LOD equation is as follows (Appendix.XII), the symbols relating to Earth and Moon are explained and their values are given in Table (1), page 119 :

$$\omega_E/\Omega_L = \text{LOM/LOD} = EX^{1.5} - FX^2 \quad (2)$$

where

X = semi-major axis of Satellite's Orbit,

E = $J_T/(BC)$,

$J_T$ = total angular momentum of he two body system,

B = $\sqrt{(G(M+m))}$,

M = mass of the Earth,

M = mass of the Moon,

C = Rotational Inertia of the Planet or the central body around the spin axis,

F = $m/(C(1 + m/M))$

LOM/LOD = 1 is the Boundary Equation where Planet and Satellite are mutually tidally locked. Satellite is in Synchronous Rotation (Satellite's Spin Period = Satellite's Orbital Period) as well as in Geo-Synchronous Rotation (Satellite's Orbital Period = Planet's Spin Period). There are two roots of the Boundary Equation : $a_{G1}$ and $a_{G2}$. These are the orbital radii of the inner and the outer Geo-Synchronous Orbits where the Satellite is in precise Keplerian Equilibrium. The Natural Satellite always begins its journey from the inner Geo-Synchronous Orbit ($a_{G1}$) when it starts as a captured body.

When the Satellite is born out of gravitational accretion of impact generated debris then there are two cases :

$a_R > a_{G1}$ then the journey begins an outward expanding spiral path from $a_R$

$a_R < a_{G1}$ then the journey begins an inward contracting spiral path from $a_R$

where

$a_R$ = Roche's radius [Canup et al 2001] = $2.45(\rho_E/\rho_L)^{1/3}$,

$R_E$ = Radius of the Earth or of the central body,

$\rho_E$ = density of Earth or of the central body = 5.5 gms/c.c.,

$\rho_L$ = density of Moon or of the orbiting body = 3.34 gms/c.c.,

Any perturbation such as solar wind, cosmic particles, star dust or radiation pressure perturbs the Natural Satellite to a new orbit within or beyond $a_{G1}$. In both cases it is launched on a gravitational runaway path.

If the new orbital radius $r < a_{G1}$, then LOM/LOD < 1 and Satellite tidal drag accelerates the spin of the Planet as a result by the law of conservation of angular momentum and energy, Satellite is pulled on an inward collapsing spiral trajectory to its certain doom. This the case with Phobos, a Satellite of Mars.

If the new orbital radius $r > a_{G1}$, then LOM/LOD >1 and the Satellite tidal drag retards the spin of the Planet as a result the transfer of energy and angular momentum is from Planet to the Satellite and latter is launched on an expanding spiral path towards the outer Geo-Synchronous Orbit.

Initially because of conservation of energy the Satellite experiences a powerful outward push which we all GRAVITATIONAL RUNAWAY or SLING-

SHOT EFFECT [Dukla, cacioppo & Gangopadhyaya 2004, Jones 2005, Epstein 2005, Cook 2005]. Subsequently the increasing value of LOM/LOD as a function of orbital radii leads to increasing tidal frictional dissipation and the maximum velocity of recession starts decreasing unit it becomes zero at the outer Geo-Synchronous Orbit. The radial acceleration profile and the radial velocity profile are shown in Figure (1) and (2). As seen from Fig. (1), the sling-shot effect is impulsive in nature. Once the sling-shot ends, the satellite is let to go free and the Satellite coasts on its own along an expanding spiral orbit until it reaches the outer Geo-Synchronous Orbit.

This general theory is equally applicable to Artificial Man-Made Satellites excepting m is negligible in man made satellites..

For Artificial Satellites :
m = 0 and M = 5.9742E24 kg;
Therefore
F = 0 and $E = \{C\sqrt{(GM)}\}/(J_{spin})_E$ ;
$C = 80.2 \times 10^{36} kg-m^2$; $(J_{spin})_E = C\omega = C \times (2\pi)/T$ ;
Therefore
$E = \{T\sqrt{(GM)}\}/(2\pi)$
Substituting the above values in Eq. (1.2) :
$Ea^{1.5} = \omega_E/\Omega_L = LOM/LOD$
For Geosynchronism :
$Ea^{1.5} = \omega_E/\Omega_L = LOM/LOD = 1$
Therefore
Geo-synchronous Orbit $= a^* = [(GMT^2)/(2\pi)^2]^{1/3}$ ;
Where T = Sidereal Day = 86164 seconds
Substituting the numerical values :

Geo-synchronous Orbit $= a^* = 42,165$ km. with respect to the centre of the Earth.

Therefore Geo-synchronous Orbit with respect to the surface of the Earth is :
$a^{**} = (42,165 - 6,378) = 35,787 Km \approx 36,000 km$.

The second Geo-synchronous Orbit is at infinity and time constant of evolution is infinite. Therefore in spite of the fact that the inner Geo-Synchronous Orbit is unstable the artificial Satellite remains stay put in the inner Geo-Synchronous Orbit because it never gets a chance to evolve out of it. Geo-Synchronous Orbits will be referred to as Clarke's Orbit in memory of Sir Arthur C. Clarke(1917-2008).

## 2. THE BENCHMARKS ADOPTED FOR CHECKING AND TESTING EARTH-MOON THEORETICAL SYSTEM.

In the present analysis of Earth-Moon system, for mathematical simplicity, the system is considered to be a two body rotating system throughout is evolutionary history of 4.53Gy which now is 4.467Gy according to revised data..

At 4.56Gya, the accretion of the slowly rotating Solar Nebula led to the formation of Sun and its eight planets in the plane of disc of accretion [Maddox 1994]. The age of the metallic core of Earth indicates that in very early phase of Earth formation, a global melting took place leading to magmatic distillation and differentiation into a metallic core and a basaltic-granitic mantle. The date of formation of this metallic core has been authoritatively placed at 4.53 billion years B.P (revised date is 4.467Gy). [Cameron 2002, Kerr 2002]. Since this global melting could not have taken place merely by accretion or due to radioactivity, hence it is widely held that the Giant Impact between Earth and Mars-sized planetismal occurring around this juncture of 4.53 billion years B.P. led to the global melting and the subsequent distillation and differentiation of Earth into core and mantle. The impact generated debris almost simultaneously accreted into Moon at an orbital radius of Roche's limit = $3R_{Earth}$ [Ida, Shigura 1997]. Based on this scenario the first priary benchmark of 4.53 billion years B.P. is adopted. It is this tie instant at which the tidal interact on between Earth and Moon started at an orbital radius of $2.46R_{Earth} = 15,700 Km$.

Moon was possibly formed by accretion from a circumterrestrial disk or debris generated by a giant impact on Earth [Ida, Shigura 1997] about 4.53 billion years ago [Cameron 2002, Kerr 2002].

In any Planet-Satellite system there are two Geo-Synchronous Orbits-the inner one $(a_{G1})$ and the outer one $(a_{G2})$. So does our Earth-Moon system have two Geo-Synchronous orbits : one at $a_{G1} = 1.46257567 \times 10^7 m$ and the second at $a_{G2} = 5.52887898 \times 10^8 m$.

The Roche Limit is defined as :
$a_R = 2.456(\rho E/\rho L)^{1/3} R_E \sim 2.9 R_E = 18.4966066 \times 10^6 m$
$\rho E$ = density of Earth = 5.5 gms/c.c.,
$\rho L$ = density of Moon = 3.34 gms/c.c.,
and Roche Zone is defined within the range;
$0.8 - 1.35 a_R$ or $2.32 - 3.915 R_E$ i.e. within $14.7972848 \times 10^6$ to $24.9704181 \times 10^6 m$ range [Ida et al 1997].

This implies that impact generated debris will be prevented from accretion within $1.48 \times 10^7$ and those in $1.48 \times 10^7$ to $2.5 \times 10^7 m$ range also known as transitional zone will experience limited accretion growth whereas those lying beyond this zone will be unaffected by tidal forces. It is a happy coincidence that the Roche zone lies just beyond the inner Geo-Synchronous orbit of the Earth-Moon System. This implies that if accretional criteria of Canup & Esposito [1996] is satisfied along with the impact velocity condition that is the rebound velocity should be smaller than the mutual surface escape velocity then merged body formation of Moon starts within the Roche zone. The accreted Moon gradually migrates outward sweeping the remnant debris.

Hence the first primary benchmark is the inception of merged body Moon at $1.5 \times 10^7$ m from the center of Earth.

The second primary benchmark chosen for this work is the present mean Solar day length, $T_{Ep} = 24$ hours = 86400 seconds with present lunar orbital radius experimental measure being $r_{Lp} = 3.844 \times 10^8$ meters. All Earth-Moon system parameters used in this paper along with their symbols are given in Table (1) [Zeik/Gauntand ?].

There are fourteen other primary benchmarks against which the theoretical Earth-Moon system's evolution will be tested and verified.

John West Wells through the study of daily and annual bands of Coral fossils and other marine creaturs in bygone era has obtained ten benchmark [Wells 1963, Wells 1966]. These benchmarks are tabulated in Table (8):

**Table 8. The Observed lod based on the study of Coral Fossils.**

| T (yrs B.P.) | T* (yrs after the Giant Impact) | Length of obs. Solar Day $T_E^*$ (hrs) |
|---|---|---|
| 65 Ma | 4.46456G | 23.627 |
| 135 Ma | 4.39456G | 23.25 |
| 180 Ma | 4.34956G | 23.0074 |
| 230 Ma | 4.29956G | 22.7684 |
| 280 Ma | 4.24956G | 22.4765 |
| 345 Ma | 4.18456G | 22.136 |
| 380 Ma | 4.14956G | 21.9 |
| 405 Ma | 4.12456G | 21.8 |
| 500 Ma | 4.02956G | 21.27 |
| 600 Ma | 3.92956 G | 20.674 |

Leschiuta & Tavella [Leschitua & Tavella 2001] have given the estimate of the synodic month. From the synodic month we can estimate the length of the Solar Day as given in Appendix [XIII]. The results are tabulated in Table (9). [Leschitua & Tavella 2001 based on the study of marine creature fossils]

**Table 9. Observed Synodic Month**

| T (yrs. B.P.) | T* (yrs. After the Giant Impact) | Observed Synodic Month (modern days) | Estimated Solar Day (hrs). |
|---|---|---|---|
| 900 Ma (Proterozoic) | 3.62956G | 25.0 | 19.2 |
| 600Ma (Proterozoic) | 3.92956G | 26.2 | 20.7 |
| 300Ma (Carboniferous) | 4.22956G | 28.7 | 22.3 |
| 0 (Neozoic) | 4.52956G | 29.5 | 24 |

Kaula & Harris [1975] have determined the synodic month through the studies of marine creatures. The results are tabulated in Table (10)

**Table 10. Observed Synodic Month (Kaula & Harris 1975) based on the studies of Marine creatures.**

| T (yrs. B.P.) | T* (yrs. After the Giant Impact) | Observed Synodic Month (modern days) | Estimated Solar Day (hrs). |
|---|---|---|---|
| 45 Ma | 4.48456G | 29.1 | 23.566 |
| 2.8 Ga | 1.72956G | 17 | 13.67 (with modern C) <br> 16.86 <br> (with C = 9.99$\times$ $10^{37} kg-m^2$) |

Walker & Zahnle [1986] through the study of Australian Banded Iron Formation also known as Cyclic banded iron formation of the Weeli Wolli Formation arrived at the Lunar Nodal Period which is also known as Saros Cycle. Based on this they arrived at Lunar Orbital Radius of $52R_E = 3.280728\times 10^8 m$ at 2.45Ga. The estimate of Solar Day in that epoch as given by Williams [2000] does not appear to be correct. From the Lengthening of Day Curve as arrived by Observed Data, the estimate of lod is 14.5 hours.

One benchmark has been provided by Charles P. Sonnett et al through the study of tidalies in ancient canals and estuaries [Sonett & Chan 1998 ]. He gives an estimate of $T_{E4} = 18.9$ hours mean solar day length at about 900 million years B.P. in Proterozoic Eon, pre-Cambrian Age.

**3. MATHEMATICAL FORMULATION OF EARTH-MOON SYSTEM AS A NON-LINEAR DYNAMIC SYSTEM.**

**Equation of Motion :**

Centripetal force on Moon = Centrifugal force due to orbiting Moon $- \varepsilon'$.
Therefore
Centripetal acceleration = centrifugal acceleration $- \varepsilon$. (3)

Where $\varepsilon$ is residual inward acceleration because of which the outward radial velocity is being slowed until it will be zero.

$\varepsilon = 2.220422E-6 \, m/sec^2$.

This implies an inward residual acceleration, which is retarding the outward radial velocity of Moon. As Moon spirals out, Earth's spin period (24 hours presently) and Lunar orbital period (27.3 solar days) differential decreases hence the retarding tidal drag on Earth as well as the residual inward acceleration of Moon, both decrease.

Therefore the residual inward acceleration is assumed to be :

$$\varepsilon = k[1 - r_L/a_{G2}] \qquad (4)$$

This equation implies a decrease in Moon's Recession Velocity with a monotonic decrease in the deceleration rate as given by Eq. (1.4), where $a_{G2}$ is the terminal point of the expanding lunar orbital radius and is also the outer Geo-Synchronous Lunar Orbital Radius $= a_{G2} = 5.5288789E8 \, m$.

It is logical to assume that Tidal Torque is a function of the differential of the angular periods of Moon and Earth on one side (i.e. of LOM/LOD) and that of the structure factor of Earth on the other side.

If LOM/LOD = 1, Tidal Torque is zero and Moon is in Geo-Synchronous Orbit.

If LOM/LOD > 1, Tidal Torque is slowing down the spin of Earth leading to transfer of angular momentum from Earth to Moon and pushing Moon in outward expanding spiral orbit.

If LOM/LOD < 1, Tidal Torque is spinning Earth faster leading to transfer of angular momentum from Moon to Earth and thereby Moon falling into a collapsing spiral orbit doomed to be crashed into Earth.

Therefore Lunar tidal torque on Earth is assumed to be :

$$\text{Tidal torque} = [(K/X^M)[(lom/lod)-1] \qquad (5)$$

Structure factor $(K/X^M)$ and exponent M are to be determined from the present value of the Tidal Torque and the age of Moon which has been assumed to be 4.53 billion years.

Based on Lunar Laser Ranging Experiments [Walker & Zahnle 1986, Yen 1999], Moon's Velocity of

$$dX/dt = 3.8 \, cm/yr \qquad (6)$$

From the Appendix (XIV), Eq.(XIV.29):

$$_L\dot{X}^2 = \sqrt{GE(1+m/E)}[X^{1/2} - AX^{5/2} + AX^{7/2}/Y]$$

$$J_{orb} = I \times _L\dot{} = \{[m/(1+m/E)]X^2\}_L\dot{}$$
$$= (mE/(m+E))\sqrt{GE(1+m/E)}[X^{1/2} - AX^{5/2} + AX^{7/2}/Y]$$

$$\mathbf{J_{orb} = (mE/(m+E))\sqrt{GE(1+m/E)}[X^{1/2} - AX^{5/2} + AX^{7/2}/Y]} \tag{7}$$

X = Lunar Orbital Radius;
Y = $a_{G2}$.
$(mE/(m+E))\sqrt{GE(1+m/E)} = B' = 1.457917826 \times 10^{30}$ Kg-m$^{3/2}$/sec.

Therefore
$$\mathbf{J_{orb} = B'[X^{1/2} - AX^{5/2} + AX^{7/2}/Y]} \tag{8}$$

Where $B' = 1.457917826 \times 10^{30} Kg - m^{3/2}/\sec$
$A = 9.142570518 \times 10^{-21} m^{-2}$
$Y = 5.528878911 \times 10^8 m$
$\tau = dJ_{orb}/dt$

Differentiating Eq (8) with respect to time we get the magnitude of the torque:
$$\tau = B'[1/(2X^{1/2}) - 5AX^{3/2}/2 + 7AX^{5/2}/(2Y)]dX/dt \tag{9}$$

From the present Velocity of Recession, the present Tidal Torque is calculated from Eq. (9) to be :
$$(\tau)_p = 4.47633261 \times 10^{16} \text{ N-m} \tag{10}$$

Equating Eq. (5) and (10),
$$(\tau)_p = (K/X^M)[(LOM/LOD) - 1] = 4.47633261 \times 10^{16} N - m$$
$$K = ((\tau)_p \times X^M)/[(LOM/LOD) - 1] \tag{11}$$

Therefore taking the present epoch parameters in Eq. (11) :
$$K = (4.47633261E16 N - m \times X^M)/[27.3 - 1] = 1.81557 \times 10^{33} N - m \quad \& \quad M = 2.1 \tag{12}$$

### 1.3.1. CALCULATION OF LOM/LOD

Estimate of Sidereal Period of Lunar Orbit From Appendix (XIV) :
$$LOM = (T_m)_{sidereal} = (2\pi)/[B'Y'/X^2] \tag{13}$$

Where $Y' = [X^{1/2} - AX^{5/2} + AX^{7/2}/Y]$
$B = 2.00843330 \times 10^7 \ m^{3/2}/sec;$
$(r_L) = X;$

Estimate of Solar Day length from Appendix (XIV),

$$LOD = T_E = ((2\pi C)/3600/[J_T - (B/X^2)(D' + E'X^2)Y'] \tag{14}$$

Where $J_T = 3.44048888 \times 10^{34} \ Kg-m^2/sec,$
$\quad$ C = moment of inertia of Earth around spin axis
$\quad\quad = 8.02 \times 10^{37} \ Kg-m^2,$
$\quad B = 2.00843330 \times 10^7 \ m^{3/2}/sec$
$\quad$ X = Lunar orbital radius of the given epoch,
$\quad D' = 8.87859824 \times 10^{34} \ Kg-m^2,$
$\quad E' = 7.25898053 \times 10^{22} \ Kg,$
$\quad Y' = (X^{1/2} - AX^{5/2} + AX^{7/2}/Y),$
$\quad A = 9.14257051 \times 10^{-22} m^{-2},$
$\quad Y = 5.5288789 \times 10^8 m$

Therefore $LOM/LOD = T_m/T_E = (1/C)[(MX^2)/(YB) - D' - E'X^2] \quad (15)$

## 1.3.2. THE BOUNDARY EQUATION FOR DETERMINING THE INNER AND OUTER GEO-SYNCHRONOUS ORBITS OF EARTH-MOON SYSTEM.

The Boundary Equation which leads to the two GEOSYNCHRONOUS ORBITS $a_{G1}$ and $a_{G2}$ is defined as :

$$LOM/LOD - 1 = 0 \tag{16}$$

The roots of Eq. (1.16) are $a_{G1}$ and $a_{G2}$.
Substituting Eq. (1.15) in Eq. (1.16) we get :

$$(1/C)[(MX^2)/(YB) - D' - E'X^2] = 1 \tag{17}$$

This has two roots :
Inner Geo-Synchronous Orbital Radius = $a_{G1} = 1.46257566 \times 10^7 m$ (17.a)
Outer Geo-Synchronous Orbital Radius = $a_{G2} = 5.52887891 \times 10^8 m$.

Eq. (15) conveys a lot of information about the evolving spiral orbit of Moon. In the following Table (11) we clearly see the complete behavior of Moon.

Table (11). Tabulation of LOM/LOD with Lunar Orbital Radius (X).

| X(*10⁷)m | 0.7 | 1 | 1.4 | 1.46 | $a_{G1}$ | 1.5 | 1.6 | 1.7 | 1.8 | 1.9 | 2.0 | 3.0 |
|---|---|---|---|---|---|---|---|---|---|---|---|---|

| Lom/lod | 0.350 | 0.583 | 0.94 | 0.997 | 1.00 | 1.036 | 1.134 | 1.23 | 1.336 | 1.44 | 1.54 | 2.69 |

| 5 | 10 | 20 | 30 | 31 | 31.5 | 32.0 | 38.44 | 40 | 50 | $a_{G2}$ | 56 |
|---|----|----|----|----|------|------|-------|----|----|----------|----|
| 5.288 | 12.3 | 24.22 | 29.567 | 29.644 | 29.6497 | 29.63 | 27.3 | 26.1 | 12.578 | 1.00 | -0.8 |

Inspecting the Table (11) we see that Moon is permitted only from $a_{G1}$ to $a_{G2}$. Moon is never permitted to go beyond $a_{G2}$ because an untenable value of LOM/LOD occurs.

As explained before, Moon is born as a merged body beyond $a_{G1}$ therefore it is launched on an outward spiral path. On reaching the outer Geo-Synchronous Orbit it is deflected back on an inward spiral path.

If a satellite is born within the inner Geo-Synchronous Orbit then it is destined to go on an inward spiral path and eventually crash into the parent planet as is the case for Phobos and all Hot Jupiters which are in scorchingly tight orbit and which are doomed to crash into their respective host stars..

Substituting the numerical values of the different parameters in Eq. (15), we obtain the plot of LOM/LOD vs Lunar Orbital Radius.

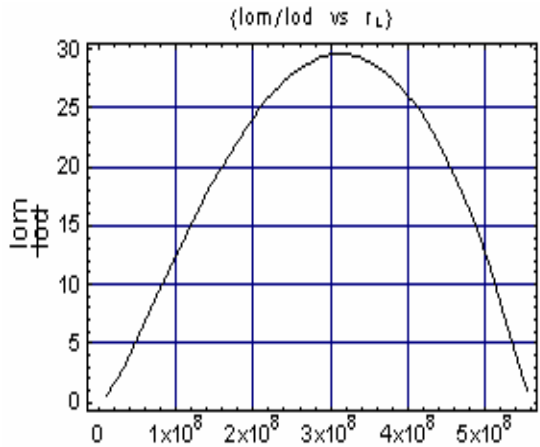

**Figure 3. The Profile of LOM/LOD vs rL.**

The Table (11) and the Fig. (3) clearly point out that the empirical form of the tidal torque should be taken as Eq. (5) and with Eq. (14) substituted in Eq. (5) we get :

$$\text{Tidal Torque} = [(K/X^M)[(lom/lod)-1]$$
$$= (K/X^M)[(1/C)\{(MX^2)/(Y'B) - D' - E'X^2\}-1] \qquad (18)$$

The algebric sign of Eq. (18) gives the direction of angular momentum transfer hence it describes outward expanding instability or inward collapsing instability.

Contemporary Solar Day in hours is calculated at $a_{G1}$ and $a_{G2}$ from Eq. (14)

$T_E$ at Inner Geo-Synchronous orbit is 4.86 hours.

$T_E$ at Outer Geo-Synchronous Orbit is 47.07 Modern Solar Days. This is in correspondence with Darwin's analysis.

### 1.3.3. TIME INTEGRAL EQUATION OF MOON'S SPIRAL TRAJECTORY.

Equating Eq. (5) and (9) we obtain the following :

$$B\ [1/(2X^{1/2}) - 5AX^{3/2}/2 + 7AX^{5/2}/(2Y)]dX/dt = [(K/X^M)][(lom/lod)-1] \quad (19)$$

Rearranging the terms, Moon's Recession velocity is obtained :

$$dX/dt = [(K/X^M)[(1/C)\{(MX^2)/(Y\ B) - D - E\ X^2\} - 1]] / [B\ /2[1/X^{1/2} - 5AX^{3/2} + 7AX^{5/2}/Y]] \quad (20)$$

Here $B' = 1.4579 \times 10^{30}\ (Kg - m^{3/2})/sec$.

Eq. (20) gives Recession Velocity in m/sec.

Let $F = B'/(2 * 31.556908 \times 10^6\ sec/SolarYear)$

$= 2.30998 \times 10^{22}\ ((Kg - m^{3/2} - SolarYear)/sec^2)$

Therefore,

$$dX/dt = [(K/X^M)\ [(1/C)\{(MX^2)/(Y'B) - D' - E'X^2\} - 1]] / [F\ [1/X^{1/2} - 5AX^{1/2} - 5AX^{3/2} + 7AX^{5/2}/Y]] \quad (21)$$

Eq. (21) gives Moon's Recession Velocity in m/year.

Substituting the numerical values for the parameters in Eq. (21), the Profile of Recession Velocity is plotted :

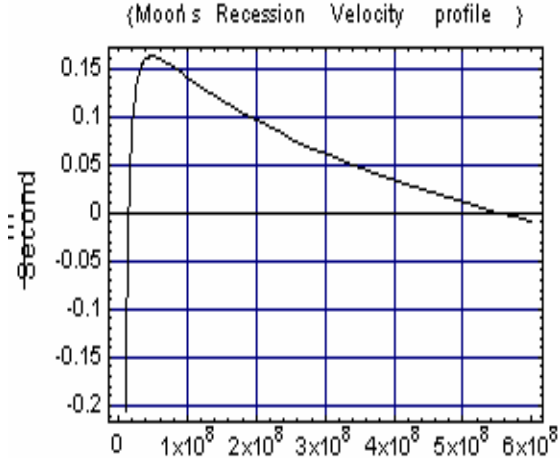

**Figure 4. The profile of Moon's Velocity (m/yr) of Recession vs rL(m)**

**Table (12). Moon's Velocity of Recession vs rL**

| $rL (*10^7)m$ | 1.0 | 1.1 | 1.2 | 1.3 | 1.4 | 1.46257569 |
|---|---|---|---|---|---|---|
| dX/dt (cm/yr) | -20.6 | -14.11 | -9.02 | -4.9837 | | 0 |

| $rL (*10^8)m$ | 1.844 | 2.844 | 3.844 | 4.844 | 5.844 |
|---|---|---|---|---|---|
| dX/dt (cm/yr) | 10.171 | 6.62 | 3.8 | 1.43 | -0.006174 |

As is clear from the Table (12) as well as from the Figure (4), Velocity of Recession is positive within the region bounded by $a_{G1}$ and $a_{G2}$. In the region beyond $a_{G2}$ the velocity of Recession is negative as well as in the region within $a_{G1}$ also it negative as already discussed.

The Figure and Table also give the profile of Velocity of Recession required to achieve an evolutionary span of 4.53 billion years. It is clear that Moon has been receding much faster in the past and is monotonically decreasing as it recedes as it recedes farther and farther from the parent Planet until it reaches the Outer Geo Synchronous orbit where radial velocity becomes zero and subsequently it starts spiraling inward.

Now that we have a revised Age of 4.467Gyrs, exponent M and structure factor ($K/X^M$) have to be suitably modified so that the outward spiral journey from 15,700 km to the present lunar orbital position 384,400 km takes the time of 4.467 Gyrs.

Rearranging and integrating both sides of Eq. (21) we get,

∫ dt(limits,$t_0$,$t_p$)

$= B/2 \int [(1/X^{1/2} - 5AX^{3/2} + 7AX^{5/2}/Y)/((K/X^{2.1})[(1/C)\{(MX^2)/(Y B) - D - E X^2\} - 1])]dX(\text{limits}, X_0, X_p)$

$(t_p - t_0)/(31.5569088E6 \text{sec/solar yr.}) =$
$B/(2*31.5569088E6) \int [[(1/X^{1/2} - 5AX^{3/2} + 7AX^{5/2}/Y)/((K/X^{2.1})[(1/C)\{(MX^2)/(Y B) - D - E X^2\} - 1])]dX, X_0, X_p]$
(22)

where $t_0 = 0.03 b$. Yrs = instant of Giant Impact after the birth of our Solar Nebula.

$(t_p) = 4.56 b$. Yrs = the present age of Earth.

$X_0 = 15{,}700 km = 1.57 \times 10^7 m \cong$ the distance from the center of Earth at which the Giant Impact generated debris accreted into partial Moon and significant tidal interaction began between Earth and Moon.

$X_p = 3.844 \times 10^8 m =$ the present orbital radius of Moon.

By determining the above integral between $X_0$ and the present orbital radius, the present age of Moon is determined which by observation has been determined to be 4.53Gyrs.

Expressing Eq. (22) in terms of Fx-570W constants,

$[(t_p - t_0)/31.5569088E6] \text{solar years} =$
$F \int [[(1/X^{1/2} - 5AX^{3/2} + 7AX^{5/2}/Y)/((K/X^{2.1})[(1/C)\{(MX^2)/(Y B) - D - E X^2\} - 1])]dX, X_0, X_p]$
(23)

When $X_0 = 1.57 \times 10^7 m$, then the evolutionary period is $4.52956 \times 10^9$ yrs. That is it takes($4.529557176695456 09\text{`*^9}$ yrs) = 4.52956 billion years evolutionary time span to cover the outward spiral journey from 15,700 Km to the present 3,84,400 Km Lunar Orbital Radius

If the integration is carried out from 10,000Km to 14,500 Km we get a negative time($-1.22772988112015557\text{`*^8}$) yrs which implies that journey started from 14,500Km and ended at 10,000 Km in the time span of 122Myrs.

Eq. (23) can be interpolated to any epoch in the past or to any future epoch. If the orbital radius in any epoch is known then by integrating between the limits $X_0$ (15,700Km) and $X_n$ (Orbital radius of Moon in the unknown epoch), the chronological date of the unknown epoch can be pinned down. If the epoch is known then through several iterations of Eq. (1.23) the correct lunar orbital radius can be arrived at.

### (1.3.4) DETERMINATION OF LENGTHENING OF DAY CURVE

The epochs for which observed data or estimated data of Solar Day is available for those epochs the Lunar Orbital Radius ($r_L$) is accurately estimated and using the theoretical values of ($r_L$), the theoretical value of Solar Day Length is calculated and tabulated alongside the observed values of Solar Day. All these data are tabulated in Table (13)

**Table (13). A comparative study of the Observed and Theoretical Values of the Solar Day Length in different Epochs.**

| t (yrs. B.P) | t* (Gyrs.) after the Giant Impact | $r_L$ ($*10^8 m$) | $T_E$ (hrs) (Theoretical) | $T_E^*$ (hrs) (Observed) |
|---|---|---|---|---|
| 0 | 4.52956G | 3.844 | 24 | 24 |
| 65Ma | 4.46456G | 3.819095 | 23.62 | 23.627 |
| 135Ma | 4.39456G | 3.7918 | 23.2247 | 23.25 |
| 180Ma | 4.34956G | 3.7739 | 22.97 | 23.0074 |
| 230Ma | 4.29956 | 3.75396 | 22.69 | 22.7684 |
| 280Ma | 4.24956 | 3.73366 | 22.4125 | 22.4765 |
| 300Ma | 4.22956 | 3.72547 | 22.3 | 22.3 |
| 345Ma | 4.18456 | 3.70688 | 22.055 | 22.136 |
| 380Ma | 4.14956 | 3.69227 | 21.86 | 21.9 |
| 405Ma | 4.1256 | 3.68175 | 21.73 | 21.8 |
| 500Ma | 4.02956 | 3.64114 | 21.22 | 21.27 |
| 600Ma | 3.92956 | 3.59726 | 20.69 | 20.674 |
| 900Ma | 3.62956 | 3.45832 | 19.16 | 18.9 |
| 2.45Ga | 2.07956 | 2.51298 | 12.387 | 14.5* |
| 2.8Ga | 1.72956 | 2.22665 | 11.06 | 13.859 |

*From the graph of Observed Values of Length of Solar Day.

The graph giving the observed values of Solar Day vs t(yrs.) after the Giant Impact is obtained :

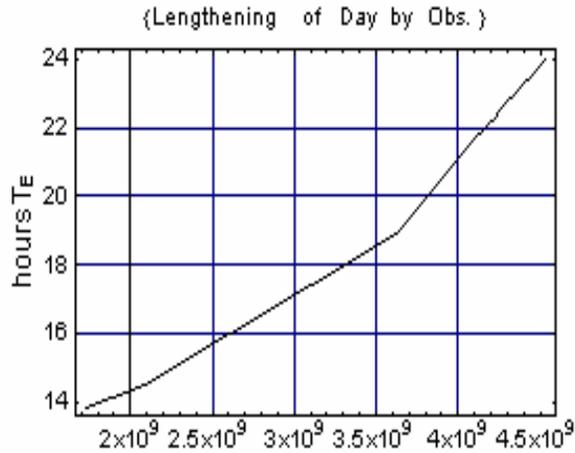

**Figure 5. Lengthening of Day Curve by Observation**

**1.3.4.i. THEORETICAL FORMULATION OF LENGTHENING OF DAY CURVE ASSUMING CONSTANT MOMENT OF INERTIA C.**

To obtain the theoretical curve for the lengthening of day we need Lunar Orbital Expansion Curve. By trial and error we arrive at the following expression which accurately fits the theoretical values of $r_L$ tabulated in Table (13).

Lunar Orbital Radius = $r_L$ =
5.52887891E8 – (5.5288789E8 – 0.157E8) Exp [-t/(2E9)] – (1.6E8) Exp [-t/(14E9) + (1.6E8)]
Exp [-t/(1.15E9)]     (24)

The Expansion Curve is obtained based on Eq. (24) in Fig (6)

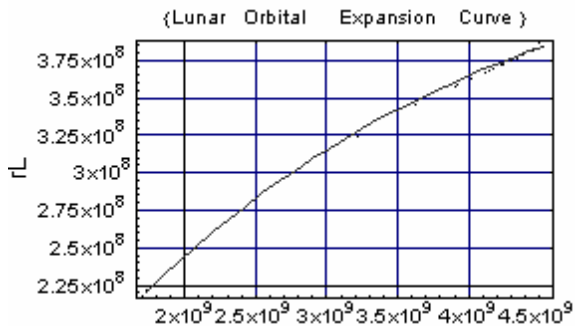

**Figure 6. Lunar Orbital Radius Expansion Curve based on Eq. (24)**

The Lunar Orbital Radius Expansion Curve based on Table (13) is obtained in Fig (7) :

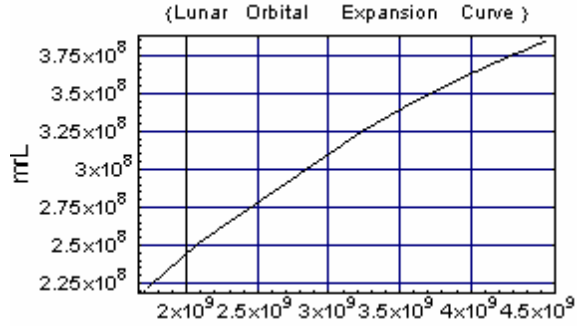

**Figure 7. Lunar Orbital Expansion Curve based on Table (13)**

Next through SHOW command we compare the two curves-one obtained through the empirical Eq. (24) and the other obtained based on Table (13).

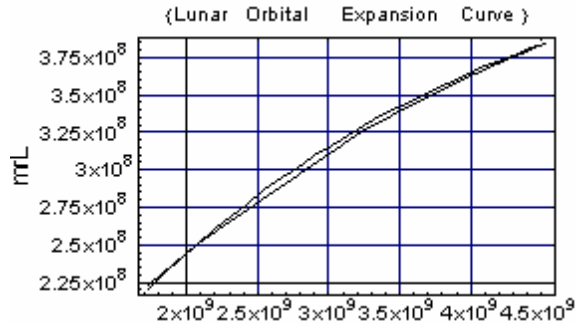

**Figure 8. Superposition of the two curves, one by Eq. (24) and the other based on the Table (13).**

As seen from the superposition of the two curves, the empirical Eq. (24) accurately represents the evolving Lunar Orbital Radii.

By substituting Eq. (24) in Eq. (14), we get LOD as a function of time. The plot of LOD vs time is given in Fig. (9).

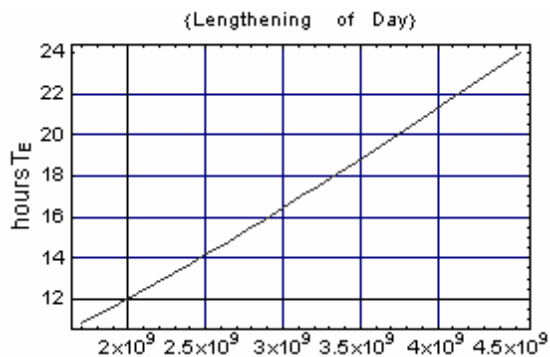

**Figure 9. Lengthening of day curve by Theory assuming constant C.**

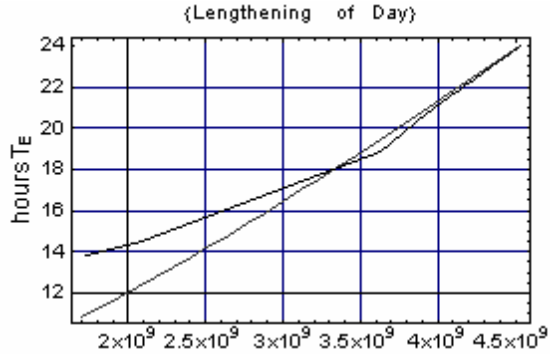

**Figure 10. Superposition of the two curves, one by observation and the other by calculation, with constant C.**

As seen from the superposition of the two lengthening of day curves, there is remarkable match between Observation and Theory in the recent past after the Pre-Cambrian Explosion of plant and animal life but in the remote past, particularly in early Archean Eon, Earth seems to be spinning much slower than predicted by theory. This implies that rotational inertia was much higher than what has been assumed in this analysis. In fact there are evidence to show that early Earth was much less stratified as compared to modern Earth. It was more like Venus [Allegre, Calnde 1994, Taylor, Rose & Mclennan 1996].

In Eq. (13) as well as in Eq. (24) C, the Principal Moment of Inertia, has been assumed to be constant whereas infact it was evolving since the Giant Impact [Runcorn 1966].

In the first phase of planet formation, Earth was an undifferentiated mass of gas, rocks and metals much like Venus. At the point of Giant Impact, the impactor caused a massive heating which led to melting and magmatic formation of total Earth. The heavier metals, Iron and Nickel, formed dense blobs and then settled down to form the metallic core and lighter rocky materials formed the mantle. The mantle consisted of Basalt and Sodium rich Granite.

Due to Giant Impact, Earth gained extra angular momentum. This led to a very short spin period of 6 hours. It has been calculated that the oblateness at the inception must have been 1% [Appendix (VIII), Kamble 1966] whereas the modern oblateness is 0.3%. Taking these two factors into account C of Earth must have been much higher than the modern value of $8.02 \times 10^{37}$ kg–m$^2$. In this paper the early C has been taken as $9.9 \times 10^{37}$ kg–m$^2$.

After Achaean Eon the general cooling of Earth over a period of 2 billion years led to slower plate-tectonic movement. The 100 continental-oceanic plates coalesced into 12 plates initially and into 13 plates subsequently. The slower plate tectonic engine led to deep recycling of the continental crust and hence to complete magmatic distillation and differentiation of the internal structure into multi-layered onion like structure. Thus at the boundary of Archean Eon and Proterozoic Eon a definite transition occurred in the internal structure.

Before this boundary, the mantle and the outer crust was less differentiated. It was composed of a mixture of Basalt and Sodium-rich granite. After this boundary a slower plate-tectonic dynamo helped create the onion-like internal structure with sharply differentiated basaltic mantle and potassium-rich granitic crust. This highly heterogenous internal structure and less oblate geometry leads to the modern value of C equal to $8.02 \times 10^{37}$ Kg–m$^2$.

The form the evolving C is as follows :
f[(t-2E9)_]:=If[(t-2E9)>0,1,0]
{9.9E37-(9.9E37-8.02E37)}{1-Exp[-t/16E9]}-f[(t-2E9)_](1.4E37){1-Exp[-t/(0.5E9)]}}  (25)

Here f[(t-2E9)_] is defined as a step function which is 0 before 2 billion years and is Unity 2 billion years and at greater times.

The profile of evolution of C with time is obtained in Fig. (11) :

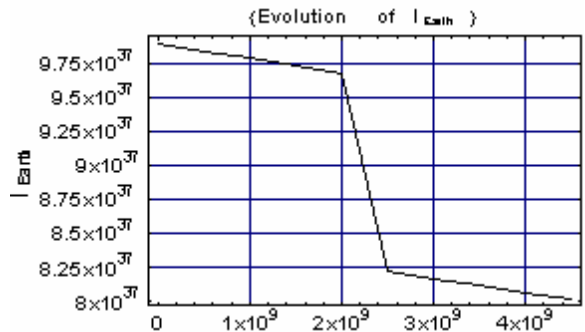

**Figure 11. The profile of the assumed evolving C.**

**1.3.4.ii. THEORETICAL FORMULATION OF LENGTHENING OF DAY CURVE ASSUMING EVOLVING MOMENT OF INERTIA.**

By substituting Eq. (25) in LOD expression as a function of time we get the desired plot. The Plot is given in Fig. (12).

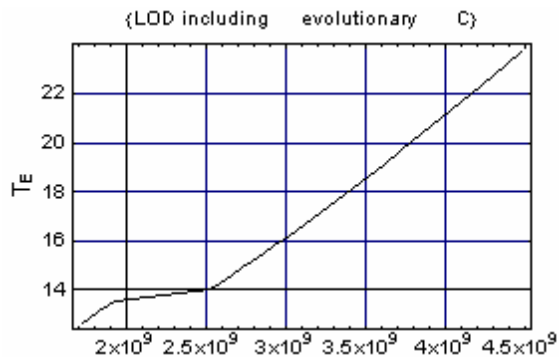

**Figure 12 Theoretical lengthening of day curve with evolving C.**

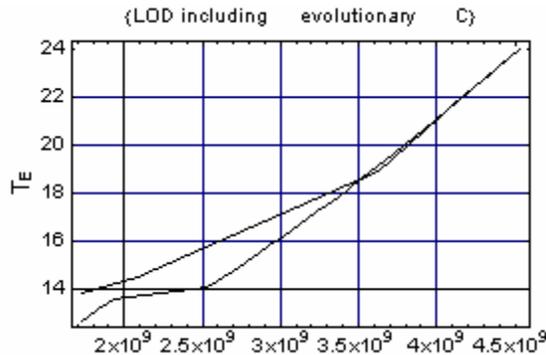

**Figure 13. Superposition of the observed curve and theoretical lengthening of day curve with evolving C.**

As can be seen in Fig. (13), there is a much closer fit except for a large deviation at 2.5Gyrs after the Giant impact. This is due to step change in Moment of Inertia, C, at 2Gyrs after the Giant Impact. It would have been more realistic to assume a gradual change in C at the boundary of Archean and Proterozoic Eon. This correction will be made in a sequel paper.

**1.4. GEOPHYSICAL AND GEOGRAPHICAL IMPLICATIONS OF THE SCATTER OF THE OBSERVED DATA AROUND THE THEORETICAL CURVE.**

Kuala and Harris made the earliest observation of the synodic month in Archean Eon about 2.8Ga from which the Mean Solar Day is deduced to be 13.859 hrs. Here the observed Mean Solar Day of 13.859 hrs is much longer than the theoretical value of 11.42 hrs. This deviation can be explained in terms of the internal structure and geometrical shape of the Earth as already discussed in Section (1.3.4.i). As seen by assuming an evolving C as given by Eq. (25) and Fig. (13) we obtain a much closer fit with the observed curve.

In the Pre-Cambrian Era lod observed is shorter than the theoretical value in the period 900Ma to 600Ma. This could be due to post-glacial rebound [Dickey, Marcus 2002, Yan, Zhour et al 2002] but we have no record of an Ice Age setting earlier than 900Ma. Therefore post glacial rebound is discounted. It is more likely due to even distribution of continental plates around the globe earlier and ,in late Pre-Cambrian period, Earth's continental plates were coalesced into a single super continent between 30 degree North and 60 degree South [Allegre, Calnde & Schneider 1994, McMenamin 1987]. This led to lighter rotational inertia leading to faster rotation as observed.

In Cambrian Era and later observed values are equal or only slightly larger than the theoretical values.

In these periods the continental plates are continuously redistributing and the Ice Ages are setting and thawing. Since 400Ma mountain building and erosion started. Prior to Himalayan, Alps, Rocky and Andes, there were many older ranges which were built and subsequently eroded. The global wind pattern interaction with the mountain ranges will also bring short term changes in lod. A combination of these factors have compensated to give less than 2% deviation between observed and theoretical.

The main conclusion of the above discussion is that the deviation of observed Mean Solar Day with respect to the theoretical curve as plotted in Figure (13) could be a reliable indicator of geographical and geophysical condition.

### 1.5. PROPOSAL OF EARLY WARNING SYSTEM FOR IMPENDING EARTHWQUAKE AND SUDDEN VOLCANIC ERUPTIONS.

In the last section we have seen that the observed data in its respective age can be correctly interpreted and justified if second order effects are included in the theoretical analysis. These second order effects are the internal structure of Earth, drifting continental plates, setting and thawing of Ice-Age, interaction between global wind pattern and emerging mountain ranges, the effect of oceanic currents and core-mantle magnetic coupling [Leschiutta & Tavella 2001, Yan, Zhour et al 2002, Stephenson 2003, Ohta et al 2008, Appendix IX]

This implies that the internal geodynamics and external geography (Ice Age, El-Nino and Wind pattern interaction with mountain range) will be reflected in Solar day plot with time and in outward spiral path of lunar orbital radius. The theoretical graphs will be plotted using Equation (14) & (23). Observational curves will be plotted based on the data obtained for the Solar Day and Lunar Orbital Radius on day to day basis. These two data are available from Greenwitch Time-Keeper and Kitt's Peak National Observatory, Tucson, Arizona. Solar Day on daily basis is being measured using Cesium-13 Atomic Clock [Cossono ?] and Lunar Orbital Radius is being being measured using Lunar Laser Ranging Experiment at Kitt's Peak National Observatory as well as Macdonald Laboratory. If on the same time scale the seismic tremors of all magnitudes, minor and major, are recorded and Chaos theory applied then a correlation between the scatter of the observed data and the tremors may emerge [Buchanan 1997, Coontz 1998, Bak ?, Swiney 1996].

Scientists have found that Earth's crustal plates behave like a sand pile [Buchnan 1997, Coontz 1998]. The internal radioactivity and convective flow of heat from the inner core to the surface of Earth is causing the Plate Tectonic dynamo. Presently the entire Earth crust is broken into thirteen oceanic and continental plates. These plates are colliding along certain lines of fault and are coming apart along mid-ocean ridges. Where plates collide, one plate goes underneath the other. This is subduction zone. At the subduction zone oceanic plate ploughs deep under the continental plate. While subducting into Earth it becomes molten and on reaching the boundary of lower mantle and outer core, the magma or the molten lava rises up through weak portions of Earth. Thus constant recycling of crustal mass of Earth is taking place and by this mechanism the primary crust has been replaced by secondary crust and secondary crust has been replaced by tertiary crust [Allegre 1994]. This is directly responsible for sudden or continuous volcanic eruptions. But apart from head on collision of crustal plates there are sideway slipping and sticking. These lateral movements give rise to rupturing faults or premonitory creep. These rupturing faults behave like sand pile [Buchanan 1997].

What is sand pile theory ?

Sand pile theory states that as we add grains of sand to a sand pile, latter continues to build up in a stable fashion until it collapses. Again the grains of sand are added until collapse takes place. The building up process is stasis and the sand pile is a non-critical mass. But gradually sand-pile becomes a universal critical class. As soon as it becomes an universal critical class, it reaches self-organized criticality [Swiney 1996]. At this point the sand pile is delicately poised that there is no way to tell whether the next grain you add will drop quietly into place or trigger a landslide. Thus there are periods of stasis punctuated by Avalanches. The frequency of Avalanche is dependent on the size of the sand-pile. Avalanche is a kind of readjustment when energy E is released.

Let $N(E)$ = No. of avalanches of size E.

E = energy released during an avalanche.

According to Sand-Pile theory [Buchanan 1997, Coontz 1998]

$$N(E) = K/(E^p);$$

**Where K is an arbitary constant**
**p is the power law.**

This power law is the signature of chaos or the critical sand pile. In critical state all sand piles are self similar. That is no matter what the size is the sand piles look identical.

The rupturing faults can be modeled as sand piles. During stasis premonitory creeps take place. At avalanche this rupturing fault spreads. The length of the rupturing fault determines the magnitude of the Earthquake. When power law is applied to rupturing faults then we have Gutenburg-Richter law.

Our contention is that rupturing faults Earthquakes as well as subduction zone Earthquakes are chaotic and they directly perturb the polar moment of inertia of Earth. Hence scatter of observed data, after eliminating the periodic variations, around the theoretical curve will also be chaotic. Furthermore in chaotic systems an event can be predicted by its characteristic precursors.

If the contention above is correct the axially spinning but slowly and surely slowing down Earth and outward spiraling Moon can be utilized as a sensitive analytical Seismograph to predict an impending Earthquake and sudden volcanic eruptions.

**CONCLUSIONS**

In this study we have successfully been able to give the theoretical formulation for the lunar tidal drag on Earth and the subsequent slowing down of the axially spinning Earth and the expanding lunar orbit. It has been contended that the plotting of the observed data of the Solar day and that of the Lunar Orbital

Radius on the theoretical curves of the same could lead to an Early Warning System for impending Earthquakes and Sudden Volcanic Eruptions. This contention has been corroborated by a NASA Press Release (?) which states, "The successful observation of the Josephson Effect in superfluid helium-4 allows measurements of very small rotation change, enabling Scientists to measure very precisely how fast Earth rotates. Monitoring Earth's rotation speed could yield information on minute movement of tectonic plates, which may eventually help predict Earthquakes." This work lays down the theoretical foundation for developing a simulation software for plotting Moon's spiral trajectory in first phase. In the second phase we will be able to study the interaction between any Planet-Satellite pair and determine what kind of instability the given satellite is experiencing-inward collapsing spiral instability or outward expanding spiral instability.

**BIBLOGRAPHY**

| | |
|---|---|
| * | Alley,C.O., Bendu, P.L., Dicke, R.H., Faller,J.E., Franken, P.A., Plotkin, M.M., Wilkinson, D.T., " Optical Radar Using a Corner Reflector on the Moon", *Journal of Geophysical Research(all sections)*, 70, 2269,(1965) |
| * | Allegre, Calnde J. & Schneider, Stephen H. "The Evolution of Earth," *Scientific American*, October, 1994. |
| * | Bak, Per "*How Nature Works*", Oxford University Press. |
| * | Benz, W., Slattery, W. L. & Cameron, A. G. W. "The Origin of the Moon & the Single Impact Hypothesis (1)," *ICARUS*, 66, 515-535, 1986. |
| * | Benz, W., Slatery, W. L. & Cameron, A. G. W. "The Origin of the Moon & the Single Impact Hypothesis (3)," *ICARUS*, 71, 30-45, 1987. |
| * | Brandon,A., ' Planetary Science: A younger Moon', *Nature,* 450, 1169-1170. (2007). |
| * | Buchanan, Mark, "One law to rule them all," *New Scientist*, 8, Nov. 1997. pp. 30-35. |
| * | Cameron, A. G. W. "Birth of a Solar System", *Nature*, Vol. 418, pp. 924-925, 29 August 2002. |
| * | Cameron, A.G.W. "The Origin of the Moon & the Single Impact Hypothesis (5)," *ICARUS*, 126, 126-137, 1997. |
| * | Canup, R. N. & Esposito, "Accretion of the Moon from an impact generated disk," *ICARUS*, 119, 427-446, 1996. |
| * | Canup, R. N. & Esposito, "Origin of Moon in a giant impact near the end of the Earth's formation," *Nature*, 412, 16$^{th}$ August,2001. |
| * | Chambers, John E. 'Planetary Accretion in the inner Solar System', *Earth and Planetary Science Letters*, 223, Issues3-4, July 2004, 241-252, |
| * | Cook, C.L. " Comment on 'Gravitational Slingshot,'by Dukla,J.J., Cacioppo, R., & Gangopadhyaya, A. [American Journal of Physics, 72(5), pp 619-621,(2004)] *American Journal of Physics*, 73(4), pp 363, April, 2005. |
| * | Coontz, Robert, "Like a bolt from the Blue", *New Scientist*, 10th Oct. 1998. pp. 36-40. |
| * | Darwin, G. H. "On the precession of a viscous spheroid and on the remote history of the Earth," *Philosophical Transactions of Royal Society of London*, Vol. 170, pp 447-530, 1879. |
| * | Darwin, G. H. "On the secular change in the elements of the orbit of a satellite revolving about a tidally distorted planet," *Philosophical Transactions of Royal Society of London,* Vol. 171, pp 713-891, 1880. |
| * | Dickey,J.O., Bender,P.L.,Faller,J.E., Newhall,X.X., et al "Lunar Laser Ranging: A Continuing Legacy of the Apollo Program", *Science,* 265, 482-490, 22 July 1994. |
| * | Dickey, J. O., Marcus, S. L., Viron, O. de & Fukimori, I., "Recent Earth Oblateness Variations : Unravelling Climate and Postglacial Rebound Effects", *Science*, Vol. 298, 6 Dec. 2002, pp. 1975-1977. |
| * | Dukla,J.J., Cacioppo, R., & Gangopadhyaya, A. " Gravitational slingshot", *American Journal of Physics*, 72(5), pp 619-621, May,2004. |
| * | Epstein, K.J. "Shortcut to the Slingshot Effect," *American Journal of Physics*, 73(4), pp 362, April, 2005. |
| * | Faller, J.E., Winer,I., Carrion, W., Johnson, T.S., Spadin,P., Robinson, L., Wampler, E.J., Wieber, D., " Laser Beam Directed at the Lunar Retro-Reflector Array: Observation of the First Returns", *Science,* 166, 99, (1969) |
| * | Garlick, Mark A. "The Fate of Earth", *Sky & Telescope*, October 2002, pp. 30-35. |
| * | Genstenkorn, H. "Uber Gezeitenreibung beim Zweikarpenproblem," *Z. Astrophysics*, 26, 245-274, 1955. |
| * | Gladman, B.Quin, D.D. Nicholson,P. & Rand, R , "Synchronous Locking of Tidally Evolving Satellites," *ICARUS,* 122, 166-192, 1996 |
| * | Goldreich, P. "History of Lunar Orbit," *Review of Geophysics*, 4. No. 4., pp. 411-439, Nov. 1966. |
| * | Goldreich,P.,Tremaine,S.,*Astrophysics J.,* 241, 425(1980) |
| * | Halliday, A.N.& Wood, B.J. 'Treatise on Geophysics' Vol 9 (ed. Schubert, G.) 13-50 (Elsevier, Amsterdam, 2007) , |
| * | Hartmann, *"The Cosmic Journey"*, Wadsworth Publishing Co. Inc. Belmont, California, 1978. |
| * | Ida, S., Canup, R. M. & Stewart, G. R., "Lunar Accretion from an impact-generated disk," Nature, 389, 353-357. 25th Sept. 1997. |
| * | Jacobsen,S.B. 'The Hf-W isotopic system and the origin of the Earth and Moon,' *Annual Review of Earth Planet,* Sci.33, 531-570 (2005), |
| * | Jones, J.B. "How does the slingshot effect work to change the orbit of spacecraft?" *Scientific American*, pp 116, November, 2005 |
| * | Jong, T. de. & Soldt. W. H. van "The Earliest known Solar Eclipse record redated", *Nature*, Vol. 338, pp. 238-240, 16th March 1989. |
| * | Kamble, Edwin C., *"Physical Science-its Structure and Development (From* |

|   | *Geometric Astronomy to the Mechanical Theory of Heat)*", Chapter 9, Section 22, M. I. T. Press, 1966. |
| * | Kaula, William K., "The Gravitational Field of the Moon", *Science,* 166, 1581-1588, No.3913, 26Dec. 1969. |
| * | Kaula, W. K. & Harris, A. "Dynamics of Lunar Origin and Orbital Evolution," *Review of Geophysics and Space Physics*, 13, 363, 1975. |
| * | Kaula, W. M. "Tidal dissipation by Solid friction and the resulting orbital evolution," *Review Geophysics*, 2, 661-684, 1964. |
| * | Kerr, R. A. "The First Rocks Whisper of their Origins", *Science*, Vol. 298, pp. 350-351, 11 October 2002. |
| * | Kipp, M. E. & Melosh, H. J. "*Origin of the Moon,*" edited by Hartmann, Phillip and Taylor, 643-647, Publishers : Lunar & Planetary Sciences, Houston. |
| * | Kleine, T., Mezger, K., Palme, H., Scherer, E. & Munker, C. 'The W isotope evolution of the bulk silicate Earth: constraints on the timing and mechanisms of core formation and accretion,' *Earth Planet Science Letters,* 228, 109-123 (2004), |
| * | Krasinsky, G. A., "Dynamical History of the Earth-Moon System", *Celestial Mechanics and Dyanical Astronomy*, 84, 27-55, 2002. |
| * | Leschiutta, S. & Tavella P., "Reckoning Time, Longitude and The History of the Earth's Rotation, Using the Moon" *Earth, Moon and Planets*, 85-86 : 225-236, 2001. |
| * | Lissauer, J.J. & Stevenson,D.J. 'Protostars and Planets V,'(eds. Reipurth, B., Jewitt,D. & Keil,K.),591-606,(University of Arizona Press), Tucson,(2007). |
| * | Macdonald, G. J. F. "Tidal Friction", *Review Geophysics*, 2, 467-541, 1964. |
| * | Maddox, J. "Future History of our Solar System", *Nature*, 372, pp.611, 15[th] Dec 1994. |
| * | McMenamin, Marks A. S. "The Emergence of Animals." *Scientific American*, April 1987. |
| * | Melosh, H. J. & Kipp, M. E. *Lunar Planetary Science Conference XX*, 685-686, 1989. |
| * | Moore, Patrick *"Passion for Astronomy"*, John Wiley & Sons? |
| * | Morrison L. V., "Tidal Deceleration of the Earth's Rotation Deduced from Astronomical observations in the Period A.D. 1600 to the Present." pp. 22-27, *Tidal Friction,* edited by Brosche & Sudermann, Springer, 1978. |
| * | Nimmo, F. & Agnor, C.B. 'Isotopic outcomes of N-body accretion simulations: Constraints on equilibration processes during large impacts from Hf/W observation,' *Earth Planet Science Letters,* 243, 26-43 (2006), |
| * | Ogihara,M.,Ida,S. & Morbidelli, A. 'Accretion of terrestrial planets from oligarchs in a turbulent disk,'*ICARUS,* 188, 522-534,(2007), |
|   | Ohta, K., Onoda, S., Hirose, K., Sinmyo, R., Shimizu, K., Stata, N., Ohisi, Y. and Yasuhara, Akira " The Electrical Conductivity of Post-Perovskite in Earth's D" Layer", Science , Vol 320, 89-91, 2008 |
| * | Raymond, S.N., Mandell, A.M. & Siggurdson, S. 'Exotic Earths forming |
| | habitable worlds with giant plant migration', *Science*, 313, 1413-1416 (2006), |
| * | Rubicam. D. P. "Tidal Friction & the Early History of Moon's Orbit", *Journal of Geophysical Research*, Vol. 80, No. 11, April 19, 1975, pp. 1537-1548. |
| * | Runcorn, S. K. "Change in the Moment of Inertia of the Earth as a result of a Growing Core", *Earth-Moon System* edited by Marsden & Cameron, Plenum Press, 1966., 82-92. |
| * | Sharma, B. K. "Theoretical Formulation of Earth-Moon System revisited," *Proceedings of Indian Science Congress 82$^{nd}$ Session*, 3$^{rd}$ January 1995 to 8$^{th}$ January 1995, Jadavpur University, Calcutta pp. 17. |
| * | Sharma, B. K. & Ishwar, B. "Lengthening of Day curve could be experiencing chaotic fluctuations with implications for Earth-Quake Prediction", *World Space Congress-2002*, 10$^{th}$-19$^{th}$ October 2002, Houston, Texas, USA, Abstract 03078, Pg 3. |
| * | Sharma, B. K. & Ishwar, B. "Planetary Satellite Dyanics : Earth-Moon, Mars-Phobos-Deimos and Pluto-Charon (Parth-I)" *35$^{th}$ COSPAR Scientific Assembly*, 18-25$^{th}$ July 2004, Paris, France |
| * | Sharma, B. K. & Ishwar, B. "A New Perspective on the Birth and Evolution of our Solar System based on Planetary Satellite Dynamics", *35$^{th}$ COSPAR Scientific Assembly*, 18-25$^{th}$ July 2004, Paris, France. |
| * | Sharma, B. K. & Ishwar, B., "Jupiter-like Exo-Solar Planets confirm the Migratory Theory of Planets" *Recent Trends in Celestial Mechanics-2004*, pp.225-231, BRA Bihar University, 1$^{st}$ – 3$^{rd}$ Novermber 2004, Muzaffarpur, Bihar.Publisher Elsiever. |
| * | Sonett, C. P. and Chan, M. A. "Neoproterozoic Earth-Moon Dynamics : rework of 900 million ago Big Cottonwood Canyon tidal laminae" *Geophysics Research Letters*, 25(4), 539-542.1998 |
| * | Stephenosn, F. R. "Historical Eclipses and Earth's Rotation", *Astronomy & Geophysics,* Vol. 44, April 2003. |
| * | Stephenson, F. R. *"Historical Eclipses and Earth's Rotation"*, Cambridge University Press, 1997. |
| * | Stephenson, F. R. and Houldon, M. A. *"Atlas of Historical Eclipse Maps"*, Cambridge University Press, 1986. |
| **\*** | Stevenson, J. David 'A planetary perspective on the deep Earth,' *Nature,* 451, 261-265,(17 January 2008), Year of Planet Earth Feature. |
| **\*** | Swiney et. al. Nature, September (Referenced in Hindu, Sept. 12, 1996, by Malcome W. Browne, N. Y. Times). |
| **\*** | Taylor, S. Rose and Mclennan Scott M. "The Evolution of Continental Crust," *Scientific American*, January, 1996. |
| **\*** | Toubol, M. , Kliene, T. Bourdon, B., Palme,H. & Wiele, R.'Late formation and prolonged differentiation of the Moon inferred from W isotopes in Lunar Metals,' *,Nature* , 450, 1206-1209 (20 December 2007); |
| **\*** | Touma, J. & Wisdom, J. "Resonances in the Early Evolution of Earth-Moon System," *The Astronomical Journal*, 115, 1655-1663, 1998, April. |
| * | Walker, J. C. G. & Zahnle, K. J. "Lunar Nodal Tide and distance to the Moon during Precambrian," *Nature*, 320, 600-602, (1986). |
| * | Ward, W. R. & Canup, R. M. "Origin of the Moon's orbital inclination from resonant disk interactions," *Nature*, 403, 741-743, 17th Feb. 2000. |
| * | Wells, John W., "*Paleontological Evidence of the Rate of the Earth's Rotation*", Earth-Moon System, edited by Marsden & Cameron, Plenum Press, 1966. pp. 70-81. |
| * | Wells, John W., *Nature*, 197, 948-950, 1963. |
| * | Williams, D. M., Kasting, J. F. & Frukes, L. A. "Low-Latitude glaciation and rapid changes in the Earth's Obliquity explained by obliquity-oblateness feedback," *Nature*, 396, 3rd Dec. 1998. |
| * | Williams, George E., "Geological Constraints on the Precambrian Hisotry of Earth's Rotation and the Moon's Orbit", *Review of Geophysics*, 38, 37-59. 1/February 2000. |
| * | Yan, X. Zhour, Y., Pan, J., Zheng, D., Fang, M., Liao, X., He, M-X., Liu. W. T. and Ding, X. " Pacific warm pool excitation, earth rotation and El nino Southern Oscillations",*Geophysical Research Letters*, Vol. 29, No 21, 2031, doi.10.1029/2002GLO15685 pp. 27-1 to 27-4, 2002. |
| * | Yen, B. "The dark side of Moon", New Scientist, pp.30-33, 30th January 1999.. |
| * | Zeik/Gauntand editors *Astronomy* (IInd Edition), Cosmic Perspective, |

| **APPENDIX (I)** | **A BRIEF HISTORY OF APPOLLO AND LUNA MISSION** |
|---|---|
| 4$^{th}$ Oct. 1957 | Sputnik (I) launched by erstwhile U.S.S.R. It carried a Radio Transmitter. |
| 3$^{rd}$ Nov. 1957 | Sputnik (II) launched. It carried a dog, laika. The space endurance of the dog, in-situ, was monitored by methods of Telemetry. |
| 17$^{th}$ April 1961 | Yuri Gagrin, the first Cosmonaut, in Space atop VOSTOK (1). Duration of stay in Space was 108 minutes. |
| May, 1961 | President of U.S.A. makes his historic Statement : "I believe this Nation should commit itself to achieving the goal, before the decade is out, of landing Man on Moon and returning him safely to Earth." |
| **APOLLO MISSIONS.** | |
| Dec. 24, 1968 | Apollo 8 carried out LUNAR ORBITAL MISSION. Moon's surface was photographed and sent back by digital technique. The minimum altitude over Moon was 115 km. |
| May 21, 1969 | Apollo 10 repeated Lunar Orbital Mission but achieved a minimum altitude of 17 km. |
| July 20, 1969 20 : 17 GMT | Apollo 11 carried out the historic mission of putting Man on Moon. Saturn V rocket launched Apollo Spacecraft on the course to Moon. The crew consisted of Mission Commander, Neil Armstrong, Pilot of the Lunar Module, Eagle, Edwin E. Aldrin, and Command Module Pilot, Michael Collins. Apollo 11 had landed in Sea of Tranquility, a low land. 22 kg ms of Moon rocks and samples brought back. Eagle while trying to land traveled westward overshooting the target by 6 km and 2 km off the track due South. Eagle was supposed to land on smooth patch in the Sea of Tranquility. Instead it landed in a most inhospitable terrain full of boulders. This actual landing site was outside the region that had been chosen for and catered for in ground simulation of landing. This mishap occurred because of sketchy knowledge of Moon's gravitational field. Mass concentrations or MASCONS cause perilun wiggle which have to be accounted for in trajectory calculation. These MASCONS are either remnants of the asteroids that hit the Moon or dense material from the interior. If these wiggles are not accounted for then trajectory calculation will go hay-wire as it did in the case of Eagle's landing. |
| Nov. 18, 1969 | Apollo 12 landed in Oceanus Procellarum (the ocean of storms). Sample of Mare Basin was brought back. The participants were Pete Conrad and Allen Bean. The landing site |

|  | was 200 meters from the landing site of Surveyor 3 probe. Apollo 12 landed exactly on target. |
|---|---|
|  | Apollo 13 mission got aborted and had to come back to Earth. |
| Feb. 2, 1971 | Apollo 14 landed on Fra Mauro (ejecta from Imburin Basin). Samples of ejecta from Imburin Basin were brought back. Situ Rossa piloted the Command Module and Alan Shepard and Ed Mitchell explored Fra Mauro highlands. |
| July 30, 1971 | Apollo 15, landed at the edge of Mare basin. In this Mission, Lunar Buggy, a roving vehicle, was used. Samples of material from Apennine Mountains, the rim of Imburin Basin, was brought back. Apennine mountains are Moon's primordial crust. Samples of Mare rocks and dust were brought back. Imburin Basin is the great canyon on Moon. David Scott and Jim Irwin spent three days on Moon. In these three days thrice they drove their Lunar Buggy and explored Moon's surface for seven hours in each drive/ |
| April 20, 1972 | Apollo 16 landed on Lunar uplands near creater Descrates. Samples of highland brought back. |
| Dec. 11, 1972 | Apollo 17, landed on Taurus mountains. Samples of a region suspected of recent volcanism was brought back. Rim Evans piloted the Command Module. Gene Cornan and geologist Jack Schmitt did 3 days of prospecting in Taurus Littrow Valley. |
|  | In all 24 men participated in Apollo Mission and 12 men, amongst these, set foot on Moon. |

**LUNA MISSIONS**

| Feb. 1966 | Luna 9 carried out soft landing Mission. It sent back detailed pictures of hard and rocky landing site. |
|---|---|
| March 1966 | Luna 10 carried out Lunar Orbital Mission. It sent back data on radiation revels and micro meteorite impacts. |
| 1970 | Luna 17 Mission carried out. Lunokhod 1, a roving vehicle, was placed on Moon's surface. Lunokhod 1 had an active life of 10 months and traveled 10.5 km. Enroute it did sample testing and sent back data by telemetry. |
| 1973 | Luna 21 mission carried out. Lunokhod 2 was placed on Moon's surface. This vehicle had a life of 4 months and traversed 37 km. Enroute this also did sample testing. By television/radio link remote control of the roving vehicle and the mechanical arm was carried out and through telemetry techniques communication maintained. |

**APPENDIX(II)　EONS, ERAS AND AGES OF THE EARTH'S GEOLOGICAL PERIODS.**

Eons, Era and Ages of Earth's Geological History.

From New Scientist, 20[th] May 1995, INSIDE SCIENCE No.81 Great World Atlas-The Reader's digest Association.

| EONS | ERAS | AGES |
|---|---|---|
| **HADEAN** | | |
| 4.56Gya-4.00Gya Last Phase of Earth formation. Earth was hot through Radioactive heating and asteroid bombardment. | | |
| **LUNAR CATACLYSM** | | |
| 4.00Gya very heavy asteroid bombardment. The last remnants of the solar nebula is swept clean by the planets. | | |
| **ARCHEAS** | | |
| 4Gya-2.5Gya **Ancient and barren of life.** Plate Tectonic Dynamo slowed down And Earth's surface took the modern pattern of plate distribution. | | |
| **PROTEROZOIC** | | |
| 2.5Gya-540Mya **Early Life** until shelled Animals appear. | | **PRECAMBRIAN** Before 540Mya A PreCambrian Explosion occurs. |

|  |  | Diversification of life. |
|---|---|---|
| **PHANEROZOIC** |  |  |
| 540Mya-present<br>**Visible life** | **PALAEZOIC**<br>540Mya-245Mya<br>Early Life in form of marine invertebrate.<br>**MESOZOIC** | **CAMBRIAN**<br>540Mya-505Mya<br>Small shelly animals, Reef builders, Tribolites. |
|  | 245Mya-65Mya<br>Age of Reptiles<br>**CENOZOIC**<br>65Mya to the present<br>Age of Mammals and broad tree leaves. | **ORDOVICIAN**<br>505Mya-433Mya<br>Shelly marine animals, Reef builders, tribolites, Nautilods. Jawless fish. |
|  |  | **SILURIAN**<br>433Mya-410Mya<br>Shelly marine animals, Tribobites, reef builders, Primitie fish, Ammonoids. Land plants appear. |
|  |  | **DEVONIAN**<br>410Mya-360Mya<br>Extensive mountain building and volcanic activity in N.W. Europe. Amphibians appars. |
|  |  | **CARBONIFEROUS**<br><br>360Mya to 286Mya<br>Clear shallow seas, chief coal forming period. Reptiles appear. |
|  |  | **PERMIAN**<br>286Mya-245Mya<br>A period of considerable Earth movement. Lofty mountains form in Europe, Asia and Eastern USA.<br>Arid in N. Hemisphere and Ice Age in S. Hemisphere.<br>Worst Mass Extinction of |
|  |  | shelly marine animals, reef builders, mammal like reptiles, ammonoids, conodents & snails. |
|  |  | **TRIASSIC**<br>245Mya-202Mya<br>Land area covered with deserts and shrub covered mountains. Marble and sandstones in warm sea.<br>Extinction of shelly marine animals, dinosaurs & ammonoids. Mammals appear. |
|  |  | **JURASSIC**<br>202Mya-144Mya<br>Mountains reduced to low hills due to wet condition. Sea advances due to the melting of Ice Age.<br>Extinction of shelly marine animals and dinosaurs.<br>Earliest birds. |
|  |  | **CRETACEOUS**<br>144Mya-65Mya<br>Rockies Ranges in USA, Andes Ranges in S. America, European Ranges begin to appear giving rise to Gulf Streams. Parts of Australia covered by Glaciers.<br>At 65Mya an Iridium-rich Asteroid strikes Gulf of Mexico leading to Mass Extinction Dinosaurs are wiped out and 96% of other form of life. |
| | | At 70Mya first flowering plants occur. |
| | | **PALEOCENE** 65Mya-58Mya Subsidence of much of Europe causes the seas to advance. Mountain Ranges which emerged in Cretaceous Age continue to grow. Vast amount of lava ejected and deposited. Glaciers exist in high mountains in W.N. America. |
| | | **EOCENE** 58Mya-36Mya Second phase of mountain folding occurred along the Himalaya Ranges. Whales and bats. |
| | | **OLIGOCENE** 36Mya-24Mya The Alps in Europe begin to from. Ice Age sets in. Sea recedes. Extensive movement of Earth's crust in America and Europe. Modern Mammals |
| | | **MIOCENE** 24May-7Mya Powerful Earth movement. Mediterranean becomes land locked. European and Asian land masses are joined together. Alps is completed. Much volcanic activity. |
| | | **PLIOCENE** 7Mya to 2Mya |

|  |  | Land subsidence and mountain building continues at reduced rate. Hominids appear. Hominids evolve into Australopethicus Afarensis with a cranial capacity of 400cc. Homo-habilis-handy man, cranial capacity of 700cc. Homo-erectus-erect man, cranial capacity of 1000cc. Homo-sapiens-wise man. Cranial capacity of 1400cc. |
|--|--|--|
|  |  | **PLEISTOCENE** 2Mya-10,000yrs ago Ice Age sets in. Periodic melting of Ice. |
|  |  | **HOLOCENE** 10,000yrs ago-present Ice Age ends and warm period begins.The Age after the Great Floods. This event is mentioned in the written records of all ancient civilizations, |

**APPENDIX(III)    THE ORIGIN OF MOON.**

Initially there were three competing theories for the origin of Moon:

(i) Fission Model- This was first proposed by George Howard Darwin in 1879. According to this hypothesis the Moon was torn apart from the Earth leaving behind Pacific basin.

(ii) Capture Model- According to thgis hypothesis our Moon was formed elsewhere in the Solar System and while making fly-by past the Earth it was captured.

(iii) Double Planet or Binary Planet or Co-formation Model- According to this hypothesis Earth-Moon were formed simultaneously in same place as twin sisters.

The study of the Lunar Samples brought during Apollo 11 to Apollo 17 Missions and during automated Luna 16 to Luna 20 Missions, the study of the data sent by the network of seismometers on the surface of Moon and through spectroscopic studies, following facts about Earth-Moon System have emerged :

(a) Oxygen Isotopes are in similar ratios on Earth and Moon suggesting a close kinship between the two;

(b) Moon is bone dry and has much less volatile materials such as Sodium, Potassium, Bismuth and Thallium. Earth has much more volatile materials in its mantle;

(c) Moon has an abundance of refractories such as Alumina, Calcium, Thorium and Rare Earth Elements (REE). REE are 50% higher on Moon than on the Earth;

(d) Magnesium Oxide is 10% higher for Moon as compared to that on Earth.

In light of these facts all the three hypotheses have been debunked. Instead in the year 1984 at the International Conference at Kona, Wawaii, by near unanimity the Giant Impact Theory was accepted.

**Re-examination of the three Models.**

(i) In Fission Model the initial rapid axial spin period of 6 hours cannot be explained. In the process of Planet formation by accretion only a slow angular spin velocity can be imparted to the planetary bodies including the Earth. According to Fission theory Moon's composition should be the same as that of crust and mantle of the Earth but that is not the case. Hence Fission Model loses ground.

(ii) Statistically capture is most unlikely. Either head on collision can take place leading to complete embeddedment of the impacting body into the core of the Earth or the colliding body will get deflected away from the Earth. From space craft entry science we know that there exists a very narrow corridor for re-entry into the Earth's atmosphere. Similarly capture of a foreign body would be a very delicate balance between conflicting bodies. The close kinship between Moon and Earth on the basis of Oxygen isotopes indicate that the colliding planetismal could not have come from very different part of the Solar System. For all these reasons Capture Model too is ruled out.

(iii) According to Double Planet Model, Earth and Moon should have similar compositions and near identical densities but the Moon has a much lighter density mainly because metallic core is insignificant by weight. Earth's density is 5.5 gms per $(cm)^3$ and Moon's density is 3.3 gms per $(cm)^{3.}$ Therefore Double Planet Model is also not possible.

To give a completely consistent theory Giant Impact Theory is invoked. According to this Theory, the shock waves from a nearby Super Nova Explosion threw a Giant Cloud of gas and dust into spin mode transforming it into a pancake spinning around the longitudinal axis. This spinning pancake became the Solar Nebula, the cradle of our Solar System. Solar Nebula was born about 4.56Gya.

The central part of the Solar Nebula collapsed into our Sun and the circumstellar disk of gas and dust became the breeding ground of the diverse orbiting objects such as Meteorites, Comets, Gas Giants, Ice Giants and Terrestrial Planets. There is no coherent theory or model to date to explain the formation of diverse planetary objects and that is one of the tasks of this thesis. But according to the present understanding this is how thee planets were formed :

The central nucleus shrank and started a nuclear fusion furnace in a period of 50 million years. This furnace became the Sun of our Solar System. The raw materials of Sun and the Solar Nebula consisted of materials recycled through several generations of star formation. To be precise the Spiral Galaxy called the Milky Way is the Third Generation Galaxy formed 10Gya.

Because of intense Solar Activity the gas was sept to the outer region and heavier dust remained in the inner part.

Gaseous part accreted to form the Gas Giants, Jupiter and Saturn, and Ice Giants, Uranus and Neptune.

In the inner part, the dust accreted to form gravels, gravels accreted to form rocks, small sized rocks accreted into large sized rocks, large sized rocks accreted into kilometer-sized planetismals, planetismals accreted into planetary embryos containing 1-10% of the Earth's mass and the collision of tens to hundreds of planetary embryos yielded the final four terrestrial planets.

By the time the Earth was fully formed, it experienced a glancing angle impact from a Mars-sized Planetismal. This Giant Impact gave an Obliquity Angle to the Earth because of which we have our four Seasons and today the Obliquity Angle is $\Phi = 23.45°$. The Giant Impact provided the additional torque to put the Earth in rapid spin mode of 6 hours diurnal period. The core of the impactor stuck into the interior of the Earth and the mantle of both the Impactor and the Earth blew up creating a circumterrestrial disc of debris containing the materials from the two mantles. From this impact generated circumterrestrial debris, our Moon was born in Roche's Zone.

The debris lying within the Roche's Zone spiraled inward and were absorbed by the Earth.

The rocky debris within Roche's zone accereted to form the Moon around the same time i.e. 4.53Gya. Because of impact heating the Earth turned into a magmatic ball. This enabled magmatic differentiation and stratified the Earth into Onion-like structure. This impact heating drove away all the volatiles and moisture. Hence Moon is bone dry, devoid of all volatiles and rich in Silicates and refractories.

The full grown Moon and Earth strated tidally interacting as a result Moon got tidally locked or entered into captured rotation phase. The tidal interaction is pushing out the Moon and slowing the Earth on its spin axis. Once Moon spirals out to the outer Geo-synchronous Orbit then Earth will also get tidally locked to the Moon.

At present the Earth's spin has slowed down from a diurnal period of 6 Hours to 24 hours and the present recession velocity of the Moon is 3.8 cm/yr.

**APPENDIX(IV)     GEOLOGICAL HISTORY OF EARTH AND MOON.**

The Earth was initially as undifferentiated as Venus is today. The impact heating and radiogenic heating had converted the Globe into a molten sphere. In the molten phase the heaviest elements settled down to form the metallic core consisting of Iron and Nickel. The remaining rocky part formed the basalt rich Mantle. The Mantle formed the primary crust. At this undifferentiated stage the principle moment of inertia C was much higher than the modern value of $C = 80.25 \times 10^{36}$ kg-m$^2$ as evidenced by Kaula & Harris regarding the Synodic Month data 2.8Gya.

Withing a time span of 100 million years to 300 million years from the inception, the outer part cooled, solidified and got covered with a shallow ocean of water. Internal part remained hot due to radiogenic heating. This led to setting up of huge convection currents physically carrying hot lava from within and bursting out into volcanic eruptions at the hot spots or at the fault lines. The existence of fault lines imply that there were numerous plates. The plate tectonic dynamo was rapid because of excessive radiogenic heating. Hence convection currents were shallow, magmatic differentiation was incomplete and the outer surface was divided into 100 plates. These plates were diverging, converging and partaking into slip-stick lateral movement. At the lines of divergence mid-oceanic ridges were formed. Volcanic eruptions were taking place and basaltic lava was flowing out to form the ocean floors. This formed the secondary crust. At the lines of convergence, oceanic plates moved underneath other plates causing subduction zones and trenches. Here the crust is destroyed and whatever crust is lost at the subduction zone is compensated by the crust formation at mid-oceanic ridges. The mechanism of the continual creation, motion and destruction of Planet's surface is known as Plate-Tectonics.

In next billion years the interiors cooled because of reduced radioactivity. The mechanism of Plate-Tectonic considerably slowed down and convection currents in the interior became deeper leading to complete magmatic differentiation. The Earth was further differentiated into Granite-rich outer Crust, the Basalt rich Mantle and inner Iron-Nickel Core. The slowing down of Plate-Tectonic dynamo, caused 100 plates to coalesce into13 major plates and 4 minor plates. These crustal plates carry the 5 oceans and the 7 continents. The mantle is divided into upper and lower mantle. The upper mantle is weakly conducting, semi-molten, plastic, weak and hot. The cool, rigid and brittle crustal plates float over the semi-molten upper Basaltic mantle.

**The major plates are :**
1. The Anhtartic Plate carries the Anartic Continent and the surrounding water mass;
2. The Pacific Plate carries the Pacific Ocean and is surrounded by the ring of fire;
3. The North American Plate carries the North American Continent and Greenland;
4. The South American Plate carries the South American Continent;
5. The African Plate carries the African Continent;
6. The Eurasian Plate carries the European Continent;

7. The Australian-Indian Plate carries the Australian Continent and Indian Subcontinent;
8. The Phillipine Plate carries Phillipine part of South East Asia;
9. The Iranian Plate carries Iran;
10. The Arabian Plate carries Arabian Peninsula;
11. The Antolian Plate carries the Black Sea;
12. The Hellenic Plate carries Turkey;
13. The Carribean Plate carries the Carribean Sea;
14. The Cocos Plate carries Central America;
15. The Scotia Plate lies between S. American Plate and Antartic Plate;
16. The Nazca Plate is sandwiched between Pacific Plate and S. American Plate;
17. The Juan De Fuca Plate is sandwiched between N. American Plate and Pacific Plate.

As mentioned earlier, at all convergent fault lines we have subduction zones and formation of trench systems. The major Trench Systems are :

a. Aleutan System due to subduction of Pacific Plate underneath N. American Plate;
b. Tonga System due to sub-duction of the Indian-Australian Plate sub-ducting underneath pacific Plate;
c. Peru_Chile System due to Nazca Plate subducting under S. American Plate;
d. Japan System due to Pacific Plate sub-ducting underneath Eurasian Plate;
e. Marina trench due to Pacific Plate sub-ducting underneath Phillipine Plate;
f. Indo-Australian Plate subducting underneath Eurasian Plate.

The formation of the Continental Plates above the sea level is the tertiary crust. This formation has been going on for last 4.26 billion years but there is a distinct change in the character of the tertiary crust around 2.5Gya which marks the boundary between Archean and Proterozoic Eons.

Before this boundary period, the Mantle and Continental crust are less differentiated composed of a mixture of Basalt and sodium-rich Granite. In this remote past as already mentioned the higher level of radio-activity was fuelling a faster plate-tectonic engine. As a result there were shallower convection currents leading to incomplete magmatic differentiation and the surface being broken in 100 separate plates.

After the Archean Eon the plate-tectonic became slower, continental plates coalesced together to form 12 plates and deep recycling of the continental crust led to sharply differentiated Basaltic Mantle and outermost Potassium-rich Granitic Crust.

None of the early Basaltic Crust, known as primary crust, survived to Modern Times. The radiogenic heat caused the complete ploughing back of the primary crust into the interior of the Earth. Similarly the secondary crust also got ploughed back and replaced by tertiary crust.

Earth and Moon have similar ratios of Oxygen Isotopes. This implies that the places of origin of the Mars-sized Impact or and the Earth were in close neighborhood.

Moon was mainly constituted of the debris of the Impact or which had less of FeO as compared to that of the Earth. Hence MgO/FeO ratio is greater by 10% for the Moon as compared to that of the Earth.

Impact collision heat and heat of accretion during Moon formation drove away all the water and volatiles from Moon and made it bone dry on one hand and refractory rich on the other. Refractory materials quickly reconvened.

Hence Moon lacks volatiles like Sodium, Potassium, Bismuth, and Thallium whereas it is rich in Aluminum, Calcium, Thorium and has 50% higher Rare Earth Elements as compared to those of the Earth.

Due to impact heating, accretion heating, core formation and sinking of the metallic core to the centre released an immense amount of heat leading to sea of magma covering the whole of Moon. Heavier minerals composed of Iron and Magnesium Silicates – Olivine and Pryoxine – sank to create the mantle and white Felspar-rich material composed of light Calcium and Aluminum Silicates floated on the top of the magma. In first two billion years highlands arose from the magma sea of the geologically active Moon. These highlands were covered by white Feldspar and highland rocks. Highland rocks are called Breccias which crystallized out from magma around 4.4Gya and top layer is light colored apotheosize composed of white feldspar. These highland feldspar are rich in Europium.

Later Mare basins emerged deficient in Europium. Mare basins consist of Basaltic bedrock rich in Titanium. This Basaltic bedrock is covered with upper Moon dust called megalith. Regolith is the loose debris which gets blown off from Moon surface by meteoritic impact and which resettles on the surface. This is as deep as 20 meters.

Samples of regolith contain small percentage of plagioclase Feldspar. Meteoritic impact on the highlands caused debris of Feldspar to fly and fall in Mare Basin.

The Basalts in Mare region are rich in olivine and pyroxene.

About 4 Gya, after the emergence of the highlands covered with white Feldspar, there was vast lava flows all over Moon's surface. During the Big Bombardment era between the narrow slot of 4Gya and 3.85Gya these Basaltic lava solidified to form the basins.

In this narrow slot there was fierce meteoritic bombardment and numerous cratering was caused specially on the highlands.

Ages of all rocks from Apollo and Luna missions fall between 3.95Gy to 3.85Gy.

**APPENDIX(V) CALCULATION OF MOMENT OF INTERTIA (C) OF UNDIFFERENTIATED AND DIFFERENTIATED EARTH.**

Moment of Inertia about the spin axis of a homogeneous sphere:

$C_{homo} = (2/5) \, M_1 R_1^2$.

Substituting the numerical values from Table(1):

$C_{homo} = 97.06 \times 10^{36}$ kg-m$^2$ ;

Calculation of Moment of Inertia (C) of Modern Day Earth which is stratified as given in University Physics by Resnik, Halliday & Walker:

| Distance from the Centre of the Earth (km) | Density of the annular shell (kg/m$^3$) | Mass of the Annular Shell (kgm) | C (kg-m$^2$) |
|---|---|---|---|
| 1288- core = $R_4$ | $13.25 \times 10^3 = \rho_4$ | $0.188 \times 10^{24}$ | $0.078 \times 10^{36}$ |
| 3477.6- lower mantle = $R_3$ | $11.00 \times 10^3 = \rho_3$ | $1.839 \times 10^{24}$ | $9.3 \times 10^{36}$ |
| 6185 - upper mantle = $R_2$ | $4.62 \times 10^3 = \rho_2$ | $3.765 \times 10^{24}$ | $66 \times 10^{36}$ |
| 6370 - crust = $R_1$ | $2.80 \times 10^3 = \rho_1$ | $0.2563 \times 10^{24}$ | $6.74 \times 10^{36}$ |
| Total mass | | $5.9783 \times 10^{24}$ | $82.118 \times 10^{36}$ |

$C = (2/5) \times (4\pi/3) [\rho_1 \{R_1^5 - R_1^5\} + \rho_2 \{R_2^5 - R_3^5\} + \rho_3 \{R_3^5 - R_4^5\} + \rho_4 \times R_4^5]$

$C = 82.696 \times 10^{36}$ kg-m$^2$.

Calculation of C of Modern day Earth based on Macropaedia:

| Depth from the surface(km) | Distance from the centre(km) | Density ($\times 10^3$ kg/m$^3$) | Classification | Moment of Inertia around the spin axis ($\times 10^{36}$ kg/m$^2$) |
|---|---|---|---|---|
| 0 | 6,371 | | | |
| 15 | 6,356 | 3.31 | Shell A | |
| 350 | 6,021 | 3.54 | Shell B | |
| 650 | 5,021 | 4.25 | Shell C | |
| 1000 | 5,371 | 4.80 | Shell C | |
| 2000 | 4,371 | 5.06 | Shell D' | |
| 2500 | 3,871 | 5.32 | Shell D" | |
| 2700 | 3,671 | 5.42 | Shell D" | |
| 2890 | 3,481 | 5.64 | Gutenburg discontinutiry | |
| 2890 | 3,481 | 9.91 | Gutenburg discontinutiry | |
| 3000 | 3,371 | 10.08 | Shell E | |
| 3500 | 2,871 | 10.79 | Shell E | |
| 4000 | 2,371 | 11.35 | Shell E | |
| 4500 | 1,871 | 11.79 | Shell E | |
| 5000 | 1,371 | 12.56 | Shell E | |
| 5160 | 1,211 | 12.70 | Shell F | |
| 5500 | 871 | 12.85 | Shell G | |
| 6000 | 371 | 12.98 | Shell G | |
| 6321 | 000 | 13.00 | Shell G | |

Shell A – Crust;
Shell B & C – Upper Mantle (molten);
Shell D' & D" – Lower Mantle;
Shell E – Outer Core (molten);
Shell F – Transition Region;
Shell G – Inner Core.
From these values C for the stratified Earth is calculated to be :
$88.2 \times 10^{36}$ kg-m$^2$;

The moment of Inertia
Obtained by Astronomical satellites is C = $80.2 \times 10^{36}$ kg-m$^2$;

**APPENDIX(VI)   CHANGES IN MOMENT OF INERTIA (C) OF THE EARTH DUE TO PLATE-TECTONIC MOVEMENTS.**

The outer crust is taken to be a spherical shell.
The outer radius of this shell = $6.371 \times 10^6$ m;
Assuming average thickness of the crustal shell – 15km
The inner radius of this shell = $(6.371-0.015) \times 10^6$ m = $6.35^6 \times 10^6$ m;
The average density of the crustal shell = $3.3 \times 10^3$ kg/m$^3$;
Therefore
Total mass of the crustal shell = Mcrust = $25.145 \times 10^{21}$ kg;
The outer crust forms 5 km thick layer at the ocean basin and 60 km thick layer under the mountains. We can safely assume that the crustal mass constitute the 13 major plates and 4 minor plates [Appendix iv].
There are several possibilities in plate distribution.
Case (1) The crustal plates are uniformly distributed around the Globe then the Moment of Inertia due to spherical shell = $I_{symm}$ = $(2/3) M_{crust} R_E^2$ = $1.016 \times 10^{36}$ kg-m$^2$;
Case(2) The crustal plates are distributed along the equator then the Moment of Inertia of the equatorial distribution = $M_{crust} R_E^2$ = $1.016 \times 10^{36}$ kg-m$^2$;

$\Delta I = -0.338 \times 10^{36}$ kg-m$^2$;

percentage change with respect to total C = -2.21%;
Case(3) The crustal plates are distributed along the great polar circle = $0.5 \times M_{crust} R_E^2$ = $0.408 \times 10^{36}$ kg-m$^2$;

$\Delta I = -0.169 \times 10^{36}$ kg-m$^2$;

Case(4) The crustal plates are concentrated at one of the poles and subtends an angle $2\Phi$. Moment of Inertia of the crustal plate concentrated at the Polar Cap around Polar Axis = 0.25 $M_{crust}$ $R_E^2$ [3-(1/3)(1-Cos3$\Phi$)/(1-Cos$\Phi$)];

Land surface area = 33.33%;

Water surface area = 66.67%;

If the surface area of the cap is $S_0$

$S_0 = 2\pi R_E^2 (1-\text{Cos}\Phi)$;

The total Continental Crust area is (1/3) of the total surface area of the Globe;

Therefore $S_0 = (4\pi R_E^2)/3 = 2\pi R_E^2 (1-\text{Cos}\Phi)$;

Therefore $\Phi = 70°$

Therefore Moment of Inertia due to Polar Cap distribution = 0.175$M_{crust}$ $R_E^2$ = 0.1778 × $10^{36}$ kg-m$^2$ ;

$\Delta I = -0.5 \times 10^{36}$ kg-m$^2$;

percentage change with respect to total C = -0.623%

|  | I | $\Delta I$ | Percentage change with respect to C |
|---|---|---|---|
| $I_{symm}$ | 0.677 ×$10^{36}$ kg-m$^2$ |  |  |
| $I_{equatorial}$ | 1.016 ×$10^{36}$ kg-m$^2$ | 0.3387 ×$10^{36}$ kg-m$^2$ | +0.422% |
| $I_{polar\ circle}$ | 0.508 ×$10^{36}$ kg-m$^2$ | -0.169 ×$10^{36}$ kg-m$^2$ | -0.21% |
| $I_{polar\ circle}$ | 0.1778 ×$10^{36}$ kg-m$^2$ | -0.5 ×$10^{36}$ kg-m$^2$ | -0.623% |

**APPENDIX(VII)   CHANGES IN MOMENT OF INERTIA (C) OF THE EARTH DUE TO ICE AGE.**

Land Area is (1/3) of the total surface area.

Ocean Area is (2/3) of the total surface area.

Reduction in Ocean level by 1m means displacement of water of mass :

$\Delta M = 4\pi \times R_E^2 \times \Delta h \times \rho_{water}$;

$R_E = 6.37 \times 10^6$ m;

$\Delta h = 1m$ = reduction is sea level;

$\rho_{water} = 10^3$ kg/m$^3$ ;

Substituting the numerical values we get :

$\Delta M = 340 \times 10^3$ kg;

Because of this shift of water mass from uniform distribution around the Globe to concentration on the poles can be expressed as :

$\Delta I \approx -(2/3) \times \Delta M \times R_E^2 = -9.2 \times 10^{30}$ kg-m$^2$;

At the height of Ice Age ocean level will drop by 134m;

Therefore total change in the Moment of Inertia due to shift of a large mass of water will be =

-0.0012×$10^{36}$ kg-$m^2$;

But this reduction will be compensated by the oblateness in the shape of the Earth. Oblateness in the shape will lead to increase in the the Moment of Inertia of the Earth during the Ice Age;

On the other hand there will be Post Glacial Rebound leading to overall decrease in the Moment of Inertia at the end of the Ice Age.

**APPENDIX (VIII)  THE CORRELATION BETWEEN THE EARTH'S SPIN AND THE EARTH'S OBLATENESS.**

Quote from Kamble 1966, "The Substance that respond to rapidly changing forces as rigid solids may still respond to strong, long standing forces by plastic flow if given sufficient time. Earth quake waves show that at present the Earth materials to a depth of 1800 miles react to transients as if it were a rigid body. Geologic evidence makes it clear, however, that in the enormous stretches of geologic time Earth material is subject to slow plastic flow that is undoubtedly continuing to this day. Mountain ranges are commonly composed of low density, fossil bearing, sedimentary rocks that must have been formed beneath the sea and subsequently elevated by plastic flow. The same kind of flow could have given the Earth its spherical form, even though it had never been more fluid than now."

In the light of this quote it can be safely claimed that the present oblateness of 0.3% must have been much larger in the remote past due to a much more rapidly spinning Earth.

Theoretical formulation of oblateness of a rapidly spinning Earth From the reference:

$$(e/4)(2-e)^3/(1-e)^2 = (6.209 \times 10^{-7})(R_E)^2/(T_1)^2$$

where e = oblateness = $[(R_E)_{equator} - (R_E)_{polar}] / (R_E)_{equator}$
$R_E = 6.371 \times 10^6$ m, $T_1$ = period of spin on its spin axis in seconds;
Substituting the numerical value in $R_E$, we get:

$$(e/4)[(2-e)^3/(1-e)^2] = 252 \times 10^5 / (T_1)^2$$

In the Present Age: e = 0.003 and $T_1$ = 23.9344 hours approximately satisfy the above Equation.

In the Archean eon, $T_1$ = 23.55 hours = $4.878 \times 10^4$ seconds.

e = 0.01 approximately satisfies the Oblateness-Spin equation.

Hence we can assume that Earth's Oblateness was 1% in Archean Eon about 2.8 Gya.

**APPENDIX(IX) THE SECOND ORDER EFFECTS INFLUENCING SIDEREAL DAY LENGTHENING.**

(1) Effect of Glacial Visitations on Earth's Spin.

Because of the precession of the spin axis of the Earth, because of variation of precession cone angle and because of the periodical change in eccentricity of Earth's Orbit, the Climate tends to become milder or harsher and a combination of these factors cause glacial epochs. Ever since the birth of the Earth, glacial epochs are visiting the Earth at a period of 200 to 250 million years. On an average a Glacial Epoch lasts from 1 to 3 million years. The last Glacial Period ended only 10,000 years ago leading to a massive deluge which is mentioned in the epics of several civilizations.

(2) Effect of Plate Tectonic Movement on Earth's Spin.

In 1912 the German Geologist Alfred Lothar Wegner proposed the Continental Drift Theory. In last few decades this has been finally established and today this is known as Plate Tectonic Movement.

According to this theory before the Permian Epoch i.e. 225Gya all the seven continents were joined together into a Super Continent called Pangeas.

In Triassic Epoch, 200 Gya, Pangeas broke up into Northern Part called Laurasia and the Southern Part called Gondwana. Antartica and Australia formed the third parts.

In Jurassic Epoch (135 Gya) and in Cretaceous Epoch (65Gya) South America and Africa separated from each other and Africa and India separated. That is Gondwana split into South America, Africa and India.

Subsequently North America in main separated from Europe. Africa linked up with Europe and Asian land-mass. India alsio linked up with Asian land mass. Australia drifted away from Antartica and Antartica drifted to South Pole.

Thus from 225 Mya to 65 Mya, the equator ward drifting of the continental plates caused C to increase. This added to the theoretical lengthening of day.

From 65 Mya to the present gradual north ward drift of India and Greenland, may be decreasing C hence subtracting from the secular lengthening of day.

A subtraction from the secular lengthening is indeed observed.

(3) Effect of Global Wind Paterns, Ocean Currents and that of Electromagnetic Coupling between the liquid core and the mantle.

Two meteorologists Richard Rosen and David Salstian of the University of Cambridge have established with definiteness that the Wind Patterns and Ocean Currents aiding or opposing the rotation of the Earth have the effect of subtracting or adding from the general trend of lengthening due to Moon's tidal drag respectively.

Friction between the liquid core and the upper liquid mantle has a retarding effect on Earth's spin.

Also the spinning motion of the Earth creates the circulating current in Iron-Nickel rich molten core. This circulating current creates the magnetic field. The variation in the magnetic field will always have electromagnetic induction effect which will generate secondary circulating currents in fully conducting upper mantle. These secondary circulating currents will produce a secondary magnetic field which will oppose the cause of variation in the primary magnetic field. Hence a retarding torque will be caused on Earth's axial spin.

Thus there are four second order effects which have to be quantified and to be accounted for before an exact correspondence can be established between the theoretical curve and the observed curve.

Following quote from 2002 Harold Jeffreys Lecture in October 2002 comprehensively sums up the second order effects:

*"Both external and internal mechanisms are responsible for variations in the Earth's rate of rotation. The most significant external causes are lunar and solar tides raised in the oceans and solid body of the Earth. Together with a further small solar contribution (the semi-diurnal atmospheric tide), these produce a steady increase in the LOD of about 2.3 ms/cy.*

*Internal mechanisms giving rise to changes in the LOD are more diverse. Short term effects include changing wind patterns which produce seasonal and annual variations. On the decade to centennial timescale, the most significant cause of variations is probably electromagnetic coupling between the fluid core of the Earth and the lower*

*mantle. There is reasonable correlation between observed fluctuations in the LOD over the past 150 years and core angular momentum fluctuations (Hide R. et al 2000 Geophysics Journal International, Vol. 143, pp.777) Global sea-level changes, associated with climatic variations, may also produce a significant effect on centennial and londer time scales. An additional long-term mechanism is post-glacial isotatic compensation; the ongoing rise of land that was glaciated during the last ice-age produces a slow diminution in the moment of inertia of the Earth, with consequent decrease in the LOD.*

*Seasonal and annual fluctuations have been mapped in detail only since the introduction of Atomic Time(AT) in 1955. Decadal variations in the LOD can be traced over the last four centuries or so (i.e. the telescopic period), mainly using occultations of stars by the Moon. However, any trend is difficult to detect over such a relatively short period. This is why rather crude observations made with the unaided eye in the ancient and medieval past have become so important. The substantial timescale they cover (extending back to around 700BC) enables long term trends to be determined with fair precision."*

**APPENDIX (X)    CALCULATION OF RATE OF LENGTHENING OF SIDEREAL DAY.**

Total Angular Momentum of Earth-Moon System=$(J_{spin})$Earth + $J_{orbital}$ = $J_{Total}$

$J_{Total}$ = $3.496 \times 10^{34}$ kg-m$^2$/sec

$(J_{Total})_{Earth}$ = $C\omega$ and $J_{orbital}$ = $m\sqrt{(Gma)}$

m = mass of the Moon and M=mass of the Earth.

All the symbols are as defined in the text.

Therefore                                                                                               (X.1)

$C\omega + m\sqrt{(Gma)} = 3.496 \times 10^{34}$ kg-m$^2$/sec

Differentiating Eq.(X.1) we get:                                                          (X.2)

$d\omega/dt = -(1/(2C))\{m\sqrt{(GM)}/\sqrt{a}\}(da/dt)$

In the present Age,

$\sqrt{a} = 1.9606 \times 10^4$ m$^{1/2}$ and m $\sqrt{(GMa)} = 1.4844 \times 10^{30}$ kg-m$^{3/2}$/sec

From Laser Lunar Ranging data:

$da/dt = 12.711 \times 10^{-10}$ m/sec;

From Astronomical Satellite data:

$C = 80.25 \times 10^{36}$ kg-m$^2$;

Substituting the numerical values in Eq. (X.2):

$d\omega/dt = -0.599 \times 10^{-21}$ radians per second.

$$\omega = 2\pi/T \tag{X.3}$$

Differentiating Eq. (X.3)

$d\omega/dt = -(2\pi/T^2) \times (dT/dt)$

Therefore

$$dT/dt = -(T^2/T^2) \times (dT/dt) \tag{X.4}$$

Substituting the numerical values in Eq. (X.4):

$dT/dt = 7.05 \times 10^{-13}$ second per second.

Since 1 Solar Year = $31.557 \times 10^6$ seconds;

Therefore

$dT/dt$ = 2.22s/Solar Year = 2.22ms/century.

The actual long term trend over the historical periods is :

1.7ms/century. [Historical eclipses and Earth's rotation by F. Richard Stephenson, Astronomy & Geo Physics, April 2003, Vol. 44, pp. 2.22 to 2.27.]

**APPENDIX(XI)    A BRIEF HISTORY OF LIFE.**

4.2Gya-First Organic Life emerged. The atmosphere consisted of Methane and Ammonia. Through incomplete fermentation energy was procured and $CO_2$ liberated into the atmosphere. Atmosphere was similar to that of Mars and Venus. During Archean Eon, Sun was only 75% as strong as it is today. Cold Climate was compensated by super-green house effect due to $CO_2$, $CH_3$ (Methane) and $NH_4$ (Ammonia). There was no protection against UV Rays. These lethal rays were destroying DNA material hence living organisms could not proliferate.

The most primitive method of procuring energy was anaerobic fermentation. In this sugar is split into energy, ethyl alcohol and Carbon Dioxide. Limited prokaryotic cells survived.

3.5Gya-$CO^2$ caused a mutational shift in the DNA and Chlorophyll protein was created. Chlorophyll enabled PHOTOSYNTHESIS.

Prokaryotic Cells (cells without nucleus proper) evolved into Eukaryotic Cells (cells with nucleus and DNA proper). Blue Green Algae, bacteria, protozoa and muticellular life emerged.

Now photosynthesis was available for procuring energy and for building organic building blocks from inorganic material available in the surrounding.

A byproduct of Photosynthesis was Oxygen. Oxygen developed Ozone layer thereby shielding the Earth from UV Rays. Presence of Oxygen produced a further mutational shift causing the formation of Cytochrome Protein. This enabled OXIDATION. This developed Animal Life.

Animals use Oxygen for energy and for building the organic building blocks and liberate Carbon-Dioxide whereas Plants use Carbon Dioxide and liberate Oxygen. Thus at Global level Plants and Animals resusticating each other. This enables the proliferation of all kinds of Plants and Animals life.

540Mya-Pre-Cambrian Explosion occurs-varied forms of life proliferate.

500Mya- The first jawless fish appear;

410Mya – The first land plants appear;

360Mya – The first Amphibians appear;

300Mya – The first reptiles appear;

230Mya – Dinosaurs and Mammals appear;

140Mya – The first Birds appear;

70Mya – The Flowering Plants appear;

56Mya – Whales, bats and modern Mammals appear;

5Mya-Hominids appear and they develop into Australopethicus Afraensis which have 400cc cranial capacity;

1Mya-Homohabilis, the handy man, with 700cc appear. They are tool makers.

500Kya-Homoerctus, the erect man, with 1000cc cranial capacity appears, He walks erect, knows the use of Fire and is Cave Deweller. He becomes hunter.

70kya – Homosapiens, the wise man, with 1400cc cranial capacity appears. He becomes a Herdsman, then a Farmer, then an Industrial Worker and then a Knowledge Worker.

**APPEDIX (XII)     MODELING OF EARTH-MOON SYSTEM AS A NON-LINEAR DYNAMIC SYSTEM**

Assumptions:
(1)E-M System is regarded as 2-body rotating system from its birth to its death in this   analysis. Whereas in fact it is a 3-body system including the tidal drag of Sun for Lunar orbital radius greater than ten times Earth's radius.

(2) Total angular momentum of E-M System has been assumed to be the scalar sum of the orbital angular momentum of E-M system, spin angular momentum of Earth and spin angular momemtum of Moon. The obliquity angle of Earth's spin axis with respect to the Ecliptic plane, which is α=23.45degree, and the inclination angle of Moon's orbital plane with respect to the ecliptic plane, which presently is i=5.15degree, make the total angular momentum the vector sum of the individual angular momentums.

By making the above two assumptions a certain loss of generality is involved but the calculation becomes tractable. The calculation will be further refined in a sequel paper.

## XII-A. THE CALCULATION OF THE CENTRE OF MASS OF E-M SYSTEM:

$Ed = m(r_L - d)$

Therefore **$d = (m/(m+E))r_L$**     **(XII.1)**

Substituting the numerical values of the present times:

$(d)_p = 4.67067E6$ m

p suffix means present.

### (XII-B) First Equation of motion:

Centripetal force on Moon = Centrifugal force due to orbiting Moon + ε′.

Therefore

**Centripetal acceleration = centrifugal acceleration + ε.**     **(XII.2)**

Where ε is residual radial acceleration because of which the outward radial velocity is being accelerated and subsequently retarded until it will be zero. Therefore ε can also be defined as the acceleration/deceleration rate of the velocity of recession.

At the present times,

**Centripetal acceleration = $GE/(r_L)^2 = 2.696738488E\text{-}3 \, m/sec^2$ .**     **(XII.3)**

Centrifugal acceleration = $v^2/(r_L - d)$

$= [\Omega_L(r_L - d)]^2/(r_L - d) = (\Omega_L)^2 r_L \, 1/(1+m/E)$

**Centrifugal acceleration = $(\Omega_L)^2 r_L \, 1/(1+m/E) = 2.694518066E\text{-}3 \, m/sec^2$**     **(XII.4)**

where G = 6.67E-11 Newton-$m^2$/Kg.

GE = $3.9847914 \wedge 14$ ($m^3/sec^2$);

$r_L = 3.844E8$ m.

$\Omega_L = (2\pi)/(27.3 \text{solar days/revolution} \ast 86400 \text{sec/solar day})$

$\varepsilon = 2.220422E\text{-}6 \, m/sec^2$.

This implies an inward residual acceleration which is retarding the outward radial velocity of Moon in the present epoch. As Moon spirals out, Earth's spin period (24 hours presently) and Lunar orbital period (27.3 solar days) differential decreases hence the lunar tidal drag on Earth as well as the residual inward acceleration of Moon, both decrease.

In Section(XII-D-ii) the following Boundary Equation has been derived on Keplerian considerations:

$$\Omega_E/\Omega_L = \text{lom/lod} = Ex^{1.5} - Fx^2 \qquad \text{(XII.5)}$$

**where**
$\Omega_E$ = angular spin velocity of the Earth= $2\pi/T_E$;
$T_E$ = length of the mean solar day;
$\Omega_L$ = angular orbital velocity of the Moon=$2\pi/T_m$ ;
$T_m$ = length of the sidereal month and not the synodic or lunar month;
$J_T$=3.440488884E34 Kg-m$^2$/sec = total angular momentum of Planet –Satellite System,
$E = J_T/(BC) = 2.135936694E-11$ m$^{-3/2}$,
$B = \sqrt{(GE(1+m/E))} = 2.008433303E7$ m$^{3/2}$/sec,,
E=mass of the Earth;
G=Gravitational Constant;
GE=3.9847914^14 m$^3$/sec$^2$ ,
m=mass of the Moon=7.348^22 Kg,;
C=Rotational Inertia of the Earth around the spin axis=8.02E37 Kg-m$^2$,,
F=m/(C(1+m/M))= 9.050770289E-16 m$^{-2}$ .

The roots of the Boundary Equation give the inner and outer Geosynchronous Orbit $a_{G1}$ and $a_{G2}$.

$$a_{G1} = 1.46142^7 m, \qquad a_{G2} = 5.5293^8 m. \qquad \text{(XII.6)}$$

Only at these Orbits the Satellite is in Keplerian Equilibrum where centripetal and centrifugal forces are in balance and radial acceleration and radial velocity are zero. But the Satellite is never allowed to stay in the Keplerian Orbits.

At the inner Geosynchronous Orbit slightest differential between $\Omega_E$ **and** $\Omega_L$ causes the Satellite to tumble out of the Keplerian Orbit. If the Satellite is long of $a_{G1}$ it is launched on a outward expanding spiral path and if it is short of $a_{G1}$ it is ejected onto a inward collapsing spiral path as Phobos (Martian Satellite) is launched.

If the Natural Satellite is formed by the capture of Asteroid just beyond $a_{G1}$ it is launched on a outward spiral path. The Natural Satellite experiences the maximum outward acceleration at $r_L = x_1$ . Thereafter the maximum outward acceleration reduces to null point at $r_L = x_2$. At this point recession velocity is maximum $V_{max}$ Thereafter acceleration becomes negative and recession velocity starts decelerating until it reaches the outward Geo-Synchronous Orbit where the recession velocity is zero. At the outer Geosynchronous Orbit, Solar Tides perturb the Satellite out of the Keplerian Orbit and once more the Satellite is launched onto a non-Keplarian journey but this time on an inward collapsing spiral path. Thus when the satellite perturbed out of the inner Geo-synchronous Orbit it experiences a Gravitational Runaway Phase from $a_{G1}$ to $x_2$. Thereafter there is no transfer of spin rotational energy from the Earth to the Moon. By virtue of the initial Gravitational Boost it continues to coast on its own. $a_{G1}$ to $x_2$ is Gravitational Boost phase and from $x_2$ to $a_{G2}$ is coasting phase along an outward spiral path climbing up the Gravitational Potential Well created by the Earth.

Charon is in Outer Geo-Synchronous Orbit around Pluto and the system is in stable equilibrium [Sharma 2004]. Pluto-Charon is an exception because of 90degree obliquity of Pluto and almost transverse Orbital Plane of Charon, transverse to the Ecliptic plane. Charon continues to be in stable equilibrium in the outer Geo-Synchronous Orbit.

If the Natural Satellite is formed by accretion from Impact Generated Debris, the merged body formed at the edge of Roche's zone [Ida et al 1997] recoils from inner debris disk, inner to $a_{G1}$, which is rotating twice or thrice as fast as the Natural Satellite[Ward et al 2000]. This recoil also leads to a rapid build up of recession velocity of the Natural Satellite. Subsequently with Orbital Evolution the recession velocity decelerates in much the same way as described before.

This discussion tells us that the radial velocity of a Satellite should be of the form:

$$dr_L/dt = [lom/lod - 1] = K_1[Ex^{1.5} - Fx^2 - 1] \quad (XII.7)$$
$$\text{where } K_1 = 4.58924 \wedge (-11) \text{ m/sec.}$$
$$E = 2.135936694 \wedge (-11)(1/m \wedge 1.5),$$
$$F = 9.050770289 \wedge (-16)(1/m \wedge 2).$$

Plot of Eq.(XII.7) is given in Figure(XII.1)

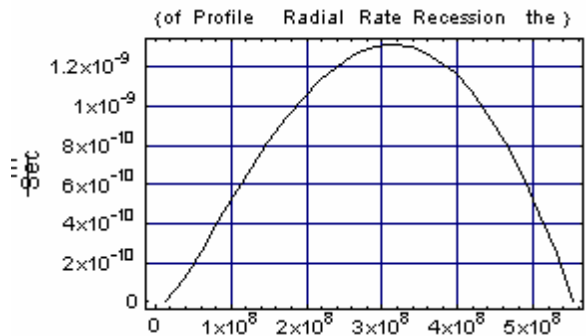

**Figure(XII.1) The general form of the profile of Recession Velocity of a given Satellite. X axis- $r_L$ (m) & Y axis-recession velocity(m/sec)**

Differentiating Eq.(XII.7) we obtain the following expression for the residual acceleration experienced by the Satellite:

$$= d^2r/dt^2 = K_2(1.5Er^{0.5} - 2Fr)(Er^{1.5} - Fr^2 - 1)$$
(XII.8)

where $E = M/(BC) = 2.135936694 \wedge (-11)$ $(1/m^{3/2})$
$F = m/(C(1+F)) = 9.050770289 \wedge (-16)$ $(1/m^2)$

The residual acceleration in the present epoch is:
$= -2.220422 \wedge (-6)$ m/sec$^2$ . (negative sign shows that it is directed towards the centre of the Planet Satellite System and is retarding the recession velocity).

The constant of proportionality $K_2$ is determined to be :

$$(-2.220422 \times 10^{-6}) \div ((3.20390504099999962\wedge\text{-}11\, x^{0.5} - 1.81015405780000016\wedge\text{-}15\, x)$$
$$(2.135936694 \times 10^{-11} \times x^{1.5} - 9.050770289 \times 10^{-16} \times x^2 - 1)) /.$$
$$x \to 3.844 \times 10^8$$

$K_2 = 1.2506471567643087$ (m$^2$/sec$^2$)

The profile of the residual acceleration experienced by The Satellite during its total evolutionary span is given in Figure(XII.2).

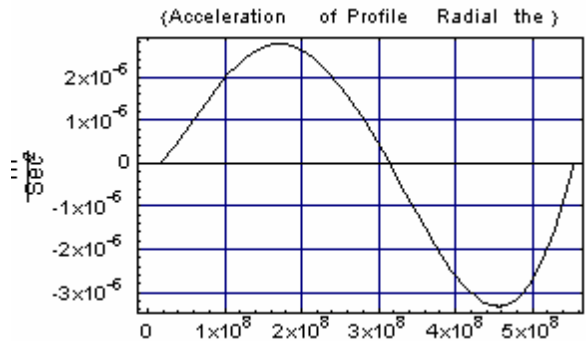

**Figure (XII.2) The Profile of Radial Acceleration experienced by the Satellite during its evolutionary span.**
X axis- $r_L$ (m), Y axis- radial acceleration (m/sec$^2$).

The equation implies an initial boost to Satellite's Recession Velocity because of Gravitational Runaway Phase. Early in the evolutionary history it reaches a maxima. Thereafter the Recession Velocity starts decelerating. Eventually the Recession velocity decelerates to a zero velocity.

The profiles give the general pattern of the Recession Velocity and the Residual Radial Acceleration but are not in exact conformity with the physical condition. The reason being that Eq.(XII.8) experiences a long drawn Gravitational Runaway Phase in the evolutionary history whereas in fact it is an impulsive phenomena and over a short distance of $(x_2 - a_{G1})$ the Runaway Phase gets damped out due to tidal dissipation .For mathematical tractibility we will use Eq.(XII.8) as

">64

the expression for residual acceleration of the satellite. Even though Eq.(XII.8) is not an accurate representation of residual acceleration the inaccuracy in the form gets corrected in the course of analysis.

This Equation also implies that the forces are in exact equilibrium at both the Geo-synchronous orbits . The equilibrium at the inner Geo-Synchronous orbit is never stable whereas the equilibrium at the outer Geo-Synchronous may be stable.

Geo-synchronous by definition means that at these two Lunar Orbital Radii the Moon's Sidereal Orbital Revolution Period is equal to Earth's Axial Spin Period. Also the Satellite is in captured rotation. That is the Satellite's spin period and orbital period are equal and the satellite shows the same face to the Planet all the time. In other words at the two Geo-Synchronous orbits the bodies are mutually interlocked

For the given Globe and Orbit parameters of E-M System, $a_{G1}$ and $a_{G2}$ are $1.46142*10^7$m and $5.52933*10^8$ m respectively.

$a_{G2}$ is the terminal point of the evolving lunar orbital radius .

These two Geo-synchronous Orbital Radii are calculated from the Boundary Equation which has been derived from the analytical relationship of lom/lod where lom is the Sidereal Orbital Period of Moon and lod is length of day or the Mean Solar Day .

Lom/lod is less than UNITY at Orbital Radii less than $a_{G1,}$

Lom/lod is UNITY at $a_{G1,}$

Lom/lod is greater than UNITY at Orbital Radii greater than $a_{G1,}$

Lom/lod is 27.3 presently,

Lom/lod will again be UNITY at $a_{G2,}$

Lom/lod will be negative beyond $a_{G2}$ .

Since negative value of lom/lod is physically untenable hence a natural satellite is forbidden from entering the spatial region beyond $a_{G2 .}$

$a_{G2}$ also defines the gravitational sphere of influence of a primary body. As long as the secondary body is within this gravitational sphere of influence,

(a) it will be considered to be the SATELLITE of the primary if primary is a PLANET

(b)and the secondary will be considered to be PLANET if the primary is a STAR

(c)and the secondary will be STAR PAIR if secondary mass is greater than $13m_{Jup}$ . In all stars, star pairs occur. Neutron Star forms a star pair and Black Hole forms a star pair.

[This raises the question as to what is a Planet? {Extra-Solar Planets: Constraints for Planet Formation Models}
International Astronomical Union has proposed a working definition
www.sciencemag.org/cg/eletters/291/5508/1487b?ck=nck
www.kepler.arc.nasa.gov

- (i) Objects with masses below the limiting mass for thermonuclear fusion of deuterium, currently calculated to be near $13m_{jup}$ for objects of solar metallicity [Saumon et al, *Astrophys. J.* **460,** 993(1996)] that orbit stars or solar remnants are planets( no matter how they form). The minimum mass or size required required for an object to be considered a planet should be the same as that used in the solar system;

(ii) Substellar objects with masses above the limiting mass for thermonuclear fusion of deuterium are brown dwarfs;

(iii) Free floating objects ( here free floating implies the secondary bodies which have come out of the gravitational sphere of influence of the primary as defined by $a_{G2}$ ) in young star clusters with masses below the limiting mass for thermonuclear fusion of deuterium are not planets but are lost planets.]

The Lunar tidal braking torque on Earth is assumed to be:

**idal torque= = [(K/X$^M$)[(lom/lod)-1]= (K/X$^M$) [EX$^2$/Y -FX$^2$-1] (XII.9)**
**Where Earth's Structure factor=(K/X$^M$),**
$Y = [\sqrt{X}-1.5EDX^3 +2FDX^{3.5}+1.5E^2DX^{4.5}-3.5DEFX^5+2DF^2X^{5.5}]$,
$E=M/(BC)=2.135936694\wedge(-11)$ (1/m$^{3/2}$),
$F=m/(C(1+m/E)= 9.050770289\wedge(-16)$ (1/m$^2$),
$D=K_2/(2GE) = 1.569275528\wedge(-15)$ (1/m),
$K_2= 1.250647157$ (m/sec)$^2$,
$GE=3.9847914\wedge(14)$ (m$^3$/sec$^2$)
$J_T=M=$ total angular momentum of Planet-Satellite System (here it is E-M System) 3.440488884E34Kg-m$^2$/sec,
C=moment of inertia of Earth around spin axis=8.02E37Kg-m$^2$,
B=2.008433303E7m$^{3/2}$/sec,
X=Lunar orbital radius of the given epoch,
m= 7.348$\wedge$22 Kg,
K= 1.018106946$\wedge$43 (Kg-m$^{(M+2)}$/sec$^2$)

From Keplerian Approximation:

Lom/lod= Ex$^{1.5}$ – Fx$^2$

From Non-Keplerian consideration:

Lom/lod= Ex$^2$/Y – Fx$^2$
Where $Y =[\sqrt{X}-1.5EDX^3 +2FDX^{3.5}+1.5E^2DX^{4.5}-3.5DEFX^5+2DF^2X^{5.5}]$,

In Eq(XII.9) the second form of lom/lod based on Non-Keplerian consideration has been adopted hence it accurately describes the evolving Satellite but as we will see there is little difference between the two except while calculating lom. The first form gives lom/lod=27.2396 whereas the second form give 27.3056

Multiplication by the Earth's Structure Factor helps shift the maxima of radial acceleration profile to the Gravitational Resonance Point $x_1$ and the maxima of Recession Velocity profile to the end of Runaway Phase namely $x_2$. The form of tidal torque as given by Eq.(XII.9) is in conformity with the physical reality as we will see subsequently.

Exponent **M** of the Structure Factor **(K/X$^M$)** is determined from the extent of Runaway phase $x_2$.

K is determined from the present rate of recession of Moon. In cases where present rate of recession is unknown there $V_{max}$ at $x_2$ will help determine K. The end of the Gravitational Runaway Phase $x_2$ will be determined from Energy Equations and ,with the help of Runaway Phase $x = x_2 - a_{G1}$ and Boost Factor, $V_{max}$ can be determined. Thus there is a single underlying theoretical formulation for understanding and solving all the parameters of an evolving two body system.

The form of the function representing Tidal Torque implies that:

(a) in the range of $a_{G1}$ to $a_{G2}$, Tidal Torque is positive and is retarding the spin of the Earth and angular momentum is transferred from Earth to Moon leading to outward evolving spiral path of the Moon and

(b) for Orbital Radii less than $a_{G1}$ and greater than $a_{G2}$, Tidal Torque is negative and is accelerating the spin of Earth and angular momentum is transferred from Moon to Earth. The reversal of transfer leads to an inward or collapsing spiral path of the Moon.

## XII-C. CALCULATION OF THE INITIAL POINT OF THE SPIRAL TRAJECTORY.

Moon was possibly formed by accretion from a circumterrestrial disk or debris generated by a giant impact on Earth [ Ida. S, Canup. R.M.& Stewart G.R. 1997] about 4.53 billion years ago [Kerr, R. A. 2002 , Cameron, A.G.W. 2002]. This Time Scale has been revised to 4.476Gy BP.

In any Planet-Satellite system there are two Geo-Synchronous Orbits-the inner one($a_{G1}$) and the outer one($a_{G2}$). So does our Earth-Moon system have two Geo-Synchronous orbits: one at $a_{G1} = 1.46142 \wedge 7$ m and the second at $a_{G2} = 5.52887891 \wedge 8$ m. The proof of this comes later.

The Roche Limit is defined as :

$$a_R = 2.456(\rho_E/\rho_L)^{1/3} R_E \sim 2.9 R_E = 18.496606 \wedge 6 \text{ m}$$
$$\rho_E = \text{density of Earth} = 5.5 \text{gms/c.c.,}$$
$$\rho_L = \text{density of Moon} = 3.34 \text{gms/c.c.,}$$

and Roche Zone is defined within the range of 0.8-1.35$a_R$ i.e. within $14.7972848 \wedge 6$ to $24.9704181 \wedge 6$m range [Ida et al 1997]. This implies that impact generated debris will be prevented from accretion within $1.48 \wedge 7$ and those within $1.48 \wedge 7$ to $2.5 \wedge 7$m range also known as transitional zone will experience limited accretion growth whereas those lying beyond this zone will be unaffected by tidal forces. It is a happy coincidence that the Roche zone lies just beyond the inner Geo-Synchronous orbit of the Earth-Moon System. The inner Geo-Synchronous orbit is also known as Cassini State 1[Gladman, B.Quin, D.D. Nicholson,P. & Rand, R 1996]. This implies that if accretional criteria of Canup & Esposito [1996] is satisfied along with the impact velocity condition that is the rebound velocity should be smaller than the mutual surface escape velocity then merged body formation of Moon starts within the Roche zone. The accreted Moon gradually migrates outward, because of the differential between lom and lod, sweeping the remnant debris.

Since the Roche Zone lies beyond $a_{G1}$ therefore in Earth-Moon system the newly formed Moon lies beyond the first Geo-Synchronous Orbit which implies a slower Lunar orbital rotation as compared to Earth's axial spin hence by decelerating the axial spin of Earth, Moon spirals out by transfer of angular momentum from Earth's axial spin to Moon's orbital rotation.

If the Roche Zone was situated within $a_{G1}$ then the merged body's orbital rotation rate would have been faster than Earth's axial spin and hence the merged body would have speeded up the spinning Earth leading to angular momentum transfer from Moon to Earth hence an inward collapsing spiral path of Moon..

The final position of fully formed Moon lies in the range $0.9\text{-}1.6 a_R$ i.e.16.65E6m-29.6E6m.

The final scenario assumed in this paper is the following:

Around 4.56Ga(billion years ago) a Supernova Explosion in the neighborhood of our Solar System led to the birth our Solar Nebula[Maddox 1994]. The central part of the nebula condensed to form the Sun and the remaining disc of gas and dust accreted to form the eight planets, Kuiper Belt Objects, asteroids, meteorites and comets. In about 30 million years the Earth was fully formed and about the same time Mars sized planetismal struck the fully formed Earth . This event occurred about 4.53 Ga. From the impact generated circumterrestrial debris Moon was partially formed and significant tidal interaction began at a distance of 15,700Km . Since 15,700Km is beyond the inner Geo-Synchronous Orbit, there is significant differential between Earth's spin rate and Lunar orbital rate. The slowly orbiting Moon causes a tidal drag torque on a much faster spinning Earth leading to the angular momentum transfer from Earth to Moon. This launches Moon on a outward spiral path. The evolutionary period since the beginning of Moon's journey at a Lunar Orbital Radius of 15,700Km to the present position of 384,400 Km should come out to be 4.53 billion years since that is the tentative age of Moon. In this analysis the age comes out to be 4.52956 billion years.

Till date no Researcher has been able to arrive anywhere near this time span. The most recent work by Krasinsky [2002] arrives at 1.9 billion years.

Assuming a constant rate of tidal dissipation as suggested by the present rate of Lunar Recession of 3.8cm/yr leads to an even shorter age of 1.5 billion years [Williams, 2000].

If a rate of tidal dissipation consistent with 3.16cm/yr is assumed then the age comes out to be 1.9 billion years. This rate of Lunar Recession is suggested by paleontological data provided by Lambeck [1980] for Phanerozoic period specimens[Williams,2000].

A rate of tidal dissipation consistent with 2.17cm/yr which possibly occurred 620Ma leads to an evolutionary age of 3 billion years[Williams,2000].

A rate of tidal dissipation consistent with 1.04 cm/yr leads to an age of 8-9 billion years which is impossible[Williams,2000].

So a mathematical analysis based on assumed rate of tidal dissipation leads to this pitfall of untenable evolutionary period of Earth- Moon system.

In the present non-linear dynamical approach nothing has been assumed.

Sometime in remote future at Orbital radius $a_{G2}$ = 5.52933E8 m, the terminal point of the outward spiral lunar path, Moon will become geosynchronous, i.e. Moon's orbital period=Earth's spin period=47.07 present solar days.

[This calculation will come later].Moon will be in a geostationary orbit.

At this point Sun's tidal drag will try to further slow down Earth to a synchronous or captured rotation orbit whereby Earth's spin period =Earth's orbital period=365.25 solar days and Earth will present the same face to Sun as Moon presents the same face to Earth today. But simultaneously Moon will try to speed up Earth's spin. Because of the proximity of Moon the net effect will be the speeding up of Earth's spin and a net transfer of angular momentum to Earth. To conserve the angular momentum of the system, Moon's orbital momentum will be forced to decrease and Moon will start spiraling inward until it spirals down to Earth and crashes. But much before this celestial crash, Sun will have exhausted its nuclear fuel , expanded into a Red Giant and devoured the inner terrestrial planets including Earth. This will happen 5 billion years [Garlick,2002] in future therefore Moon will never get a chance to reach the outer Geo-synchronous orbit even. So there is no possibility of Moon falling into inward collapsing spiral path.

**XII-C-i.          CALCULATION OF TOTAL ANGULAR MOMENTUM.**

$J_T = (J_{spin})_E + (J_{spin})_m + (J_{orbit})$
          **(XII.10)**

$(J_{orbit}) = Ed^2\Omega_L + m(r_L-d)^2\Omega_L$
          **(XII.11)**

Substituting Eq.(1) in Eq.(8),

$J_{orbit} = (m/(1+m/E))(r_L)^2\Omega_L$
          **(XII.12)**

Substituting the present day parameters( Table.I.1),
$\Omega_L, r_L$ and E and m/E=0.0123,

$[J_{orbit}]_p = 2.857234392E34$ Kg-m$^2$/sec.
(XII.13)

$[J_{spin}]_E)_p = C[\omega_E]_p = 0.5832308584E34$ Kg-m$^2$/sec.
(XII.14)

where $C = 8.02E37$ Kg-m$^2$, $[\omega_E] = 2\pi/(1$ solar day*86400secs/solar day)

$[J_{spin}]_m)_p = (2/5)m(R_m)^2[\omega_m]_p = 2.363359753E-5E34$ Kg-m$^2$/sec
(XII.15)

Where $R_m = 1.738E6$ m,
$[\omega_m]_p = (2\pi)/(27.32$ solar days/rotation*86400secs/solar day)

Substituting Eq.(XII.13),(XII.14) and (XII.15) in (XII.10),

$J_T = 3.440488884E34$ Kg-m$^2$/sec
(XII.16)

Since lunar spin angular momentum constitutes an insignificant component of the total angular component therefore $(J_{spin})_m$ will be neglected throughout the analysis and $J_T$ will be expressed as follows:

$$J_T = C\omega_E + (m/(1+m/E))(r_L)^2\Omega_L = C\omega_E + (m/(1+m/E))(r_L)^2\Omega_L$$
(XII.18)

### XII-C-ii. CALCULATION OF THE SIDEREAL PERIOD $(T_M)_{SIDEREAL}$:

Rewriting Eq.(XII.2),
$GE/(r_L)^2 = (\Omega_L)^2 r_L/(1+m/E) -$
$(\Omega_L)^2 = GE(1+m/E)/(r_L)^3 + ((1+m/E)/r_L)$
$(\Omega_L)^2 = GE(1+m/E)/(r_L)^3[1 + (r_L)^2/(GE)]$

Multiplying both sides by $(r_L)^4$:
$(\Omega_L)^2(r_L)^4 = (GE(1+m/E)r_L)[1 + (r_L)^2/(GE)]$

Substituting the analytical form of the residual acceleration from Eq.(XII.5) in the above equation we get,

$(\Omega_L)^2(r_L)^4 = G(E+m)r_L[1 + (K_2/(GE))(r_L)^2 (1.5Er^{0.5} - 2Fr)(Er^{1.5} - Fr^2 - 1)]$

Square rooting both sides:

$$\Omega_L(r_L)^2 = \sqrt{(G(E+m)r_L)}[1 + (K_2/(GE))(r_L)^2(1.5Er^{0.5}-2Fr)(Er^{1.5}-Fr^2-1)]^{1/2}$$
(XII.19)

Let $D = (K_2/(2GE)) = 1.569275528 \wedge (-15)$ (1/m)
Where $K_2 = 1.250647157$ (m/sec)$^2$,
$GE = 3.9847914 \wedge 14$ (m$^3$/sec$^2$),
$= D(r_L)^2(1.5Er^{0.5}-2Fr)(Er^{1.5}-Fr^2-1)$,
Let $B = \sqrt{(GE(1+m/E))} = 2.008433303 \wedge 7$ (m$^{3/2}$/sec),

The value of Delta is tabulated in Table(XII.1);

**Table XII. 1. Values of Delta for the whole range of rL.**

| $r_L(*10^8)$m | $a_{G1}$ | 3.68 | 3.844 | 4 | 4.8 | 5.0 | 5.3 | $a_{G2}$ | 5.6 |
|---|---|---|---|---|---|---|---|---|---|
| $(*10^{-6})$ | 0 | -302.46 | -411.69 | -524 | -910 | -853 | -526 | 0 | +223 |

Thus it is correct to state that:
At all times $(K_2/(GE))(r_L)^2(1.5Er^{0.5}-2Fr)(Er^{1.5}-Fr^2-1) \ll 1$;

Therefore it is valid to apply Binomial expansion to Eq.(XII.16) and neglect the terms of higher order of smallness.
Therefore
$$\Omega_L(r_L)^2 = B\sqrt{r_L}[1 + D(r_L)^2(1.5Er_L^{0.5}-2Fr_L)(Er_L^{1.5}-Fr_L^2-1)]$$

or $\Omega_L(r_L)^2 = B\sqrt{r_L}[1 + D(r_L)^2(-1.5Er_L^{1/2}+2Fr_L + 1.5E^2r_L^2 - 3.5EFr_L^{2.5} + 2F^2r_L^3]$

or $\Omega_L(r_L)^2 = B[\sqrt{r_L} - 1.5EDr_L^3 + 2FDr_L^{3.5} + 1.5E^2Dr_L^{4.5} - 3.5EFDr_L^5 + 2F^2Dr_L^{5.5}]$

$$\Omega_L(r_L)^2 = B[\sqrt{r_L} - 1.5EDr_L^3 + 2FDr_L^{3.5} + 1.5E^2Dr_L^{4.5} - 3.5EFDr_L^5 + 2F^2Dr_L^{5.5}]$$
(XII.20)

Let $Y = [\sqrt{r_L} - 1.5EDr_L^3 + 2FDr_L^{3.5} + 1.5E^2Dr_L^{4.5} - 3.5EFDr_L^5 + 2F^2Dr_L^{5.5}]$

$X = r_L$.
Substituting these new symbols we get:

$$\Omega_L = BY/X^2$$
(XII.21)

**Therefore** $\Omega_L = [BY/X^2] = 2\pi/(T_m)_{sidereal}$ (XII.22)

Where $Y = [\sqrt{X} - 1.5EDX^3 + 2FDX^{3.5} + 1.5E^2DX^{4.5} - 3.5EFDX^5 + 2F^2DX^{5.5}]$
$B = 2.008433303E7 \, m^{3/2}/sec$,
$E = M/(BC) = 2.135936694E(-11) \, (1/m^{3/2})$,
$F = m/(C(1+m/E)) = 9.050770289E(-16) \, (1/m^2)$,
$D = K_2/(2GE) = 1.569275528E(-15) \, (1/m)$,

## XII-D-i. CALCULATION OF THE SOLAR DAY ON EARTH AND LOM/LOD.

Rewriting Eq.(XII.9),(XII.11) and (XII.12), the general expression of the total angular momentum is:

$J_T = 3.440488884E34 \, Kg\text{-}m^2/sec$
$= (J_{spin})_E + (J_{spin})_m + J_{orb} = C\omega_E + 0.4*0.0123E*R_m^2\omega_L + 0.0123E(r_L)^2(\Omega_L)/1.0123$ (XII.23)

Since Moon is in captured rotation or synchronous rotation therefore $\omega_L = \Omega_L$ and as mentioned before the spin angular momentum will be neglected in all analysis.

$$J_T = \omega_E C + (m/(1+m/E))\Omega_L r_L^2 \quad (XII.24)$$

Substituting Eq.(XII.20) in Eq.(XII.24),

$J_T = \omega_E C + (m/(1+m/E))B(\sqrt{X} - 1.5EDX^3 + 2FDX^{3.5} + 1.5E^2DX^{4.5} - 3.5EFDX^5 + 2F^2DX^{5.5})$ (XII.25)

$\omega_E = \{J_T - (m/(1+m/E))B(\sqrt{X} - 1.5EDX^3 + 2FDX^{3.5} + 1.5E^2DX^{4.5} - 3.5EFDX^5 + 2F^2DX^{5.5})\}/C = 2\pi/T_E$ (XII.26)

Dividing Eq.(XII.26) by Eq.(XII.22)

$$\omega_E/\Omega_L = lom/lod = (J_T - (m/(1+m/E))BY)/C)/(BY/X^2)$$

Simplifying the above Equation;

$$_E/_L = (X^2/BC)[J_T/Y - mB/(1+m/E)]$$

Taking $M = J_T$ and writing the full form of $Y$,

$$_E/_L = (X^2/BC)[M/(\sqrt{X} - 1.5EDX^3 + 2FDX^{3.5} + 1.5E^2DX^{4.5} - 3.5EFDX^5 + 2F^2DX^{5.5}) - mB/(1+m/E)]$$

$$_E/_L = EX^2/(\sqrt{X} - 1.5EDX^3 + 2FDX^{3.5} + 1.5E^2DX^{4.5} - 3.5EFDX^5 + 2F^2DX^{5.5}) - FX^2$$
$$(XII.27)$$

Where $M = 3.440488884E34 \text{ Kg-m}^2/\text{sec}$,
 $C$ = moment of inertia of Earth around spin axis = $8.02E37 \text{ Kg-m}^2$,
 $B = \sqrt{(GE(1+m/E))} = 2.008433303E7 \text{ m}^{3/2}/\text{sec}$,
 $GE = 3.9847914\wedge 14 \text{ m}^3/\text{sec}^2$,
 $X$ = Lunar orbital radus of the given epoch,
 $E = M/(BC) = 2.135936694\wedge(-11) \ (1/\text{m}^{3/2})$,
 $F = m/(C(1+m/E)) = 9.050770289\wedge(-16) \ (1/\text{m}^2)$,
 $D = K_2/(2GE) = 1.569275528\wedge(-15) \ (1/\text{m})$,
 $K_2 = 1.250647157 \ (\text{m/sec})^2$,
 $GE = 3.9847914\wedge(14) \ (\text{m}^3/\text{sec}^2)$
 $m$ = mass of the Moon = $7.348\wedge 22$ Kg,
 $m/(1+m/E) = 7.258717771\wedge 22$ Kg.

Solving Eq.(XII.23) we get the following set of values for $_E/_L$ = lom/lod which are tabulated in Table(XII.2).

**Table XII.2. Values of lom/lod for the whole range of rL.**

| rL(*10$^8$)m | 0.12 | a$_{G1}$=0.146142 | 3.844 | a$_{G2}$=3.844 | 6 |
|---|---|---|---|---|---|
| $_E/_L$=lom/lod | 0.75759 | 1 | 27.3059 | 0.999957 | -12.6 |

As is evident from the Table:
 lom/lod is less than Unity for $r_L < a_{G1}$,
 lom/lod is 1.000001905 for $a_{G1}$,
 lom/lod is greater than Unity for $a_{G1} < r_L < a_{G2}$,
 lom/lod is 27.305933 for the present epoch and
 lom/lod is negative for $r_L > a_{G2}$.

Negative lom/lod is physically untenable hence spatial region $r_L > a_{G2}$ is forbidden space for Natural Satellite trajectory. As soon as a Natural Satellite reaches the Outer Geo-synchronous Orbit it either continues to orbit in the Outer Geo-synchronous Orbit or it is perturbed out of the equilibrium condition because of Solar tides and it is launched on a collapsing spiral orbit.
 The physical dynamics is the following:

In the Outer Geosynchronous Orbit the Orbital Period of the Satellite is equal to the Spin Period of the mother Planet hence retarding torque on the mother Planet becomes zero. If the Solar tide has no effect on the Spin of the given Planet as is the case in Pluto-Charon because , the equatorial plane of Pluto and the Orbital Plane of Charon , both are transverse to the Pluto Orbital Radius Vector then the Natural Satellite continues in stable Geo-Synchronism. But if the Solar tide has a retarding effect on the Spin of the Planet as it will have in most cases where the obliquity is less than 30 degrees, then the Planet will spin at a slower rate as compared to the orbital rate of its Natural Satellite as a result the Natural Satellite will cause an accelerating torque on the mother Planet leading to the reversal of transfer of angular momentum. The transfer of angular momentum from the Natural Satellite to the mother Planet leads to inward launching of the Natural Satellite on a collapsing spiral trajectory.

At the beginning of times, before magmatic differentiation was completed within Earth, latter was a homogeneous spherical mass of axial moment of inertia of $(2/5)ER_E^2 = 9.723880431E37 Kg$-$m^2$. After the Giant Impact, when Earth was heated into a molten mass , the first phase of magmatic differentiation took place and heavier elements precipitated to form the heaviest Iron-Nickel core whereas basaltic-granites outer mantle was formed. Up to the boundary of Archean Eon (ancient) and Proterozoic Eon(early life) magmatic differentiation and internal stratification was incomplete. Before this boundary which occurs at 2.5b.yrs.Ga.there was metallic core and outer mantle composed of basalt and sodium rich granite. In the remote past that is before this boundary because of high amount of impact heating as well as because of higher level of radio-activity, intense heat was fuelling a faster plate-tectonic engine. As a result the primary continental crust was broken into hundred crustal plates. After Archean Eon the plate tectonic engine became slower, continental plates coalesced together to form twelve plates and deep recycling of the outer crustal plates led to sharply differentiated basaltic mantle and potassium-rich granites crust. Primary crust gave way to secondary crust and secondary crust gave way to tertiary crust. Because of rapid spin of Earth, the oblateness was 1% as compared to 0.3% in the modern era. In the past after the Giant Impact, Earth was homogeneous, much less stratified and much more oblate , very much like Venus. Today in the modern era Earth has an onion like internal structure with almost one order of magnitude less oblateness. If all these factors are taken into account the axial moment of inertia was almost $10E37 Kg$-$m^2$ in the remote past. Whereas in the modern era it is only $8.02E37 Kg$-$m^2$.[Runcorn,S.K. 1966].So to obtain realistic values of Solar Day length we must use the moment of inertia value which prevailed in the given era of interest. In our calculations we should assume a moment of inertia of Earth as $10E37 Kg$-$m^2$ before 2.5Ga.and $8.02E37 Kg$-$m^2$ after 2.5Ga.In actual calculation C has been taken as $8.02E37 Kg$-$m^2$.

**I-D-ii.         THE BOUNDARY EQUATION OF EARTH-MOON SYSTEM FROM LOM/LOD EQUATION.**

The Earth-Moon System is Geo-Synchronous at two values of Lunar Orbit radius:
Once at the inception and second time at the terminal value of the spiral trajectory of Moon.
Geosynchronism implies :

$$T_E/T_L = (T_m)_{sidereal}/T_E = 1 \qquad \text{(XII.28)}$$

Rewriting Eq.(XII.27),

$$T_E/T_L = EX^2/(\sqrt{X} - 1.5EDX^3 + 2FDX^{3.5} + 1.5E^2DX^{4.5} - 3.5EFDX^5 + 2F^2DX^{5.5}) - FX^2 \qquad \text{(XII.27)}$$

The above equation is valid for the total evolutionary history of any Planet-Satellite System but at the Inner and Outer Geo-synchronous Orbits it further simplifies because of the fact that wee bit imbalance between the centripetal and centrifugal forces is completely annulled and Planet-Satellite is exactly a Keplerian System. This means:

$$1.5EDX^3 + 2FDX^{3.5} + 1.5E^2DX^{4.5} - 3.5EFDX^5 + 2F^2DX^{5.5} = 0$$

Therfore at the two Geo-Synchronous Orbits, Equation(XII.26) reduces to:

$$T_E/T_L = 1 = EX^{1.5} - FX^2 \qquad \text{(XII.29)}$$

$$E = M/(BC) = 2.135936694 \wedge (-11) \ (1/m^{3/2}),$$
$$F = m/(C(1+m/E)) = 9.050770289 \wedge (-16) \ (1/m^2),$$

Equation(XII.29) is in a convenient form since all the constants involved are readily available for any Planet-Satellite Pair in form of the Globe and Orbit parameters of the same.

The above equation is further simplified into a fourth order Polynomial with four possible roots:

$$E(\sqrt{X})^3 - F(\sqrt{X})^4 = 1 \qquad \text{(XII.30)}$$

In the present Earth-Moon System:
$M/(BC) = 2.135936694 \wedge (-11) \ (1/m^{3/2})$;
$m/((1+F)C) = 9.050770288 \wedge (-16)(1/m^2)$;
$\sqrt{X} = Z$
and Equation(XII.26) becomes:

$$2.135936694 \wedge (-11) \, Z^3 - 9.050770288 \wedge (-16) Z^4 - 1 = 0$$

**(XII.31)**

Equation(XII.31) is solved for the roots by the following Mathematica Command.

```
Solve[2.135936694×10⁻¹¹ ×Z³ − 9.050770288×10⁻¹⁶ ×Z⁴ − 1 == 0, Z]

{{Z → −1868.93593220779798` − 2966.17373211387125`I},
 {Z → −1868.93593220779798` + 2966.17373211387125`I}, {Z → 3822.84974363729901`},
 {Z → 23514.5238872957884`}}
```

The last two roots are the square roots of $a_{G1}$ and $a_{G2}$. The squaring of the roots are obtained by the following command:

```
Z² /. %

{−5.30526509038490079`*^6 + 1.10871773382370442`*^7 I,
 −5.30526509038490079`*^6 − 1.10871773382370442`*^7 I, 1.46141801624277656`*^7,
 5.5293283364620418`*^8}
```

The two roots are:

**Inner Geo-Synchronous Orbital Radius = $a_{G1}$ = 1.4614180162 $\wedge$ 7 m.**

**(XII.32)**

**Outer Geo-Synchronous Orbital Radius = $a_{G2}$ = 5.52932833646 $\wedge$ 8 m.**

### XII.D.iii.     DETERMINATION OF "$K_2$" FOR ANY PLANET-SATELLITE PAIR

Equation(XII.27) gives the actual lom/lod experienced by the given Planet-Satellite Pair whereas Eq.(XII.30) gives the Keplerian lom/lod. For any P-S Pair the actual lom/lod is known as well as from orbit parameters the Keplerian lom/lod is also known. From the ratio of the two, constant of proportionality ($K_2$) of the retardation rate of recession velocity is easily determined.

$$(\text{E}/\text{L})_{actual} = EX^2/(\sqrt{X} - 1.5EDX^3 + 2FDX^{3.5} + 1.5E^2 DX^{4.5} - 3.5EFDX^5 + 2F^2DX^{5.5}) - FX^2$$

**(XII.33)**

$$(\text{E}/\text{L})_{Keplerian} = EX^{1.5} - FX^2$$

**(XII.34)**

Subtracting Eq.(XII.34) from Eq.(XII.33) we get:

$$EX^2(1/Y - 1/\sqrt{X}) = (\text{E}/\text{L})_{actual} - (\text{E}/\text{L})_{Keplerian} =$$

**(XII.35)**

where $Y = (\sqrt{X} - 1.5EDX^3 + 2FDX^{3.5} + 1.5E^2DX^{4.5} - 3.5EFDX^5 + 2F^2DX^{5.5})$

Rearranging the terms and writing the full form we get

$$K_2 = (2GE\Delta)/[(1.5EX^3 - 2FX^{3.5} - 1.5E^2X^{4.5} + 3.5EFX^5 - 2F^2X^{5.5})(EX)(1+\Delta\sqrt{X}/(EX^2))]$$
(XII.36)

Where $\Delta = (\tau_E/\tau_L)_{actual} - (\tau_E/\tau_L)_{Keplerian} = 27.3059 - 27.2396$ for E-M System for the present epoch.

GE = 3.9847914^14 (m^3/sec^2);
E = M/(BC) = 2.135936694^(-11) (1/m^{3/2});
M = J_T = 3.440488884^34 (Kg-m^2/sec);
B = √(GE(1+m/E) = 2.008433303^7 ((m^{3/2})/sec);
C = Moment of Inertia of Earth around the spin axis = 8.02^37 Kg-m^2;
F = m/(C(1+m/E)) = 9.050770289^(-16) (1/m^2);
m = 7.348^22 Kg.

Substituting the values for the different parameters we obtain:
  $K_2 = 1.250660291$.
This is precisely what we started with. See Eq.(XII.8).
This method will be very important in determining the constant of proportionality ($K_2$) of the residual acceleration experienced by the Natural Satellites in non-Keplerian State.

Tidal torque = $\tau = [(K/X^M)[(lom/lod)-1] = (K/X^{2.1}) [EX^2/Y - FX^2 - 1]$

(XII.37)
Where Earth's Structure factor = $(K/X^{2.1})$,
  $Y = [\sqrt{X} - 1.5EDX^3 + 2FDX^{3.5} + 1.5E^2DX^{4.5} - 3.5DEFX^5 + 2DF^2X^{5.5}]$,
  E = M/(BC) = 2.135936694^(-11) (1/m^{3/2}),
  F = m/(C(1+m/E) = 9.050770289^(-16) (1/m^2),
  D = $K_2$/(2GE) = 1.569275528^(-15) (1/m),
  $K_2$ = 1.250647157 (m/sec)^2,
  GE = 3.9847914^(14) (m^3/sec^2)
  $J_T$ = M = total angular momentum of Planet-Satellite System (here it is E-M System) 3.440488884E34 Kg-m^2/sec,
  C = moment of inertia of Earth around spin axis = 8.02E37 Kg-m^2,
  B = 2.008433303E7 m^{3/2}/sec,
  X = Lunar orbital radius of the given epoch,
  m = 7.348^22 Kg,
  K = 1.018106946^43 (Kg-m^{(2.1+2)}/sec^2)

The algebric sign of Eq.(XII.37) gives the direction of angular momentum transfer hence it describes outward expanding instability or inward collapsing instability.

Within $a_{G1}$, lom/lod is less than Unity hence Eq.(XII.37) is negative and the Parent Planet Earth is being spun faster. This leads to inward collapsing instability of the Satellite Moon.

Between $a_{G1}$ and $a_{G2}$, lom/lod is greater than Unity hence Eq.(XII.37) is positive and the Parent Planet is being spun slower leading to outward spiraling instability of the Satellite.

Beyond $a_{G2}$, Eq(XII.37) is always negative leading to inward collapsing inastability. Infact Satellite is always deflected at the outer boundry hence space beyond $a_{G2}$ is unconditionally forbidden for the Satellite.

Contemporary Solar Day in hours is calculated at $a_{G1}$ and $a_{G2}$ from Eq.(I.34)

$T_E$ at Inner Geo-Synchronous Orbit is 4.855 hours.

$T_E$ at Outer Geo-Synchronous Orbit is 47.078 Modern Solar Days.

**I-D-iv. DETERMINATION OF THE STRUCTURE FACTOR (K/X^M)**

In this determination it will not be valid to use Keplerian Equations as maximum non-Keplerian disequilibrum occurs at residual acceleration maxima. Nevertheless we will use Keplerian Equations for mathematical tractibilty. With the estimate of the Structure Factor determined from Keplerian approximation we will arrive at the precise form of the Structure Factor from the age of the Moon or the given Natural Satellite.

First the Gravitational Resonance Point is determined. The recession velocity profile has an inflection point at $X_1$ and a maxima point at $X_2$. At the point of inflection maximum residual radial acceleration occurs and at maxima point of the recession velocity profile the velocity maxima occurs.

The second time derivative of recession velocity becomes zero at $X_1$ and first derivative i.e. the radial acceleration becomes zero at $X_2$.

At $X_1$, lom/lod is 2 because that is when Gravitation Resonance occurs as already discussed.

Substituting 2 in Eq.(I.34), which is Keplerian, the solution comes out to be: $X_1 = 2.4087921 \char`\^ 7$ m.

Rewriting Eq.(XII.34);
$(\omega_E/\omega_L)_{Keplerian} = (X^2/BC)[M/(X^{1/2}) - mB/(1+m/E)]$
(XII.34)
$(\omega_E/\omega_L)_{Keplerian} = 2 = EX_1^{1.5} - FX_1^2$,
(XII.38)
where $E = M/BC = 2.135936694 \char`\^ (-11)$ $(1/m^{1.5})$,
$F = mB/(1+m/E) = 9.050770289 \char`\^ (-16)$ $(1/m^2)$.

Let $= EX_1^{1.5}$ and $= FX_1^2$ then $- -1 = 1$
(XII-39)

Rewrite Eq.(XII.12):

$$J_{orbit} = (m/(1+m/E))(r_L)^2 \Omega_L$$

(XII.12)

From Eq.(XII.20),
**Keplerian value of $(r_L)^2 \Omega_L = B\sqrt{X}$,**

Substituting the above value in Eq.(I.12),
$$J_{orbit} = (m/(1+m/E)) B\sqrt{X}$$

(XII.40)

Time derivative of Eq.(I.40) is:

**$dJ/dt = [(mB)/(2(1+m/E)\sqrt{X})](dX/dt) =$ Lunar Tidal Torque retarding the spin of Earth.** (XII.41)

In Eq.(XII.9) we have assumed the appropriate form for the tidal torque:

idal torque$= = [(K/X^M)[(lom/lod)-1] = (K/X^M) [EX^2/Y - FX^2 - 1]$

(XII.42)

**Where Earth's Structure factor $= (K/X^M)$,**
   $Y = [\sqrt{X} - 1.5EDX^3 + 2FDX^{3.5} + 1.5E^2 DX^{4.5} - 3.5DEFX^5 + 2DF^2 X^{5.5}]$,
   $E = M/(BC) = 2.135936694 \wedge (-11) \; (1/m^{3/2})$,
   $F = m/(C(1+m/E)) = 9.050770289 \wedge (-16) \; (1/m^2)$,
   $D = K_2/(2GE) = 1.569275528 \wedge (-15) \; (1/m)$,
   $K_2 = 1.250647157 \; (m/sec)^2$,
   $GE = 3.9847914 \wedge (14) \; (m^3/sec^2)$
   $J_T = M =$ total angular momentum of Planet-Satellite System (here it is E-M System) $3.440488884E34 \; Kg\text{-}m^2/sec$,
   $C =$ moment of inertia of Earth around spin axis $= 8.02E37 \; Kg\text{-}m^2$,
   $B = 2.008433303E7 \; m^{3/2}/sec$,
   $X =$ Lunar orbital radius of the given epoch,
   $m = 7.348 \wedge 22 \; Kg$,
   $K = 1.018106946 \wedge 43 \; (Kg\text{-}m^{(M+2)}/sec^2)$

Equating Eq.(XII.41) and Eq.(XII.42):

$[(mB)/(2(1+m/E)\sqrt{X})](dX/dt) = (K/X^M) [EX^2/Y - FX^2 - 1]$

Therefore Recession Velocity of Moon is:

$$(dX/dt) = (K/X^M) [EX^2/Y - FX^2 - 1] / [(mB)/(2(1+m/E)\sqrt{X})]$$

(XII.43)

Let $Y = [2K(1+m/E)]/(mB)$.
Assuming Keplerian State in the region of Gravitation Resonance i.e. $Y = \sqrt{X}$ and simplifying Eq.(I.43) we get:

$$dX/dt = Y[EX^{(2-M)} - FX^{(2.5-M)} - X^{(0.5-M)}] = (Y\sqrt{X}/X^M)f(X) = \text{Recession Velocity}$$

(XII.44)

where $f(X) = EX^{1.5} - FX^2 - 1$,
and $f(X_1) = 1$ from Eq.(I.39).
At $X_2$,
**Recession Velocity is maxima** $= (Y\sqrt{X_2}/X_2^M)f(X_2)$.

First time derivative of Eq.(I.44) is the residual acceleration of the satellite,

$$d^2X/dt^2 = Y^2\sqrt{X}/X^M)f(X)f'(X)$$

(XII.45)

where $f'(X) = (1/(X^M\sqrt{X}))[(2-M)EX^{1.5} - (2.5-M)FX^2 - (0.5-M)]$

The time derivative of Eq.(XII.45) is,

$$d^3X/dt^3 = Y^2[f(X)f''(X) + f'(X)f'(X)](dX/dt)$$

(XII.46)

where $f''(X) = (1/(X^M\sqrt{X}))[(1-M)(2-M)EX^{1.5} - (1.5-M)(2.5-M)FX^2 + (0.5+M)(0.5-M)]$

(XII.47)

Because of Gravitation Resonance there is a point of inflection in the velocity profile therefore we get:

$$d^3X/dt^3 \big|_{X1} = 0,$$

Therefore:

$$f(X_1) = -(f'(X_1))^2/(f''(X_1))$$

(XII.48)

At $X_1$,  $f(X_1) = \sqrt{(X_1)}/(X_1)^M$ ;
$f'(X_1) = (1/(X_1^M \sqrt{X_1}))[(2-M) - (2.5-M) - (0.5-M)]$
$f''(X_1) = (1/(X_1^M X_1 \sqrt{X_1}))[(1-M)(2-M) - (1.5-M)(2.5-M) + (0.5+M)(0.5-M)]$

(XII.49)

Substituting Eq(XII.49) in Eq.(XII.48) and simplifying we get:

$$[(1-M)(2-M) - (1.5-M)(2.5-M) + (0.5+M)(0.5-M)] = -[(2-M) - (2.5-M) - (0.5-M)]^2$$
(XII.49)

Using the condition of Eq.(XII.39) the above Equation is simplified to the following quadratic equation:

$$2M^2 - M(13-2) + (4.25-1.75) + 0.25(7- )^2 = 0$$

(XII.50)

At $X_1$, = 2.525151 and = 0.525151.
Substituting the numerical values the two roots are:
$$M = 1.885365515 \text{ and } 4.4.$$
The valid value which gives the correct age of Moon is :

$$M = 1.885365515$$
(XII.51)

Eq.(XII.44) gives recession velocity maxima at $X_2$.
Eq.(XII.45) gives the first derivative of the velocity profile.
Therefore equating Eq.(XII.45) to zero at $X_2$ gives the following equation:

$$M = (2EX_2^{1.5} - 2.5FX_2^2 - 0.5)/(EX_2^{1.5} - FX_2^2 - 1) = 1.885365515$$
(XII.52)

The value of $X_2$ which satisfies Eq.(XII.52) is,

$$X_2 = 7.2 \char`\^ 7 m.$$

From Lunar Laser Ranging, the present Velocity of Recession of the Moon is,

$$dX/dt = 3.8 cm/yr = 1.204173704 \char`\^ (-9) m/sec$$
(XII.53)

Substituting Eq.(XII.53) in Eq.(XII.44) we get the constant of proportionality K:

$$K = (mB/(2(1+m/E)))[dX/dt_{present}/\{EX^{(2-M)} - FX^{(2.5-M)} - X^{(0.5-M)}\}]$$
(XII.54)

$$mB/(1+m/E) = B = 1.457865116 \char`\^ 30 \text{ Kg-m}^{1.5}/sec$$

Substituting the numerical values in Eq.(XII.54),

$$K = 2.615059351 \wedge 31 \, Kg\text{-}m^{M+2}/sec^2 \tag{XII.55}$$

### I-D-vi. CALCULATION OF THE NON-KEPLERIAN TIDAL TORQUE.

**Substituting Eq.(XII.20) in Eq.(XII.12) we get the following equation:**

$$J_{orbit} = (m/(1+m/E)) \, B \, [\sqrt{r_L} - 1.5 E D r_L^3 + 2 F D r_L^{3.5} + 1.5 E^2 D r_L^{4.5} - 3.5 E F D r_L^5 + 2 F^2 D r_L^{5.5}] \tag{XII.56}$$

Redefinition:
$(m/(1+m/E))\sqrt{[GE(1+m/E)]} = B = 1.457865116 \, Kg\text{-}m^{3/2}/sec.$

Therefore
$$J_{orb} = B \, [\sqrt{r_L} - 1.5 E D r_L^3 + 2 F D r_L^{3.5} + 1.5 E^2 D r_L^{4.5} - 3.5 E F D r_L^5 + 2 F^2 D r_L^{5.5}] \tag{XII.57}$$

**Where B $= 1.457917826 \wedge 30 \, Kg\text{-}m^{3/2}/sec$,**
  **$D = 1.569275528 \wedge (-15) \, (1/m)$,**
  **$E = 2.135936694 \wedge (-11) \, (1/m^{1.5})$,**
  **$F = 9.050770289 \wedge (-16) \, (1/m^2)$.**

Rewriting Eq.(XII.10),

$J_T = J_{orb} + (J_{spin})_E + (J_{spin})_m.$

For all practical purposes, $(J_{spin})_m$ is negligible. Therefore:

$$J_T = J_{orb} + (J_{spin})_E \tag{XII.58}$$

Differentiating with respect to time(t),

$$dJ_{orb}/dt + d(J_{spin})_E/dt = 0 \tag{XII.59}$$

Therefore,
$$dJ_{orb}/dt = -d(J_{spin})_E/dt \tag{XII.60}$$

Eq(XII.60) implies that to conserve total angular momentum, as Earth's angular momentum is slowing down orbital angular momentum is increasing. This also implies that there is tidal braking torque which is slowing down the spinning Earth and there is a accelerating torque in reaction which is increasing the orbital angular momentum. The two torques are equal and opposite and the magnitude of each is given by:

$$\tau = dJ_{orb}/dt$$

(XII.61)

Differentiting Eq(XII.57) with respect to time we get the magnitude of the torque:

$$\tau = B \ [1/(2X^{1/2}) - 4.5EDX^2 + 7FDX^{2.5} + 6.75DE^2X^{3.5} - 17.5DEFX^4 + 11DF^2X^{4.5}] dX/dt$$

(XII.62)

**Where B $= 1.457917826 \wedge 30$ Kg-m$^{3/2}$/sec,**
   **D $= 1.569275528 \wedge (-15)$ (1/m),**
   **E $= 2.135936694 \wedge (-11)$ (1/m$^{1.5}$),**
   **F $= 9.050770289 \wedge (-16)$ (1/m$^2$).**

Kitt's Peak Observatory and Macdonald Observatory Laser Bouncing Data give the following rate of recession of Moon:
$$(dX/dt)_p = 3.8 cm/year = 3.8E\text{-}2m/31.556909E6s = 1.204173704E\text{-}9 m/sec.$$
(XII.63)

Substituting Eq(XII.63) in Eq(XII.62),
Braking torque in the present epoch is,

$$(\tau)_p = 4.450778135E16 \text{ N-m.}$$

(XII.64)

From Eq.(XII.42):
 idal torque $= = [(K/X^M)[(lom/lod)-1] = (K/X^M) \ [EX^2/Y \ -FX^2-1]$

(XII.65)

**Where Y $= [\sqrt{X} - 1.5EDX^3 + 2FDX^{3.5} + 1.5E^2DX^{4.5} - 3.5DEFX^5 + 2DF^2X^{5.5}],$**

From Keplerian Consideration, values of K and M have been determined in Eq. (XII.55) and (XII.51).
Substitutring these values we see that they donot precisely satisfy Eq.(XII.65) and the culprit is K since it had been determined using Keplerian Eq. for Recession Velocity whereas infact it should have been determined using the Non-Keplerian formula.
K $= 2.631 \wedge 31$ Kg-m$^{(M+2)}$/sec$^2$ exactly satisfy Eq.(I.65) with the present value of
**$(\tau)_p = 4.515682667E16$ N-m.**

$(\tau)_p = (K/X^M)[(lom/lod)-1] = (K/X^M)[EX^2/Y - FX^2 - 1] = 4.515682667E16$ N-m.

Therefore the final values of the Structure Constant are :

**K= 2.631^31 Kg-m$^{(M+2)}$/sec$^2$     & M= 1.885365515.**

These values exactly fit the present age of Moon of 4.53Gyrs and the present Torque of 4.515682667^16N-m.

**I-D-iii.     TIME INTEGRAL EQUATION OF MOON'S SPIRAL TRAJECTORY.**

In Eq(XII.65)an analytical form of braking torque was assumed where all the parameters have been determined from rigiropus theory.There is no empiricism involved.

$\tau =$ idal torque$= = [(K/X^M)[(lom/lod)-1] = (K/X^M)[EX^2/Y - FX^2 - 1]$

**(XII.65)**

Where $X = r_L$.

Equating Eq.(XII.62)and (XII.65),
**B $[1/(2X^{1/2}) - 4.5EDX^2 + 7FDX^{2.5} + 6.75DE^2X^{3.5} - 17.5DEFX^4 + 11DF^2X^{4.5}]dX/dt = (K/X^M)[EX^2/Y - FX^2 - 1]$**

**(XII.66)**

Rearranging the terms , Moon's Recession Velocity is obtained:

**dX/dt= $[(K/X^M)[EX^2/Y - FX^2 - 1]]/$ B $[1/(2X^{1/2}) - 4.5EDX^2 + 7FDX^{2.5} + 6.75DE^2X^{3.5} - 17.5DEFX^4 + 11DF^2X^{4.5}]$**

**(XII.67)**

Here B $=1.457917826E30$ (Kg-m$^{3/2}$)/sec.
Eq.(XII.67) gives Recession Velocity in m/sec.
Let F=B /(2*31.5569088E6sec/Solar Year)=2.309982E22 ((Kg-m$^{3/2}$-Solar Year)/sec$^2$)
Therefore,

**dX/dt= $[(K/X^M)[EX^2/Y - FX^2 - 1]]/$ B $[1/(2X^{1/2}) - 4.5EDX^2 + 7FDX^{2.5} + 6.75DE^2X^{3.5} - 17.5DEFX^4 + 11DF^2X^{4.5}]$**

**(XII.68)**

Eq.(XII.68) gives Moon's Recession Velocity in m/year.

Substituting the numerical values for the parameters in Eq.(XII.41),the Profile of Recession Velocity is plotted.

Eq.(XII.68) gives Moon's Recession Velocity in m/year.

Substituting the numerical values for the parameters in Eq.(XII.41),the Profile of Recession Velocity is plotted.

$$\text{Plot}\Big[\Big(\big(1.81557 \times 10^{33} \times \big(((1 \div (8.02 \times 10^{37}))\,(3.440488884 \times 10^{34} \times x^2) \div (2.008433307 \times 10^{7} \times$$
$$(\sqrt{x} - 9.142570518 \times 10^{-21} \times x^{5/2} + 9.142570518 \times 10^{-21} \times x^{7/2} \div (5.52887891 \times 10^{8}))) -$$
$$8.878598241 \times 10^{34} - 7.258980539 \times 10^{22} \times x^2\big) -$$
$$1\big) \div$$
$$x^{2.1}\big)\Big) \div$$
$$\big((2.309982 \times 10^{22})\,(1 \div \sqrt{x} - 5 \times 9.142570518 \times 10^{-21} \times x^{3/2} + 7 \times 9.142570518 \times 10^{-21} \times x^{5/2} \div (5.52887891 \times 10^{8}))\big),$$
$$\{x, 1 \times 10^{7}, 5.52887891 \times 10^{8}\}, \text{GridLines} \to \text{Automatic}, \text{Frame} \to \text{True},$$
$$\text{FrameLabel} \to (m/year), \text{PlotLabel} \to \{\text{"Recession Velocity from } 1 \times 10^{7} m \text{ to the Terminal Point"}\}\Big]$$

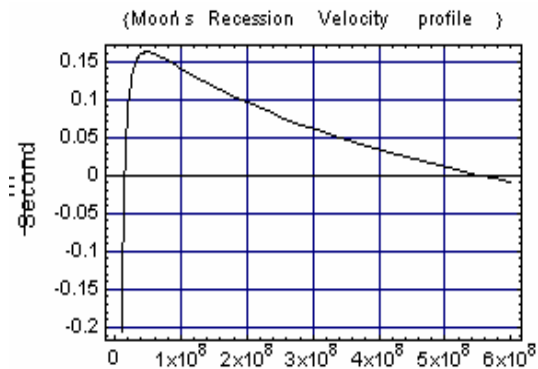

**Figure (XII.2) The profile of Moon's Velocity of Recession vs rL**

**Table XII. 3. Moon's Velocity of Recession vs rL**

| rL(*$10^7$)m | 1.0 | 1.1 | 1.2 | 1.3 | 1.4 | 1.46257569 |
|---|---|---|---|---|---|---|
| dX/dt(cm/sec) | -20.6 | -14.11 | -9.02 | -4.9837 | -1.726 | 0 |

| rL(*$10^8$)m | 1.844 | 2.844 | 3.844 | 4.844 | 5.844 |
|---|---|---|---|---|---|
| dX/dt(cm/sec) | 10.171 | 6.62 | 3.8 | 1.43 | -0.006174 |

As is clear from the Table (XII.2) as well as from the Figure(XII.2), Velocity of Recession is positive within the region bounded by $a_{G1}$ and $a_{G2}$. In the region beyond $a_{G2}$ the Velocity of Recession is negative as well as in the region within $a_{G1}$ also it is negative as already discussed.

These two Figure and Table also give the profile of Velocity of Recession required to achieve an evolutionary span of 4.53 billionn years. It is clear that Moon has been receding much faster in the past and is monotonically decreasing as it recedes farther and farther from the parent Planet until it reaches the Outer Geo Synchronous Orbit where it comes to rest and subsequently it starts spiraling inward.

Rearranging and integrating both sides we get,

$\int dt(limits, t_0, t_p)$
$= B/2 \int [(1/X^{1/2} - 5AX^{3/2} + 7AX^{5/2}/Y)/((K/X^{2.1})[(1/C)\{(MX^2)/(Y \ B\ ) - D\ - E\ X^2\} - 1])]dX(limits, X_0, X_p)$

(XII.42)

$(t_p - t_0)/(31.5569088E6 sec/solar\ yr.) =$
$B/(2*31.5569088E6) \int [[(1/X^{1/2} - 5AX^{3/2} + 7AX^{5/2}/Y)/((K/X^{2.1})[(1/C)\{(MX^2)/(Y\ B\ ) - D\ - E\ X^2\} - 1])]dX, X_0, X_p]$

**(XII.43)**

where $t_0 = 0.03$ b.yrs = instant of Giant Impact after the birth of our Solar Nebula.
$(t_p) = 4.56$ b.yrs = the present age of Earth.
$X_0 = 15,700$ km = 1.57E7m ≅ the distance from the center of Earth at which the Giant Impact generated debris accreted into partial Moon and significant tidal interaction began between Earth and Moon.
$X_p = 3.844$E8m = the present orbital radius of Moon.

By determining the above integral between $X_0$ and the present orbital radius the present age of Moon is determined which by observation has been determined to be 4.53 Ga.
Expressing Eq.(XII.43)in terms of Fx-570W constants,

**[($t_p - t_0$)/31.5569088E6]solar years =**

$$F\int [[(1/X^{1/2} - 5AX^{3/2} + 7AX^{5/2}/Y) / ((K/X^{2.1})[(1/C)\{(MX^2)/(Y'B) - D' - E'X^2\} - 1])]dX, X_0, X_p]$$
(XII.44)

where $F = B/(2*31.5569088E6) = (1.457917826E30 \text{ Kg-m}^{3/2}/\text{sec})/(2*31.5569088E6 \text{sec/solar year}) = 2.309982E22 \text{ Kg-m}^{3/2}\cdot\text{Solar Year/sec}^2$,

**$F = 2.309982E22 \text{ Kg-m}^{3/2}\cdot\text{Solar Year/sec}^2$,**
**$K = 1.702027611E15$ N-m.**
**$M = 3.440488884E34 \text{ Kg-m}^2/\text{sec}$,**

$\quad$ **C = moment of inertia of Earth around spin axis $= 8.02E37 \text{ Kg-m}^2$,**
$\quad$ **B $= 2.008433303E7 \text{ m}^{3/2}/\text{sec}$,**
$\quad$ **X = Lunar orbital radius of the given epoch,**
$\quad$ **D' $= 8.878598241E34 \text{ Kg-m}^2$,**
$\quad$ **E' $= 7.258980539E22$ Kg,**
$\quad$ **Y' $= (X^{1/2} - AX^{5/2} + AX^{7/2}/Y)$,**
$\quad$ **A $= 9.142570518E\text{-}21 \text{ m}^{-2}$,**
$\quad$ **Y $= 5.52887891E8$ m.**

When E = 1.57E7m and exponent in the structure factor is 2.1 then
The evolutionary period is 4.52956E9 yrs.

NIntegrate[
$1 \div ((1.81557 \times 10^{33} \times (((1 \div (8.02 \times 10^{37}))((3.440488884 \times 10^{34} \times x^2) \div (2.008433307 \times 10^7 \times (\sqrt{x} - 9.142570518 \times 10^{-21} \times x^{5/2}$
$+ 9.142570518 \times 10^{-21} \times x^{7/2} \div (5.52887891 \times 10^8))) - 8.878598241 \times 10^{34} -$
$7.258980539 \times 10^{22} \times x^2) - 1) \div$
$x^{2.1})) \div$
$((2.309982 \times 10^{22})(1 \div \sqrt{x} - 5 \times 9.142570518 \times 10^{-21} \times x^{3/2} + 7 \times 9.142570518 \times 10^{-21} \times x^{5/2} \div (5.52887891 \times 10^8)))),$
$\{x, 1.57 \times 10^7, 3.844 \times 10^8\}]$

$4.52955717669545609 \text{`*^9}$ Years = 4.52956 billion years evolutionary time span to cover the outward spiraling journey from 15,700Km to the present 3,84,400Km Lunar Orbital Radius

```
NIntegrate[
1÷((1.81557×10³³ × (((1÷(8.02×10³⁷)) ((3.440488884×10³⁴ × x²) ÷ (2.008433307×10⁷ × (√x − 9.142570518×10⁻²¹ × x^5
  9.142570518×10⁻²¹ × x^(7/2) ÷ (5.52887891×10⁸))) − 8.878598241×10³⁴ −
  7.258980539×10²² × x²) − 1) ÷
x^2.1)) ÷
((2.309982×10²²) (1÷√x − 5×9.142570518×10⁻²¹ × x^(3/2) + 7×9.142570518×10⁻²¹ × x^(5/2) ÷ (5.52887891×10⁸)))),
{x, 1×10⁷, 1.45×10⁷}]
```

If the integration is carried out from 10,000Km to 14,500Km we get a negative time(−1.22772988112015557`*^8 ) which implies that journey started from 14,500Km and ended at 10,000Km in the time span of 122Myrs.

−1.22772988112015557`*^8

Eq.(XII.44) can be interpolated to any epoch in the past or to any future epoch. If the orbital radius in any epoch is known then by integrating between the limits $X_0$(15,700Km)and $X_N$(Orbital radius of Moon in the unknown epoch),the chronological date of the unknown epoch can be pinned down. If the epoch is known then through several iterations of Eq.(I.44)the correct lunar orbital radius can be arrived at.

### XII-E. PALEONTOLOGICAL, PALEOTIDAL AND OTHER EVIDENCES OF THE SOLAR DAY LENGTH IN THE PAST.

John West Wells based [1963,1966]on the studies of Coral fossils and other marine creatures has given the data as tabulated in Table(XII.3).

**Table XII.4. The Observed lod based on the study of Coral Fossils.**

| t(yrs B.P.) | T*(yrs after the Giant Impact) | Length of Solar Day $T^*_E$(hrs) |
|---|---|---|
| 65Ma | 4.46456G | 23.627 |
| 135Ma | 4.39456G | 23.25 |
| 180Ma | 4.34956G | 23.0074 |
| 230Ma | 4.29956G | 22.7684 |

| 280Ma | 4.24956G | 22.4765 |
| 345Ma | 4.18456G | 22.136 |
| 380Ma | 4.14956G | 21.9 |
| 405Ma | 4.12456G | 21.8 |
| 500Ma | 4.02956G | 21.27 |
| 600Ma | 3.92956G | 20.674 |

Leschiuta & Tavella [2001] have given the estimate of the synodic month. From the synodic month we can estimate the length of the Solar Day as given in Appendix[II]. The results are tabulated in Table(I.4).

**TableXII. 5.Observed Synodic Month[Leschiuta &Tavella 2001 based on the study of marine creature fossils.**

| T(yrs. B.P.) | T*(yrs. after the Giant Impact) | Observed Synodic Month (modern days). | Estimated Solar Day (hrs). |
|---|---|---|---|
| 900Ma(Proterozoic) | 3.62956G | 25.0 | 19.2 |
| 600Ma(Proterozoic) | 3.92956G | 26.2 | 20.7 |
| 300Ma(Carboniferous) | 4.22956G | 28.7 | 22.3 |
| 0(Neozoic) | 4.52956G | 29.5 | 24 |

Kaula & Harris[1975] have determined the synodic month through the studies of marine creatures. The results are tabulated in Table(XII.5)

**Table XII. 6.Observed Synodic Month (Kaula & Harris 1975) based on the studies of Marine creatures.**

| T(yrs. B.P.) | T*(yrs. after the Giant Impact) | Observed Synodic Month (modern days). | Estimated Solar Day (hrs). |
|---|---|---|---|
| 45Ma | 4.48456G | 29.1 | 23.566 |
| 2.8Ga | 1.72956G | 17 | 13.857 |

Walker & Zahnle [1986] through the study of Australian Banded Iron Formation also known as Cyclic banded iron formation of the Weeli Wolli Formation arrived at the Lunar Nodal Period at 2.45Ga which is also known as Saros Cycle.Based on this they arrived at Lunar Orbital Radius of $52R_E$=3.280728E8m.This gives a very incorrect result for lod. Hence value of lod as estimated from the observational curve is used in the observed data column which is 14.5hours.

The epochs for which observed data or estimated data of Solar Day is available for those epochs the Lunar Orbital Radius ($r_L$) is accurately estimated and using the theoretical values of ($r_L$), the theoretical value of Solar Day Length is calculated and tabulated alongside the observed values of Solar Day. All these data are tabulated in Table(XII.6)

**Table XII. 7. A comparative study of the Observed and Theoretical Values of the Solar Day Length in different Epochs.**

| t (yrs. B.P) | t* (Gyrs.) after the Giant Impact | $r_L$ (*$10^8$ m) | $T_E$ (hrs) | $T^*_E$ (hrs) |
|---|---|---|---|---|
| 0 | 4.52956G | 3.844 | 24 | 24 |
| 65Ma | 4.46456G | 3.819095 | 23.62 | 23.627 |
| 135Ma | 4.39456G | 3.7918 | 23.2247 | 23.25 |
| 180Ma | 4.34956G | 3.7739 | 22.97 | 23.0074 |
| 230Ma | 4.29956 | 3.75396 | 22.69 | 22.7684 |
| 280Ma | 4.24956 | 3.73366 | 22.4125 | 22.4765 |
| 300Ma | 4.22956 | 3.72547 | 22.3002 | 22.3 |
| 345Ma | 4.18456 | 3.70688 | 22.055 | 22.136 |
| 380Ma | 4.14956 | 3.69227 | 21.86 | 21.9 |

| 405Ma | 4.12456 | 3.68175 | 21.73 | 21.8 |
| --- | --- | --- | --- | --- |
| 500Ma | 4.02956 | 3.64114 | 21.22 | 21.27 |
| 600Ma | 3.92956 | 3.59726 | 20.69 | 20.674 |
| 900Ma | 3.62956 | 3.45832 | 19.16 | 18.9 |
| 2.45Ga | 2.07956 | 2.51298 | 12.387 | 14.5 |
| 2.8Ga | 1.72956 | 2.22665 | 11.06 | 13.859 |

**From the graph of Observed Values of Length of Solar Day.**

By the following Mathematica Command, the graph giving the observed values of Solar Day vs t(yrs. after the Giant Impact ) is obtained:

q16 = ListPlot[{{1.72956× 10⁹, 13.859}, {2.07956× 10⁹, 14.5}, {3.22956× 10⁹, 17.25}, {3.62956× 10⁹, 19.16},
{3.90956× 10⁹, 20.59}, {3.92956× 10⁹, 20.69}, {4.02956× 10⁹, 21.22}, {4.12456× 10⁹, 21.73}, {4.14956× 10⁹, 21.86},
{4.18456× 10⁹, 22.055}, {4.22956× 10⁹, 22.3}, {4.24956× 10⁹, 22.4125}, {4.29956× 10⁹, 22.69}, {4.34956× 10⁹, 22.97},
{4.39456× 10⁹, 23.2247}, {4.46456× 10⁹, 23.62}, {4.52956× 10⁹, 24}}, PlotJoined → True, GridLines → Automatic,
Frame → True, FrameLabel → hours T_E, PlotLabel → {"Lengthening of Day by Obs."}, Prolog → AbsolutePointSize[8]]

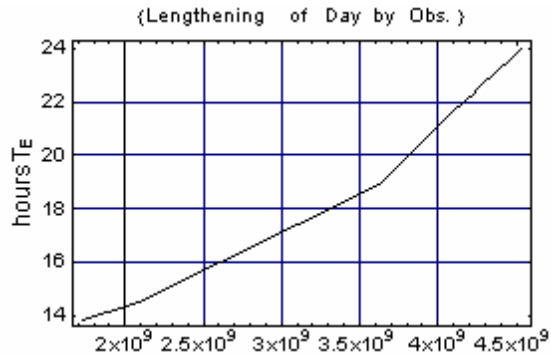

Figure XII.3. Lengthening of Day curve based on Observed values .

**XII-E-i. Theoretical Formulation of Lengthening of Day Curve assuming constant C.**

To obtain the theoretical curve for the lengthening of day we need Lunar Orbital Expansion Curve. By trial and error we arrive at the following expression which accurately fits the theoretical values of $r_L$ tabulated in Table(I.6).

**Lunar Orbital Radius = $r_L$ =**

5.52887891E8-(5.52887891E8-0.157E8)Exp[-t/(2E9)]-(1.6E8)Exp[-t/(14E9)]+(1.6E8) Exp[-t/(1.15E9)]

(XII.45)

**Through the following Plot Command the Expansion Curve is obtained based on Eq.(XII.45)**

```
q1 = Plot[5.52887891×10^8 − (5.52887891×10^8 − 0.157×10^8)Exp[−t/(2×10^9)] −
    (1.6×10^8)Exp[−t/(14×10^9)] + (1.6×10^8)Exp[−t/(1.15×10^9)], {t, 1.72956×10^9, 4.52956×10^9},
    GridLines → Automatic, Frame → True, FrameLabel → rL (m), PlotLabel → {"Lunar Orbital Expansion Curve"}]
```

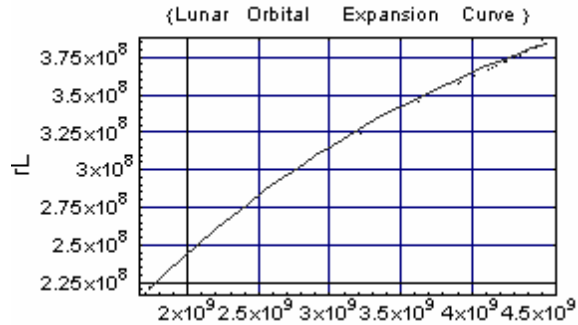

**Figure XII.4. Theoretical Lunar Orbital Expansion Curve based on Eq.(XII.45).**

By the following command we obtain the Lunar Orbital Radius Expansion Curve based on Table(XII.6)

```
q2 = ListPlot[{{1.72956×10^9, 2.22665×10^8}, {2.07956×10^9, 2.51298×10^8},
    {3.22956×10^9, 3.254235×10^8}, {3.62956×10^9, 3.45832×10^8}, {3.90956×10^9, 3.588345×10^8},
    {3.92956×10^9, 3.59726×10^8}, {4.02956×10^9, 3.64114×10^8}, {4.12456×10^9, 3.68175×10^8},
    {4.14956×10^9, 3.69227×10^8}, {4.18456×10^9, 3.70688×10^8}, {4.22956×10^9, 3.72547×10^8},
    {4.29956×10^9, 3.75396×10^8}, {4.34956×10^9, 3.7739×10^8}, {4.39456×10^9, 3.7918×10^8},
    {4.46456×10^9, 3.819095×10^8}, {4.52956×10^9, 3.844×10^8}}, PlotJoined → True, GridLines → Automatic,
    Frame → True, FrameLabel → (m)rL, PlotLabel → {"Lunar Orbital Expansion Curve"}]
```

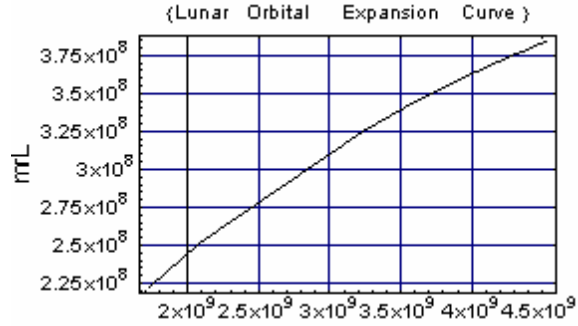

**Figure(XII.5).Lunar Orbital Expansion Curve based on Table(XII.6).**

Next through SHOW command we compare the two curves-one obtained through the empirical Eq.(XII.45) and the other obtained based on Table(XII.6).

`Show[{q1, q2}, Axes → True]`

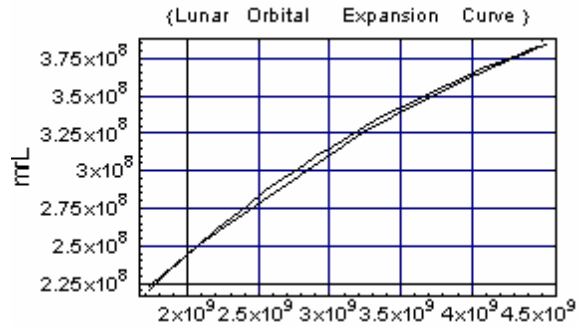

**Figure (XII.6) Superposition of the two curve-one based on Eq(XII.45) and second based on Table(XII.6).**

As seen from the superposition of the two curves, the empirical Eq.(XII.45) accurately represents the evolving Lunar Orbital Radii.

Eq(XII.45) is substituted in place of X in Eq(XII.24) to obtain the theoretical plot of Lengthening of Day curve by the following Command:

$$((2\pi \times 8.02 \times 10^{37}) \div 3600) \div (3.440488884 \times 10^{34} - (2.008433303 \times 10^7 \div x^2)(8.878598241 \times 10^{34} + 7.258980539 \times 10^{22} \times x^2)$$
$$(x^{0.5} - 9.142570518 \times 10^{-21} \times x^{2.5} + 9.142570518 \times 10^{-21} \times x^{3.5} \div (5.52887891 \times 10^8))) /.$$
$$x \to (5.52887891 \times 10^8 -$$
$$(5.52887891 \times 10^8 - 0.157 \times 10^8)\text{Exp}[-t/(2 \times 10^9)] - (1.6 \times 10^8)\text{Exp}[-t/(14 \times 10^9)] + (1.6 \times 10^8)\text{Exp}[-t/(1.15 \times 10^9)])$$

The new formulation of lod is a function of time and is checked by substituting t=4.52956E9 yrs.

$$1.39975406009945203`\wedge 35 / $$
$$(3.44048888400000052`\wedge 34 - (2.00843330299999944`\wedge 7 \, (8.87859824100000238`\wedge 34 + 7.25898053900000128`\wedge 22$$
$$(5.52887891000000131`\wedge 8 + 1.60000000000000008`\wedge 8 \, E^{-8.69565217391304301`\wedge -10 \, t} \_$$
$$5.37187891000000039`\wedge 8 \, E^{-t/2000000000} - 1.60000000000000008`\wedge 8 \, E^{-t/14000000000})\wedge 2)$$
$$((5.52887891000000131`\wedge 8 + 1.60000000000000008`\wedge 8 \, E^{-8.69565217391304301`\wedge -10 \, t} \_$$
$$5.37187891000000039`\wedge 8 \, E^{-t/2000000000} - 1.60000000000000008`\wedge 8 \, E^{-t/14000000000})\wedge 0.5` -$$
$$9.14257051799999942`\wedge -21$$
$$(5.52887891000000131`\wedge 8 + 1.60000000000000008`\wedge 8 \, E^{-8.69565217391304301`\wedge -10 \, t} \_$$
$$5.37187891000000039`\wedge 8 \, E^{-t/2000000000} - 1.60000000000000008`\wedge 8 \, E^{-t/14000000000})\wedge 2.5` +$$
$$1.65360295763829618`\wedge -29$$
$$(5.52887891000000131`\wedge 8 + 1.60000000000000008`\wedge 8 \, E^{-8.69565217391304301`\wedge -10 \, t} \_$$
$$5.37187891000000039`\wedge 8 \, E^{-t/2000000000} - 1.60000000000000008`\wedge 8 \, E^{-t/14000000000})\wedge 3.5`)) /$$
$$(5.52887891000000131`\wedge 8 + 1.60000000000000008`\wedge 8 \, E^{-8.69565217391304301`\wedge -10 \, t} \_$$
$$5.37187891000000039`\wedge 8 \, E^{-t/2000000000} - 1.60000000000000008`\wedge 8 \, E^{-t/14000000000})\wedge 2) \, /.$$
$$t \to 4.52956 \times 10^9$$

The answer is:
$24.0063927323029879`$ hrs.

By the following Plot Command the Lengthening of Day Curve is obtained by theory:

```
q13 = Plot[1.39975406009945203`^35/
    (3.44048888400000052`^34 - (2.00843330299999944`^7 (8.87859824100000238`^34 + 7.25898053900000128`^2
      (5.52887891000000131`^8 + 1.60000000000000008`^8 E^-8.69565217391304301`^-10 t _
        5.37187891000000039`^8 E^-t/2000000000 - 1.60000000000000008`^8 E^-t/14000000000)^2)
      ((5.52887891000000131`^8 + 1.60000000000000008`^8 E^-8.69565217391304301`^-10 t _
        5.37187891000000039`^8 E^-t/2000000000 - 1.60000000000000008`^8 E^-t/14000000000)^0.5` -
        9.14257051799999942`^-21
      (5.52887891000000131`^8 + 1.60000000000000008`^8 E^-8.69565217391304301`^-10 t _
        5.37187891000000039`^8 E^-t/2000000000 - 1.60000000000000008`^8 E^-t/14000000000)^2.5` +
        1.65360295763829618`^-29
      (5.52887891000000131`^8 + 1.60000000000000008`^8 E^-8.69565217391304301`^-10 t _
        5.37187891000000039`^8 E^-t/2000000000 - 1.60000000000000008`^8 E^-t/14000000000)^3.5`)) /
    (5.52887891000000131`^8 + 1.60000000000000008`^8 E^-8.69565217391304301`^-10 t _
      5.37187891000000039`^8 E^-t/2000000000 - 1.60000000000000008`^8 E^-t/14000000000)^2),
    {t, 1.7×10^9, 4.52956×10^9}, GridLines → Automatic, Frame → True, FrameLabel → hours T_E,
    PlotLabel → {"Lengthening of Day"}, PlotStyle → {{Thickness[0.001], GrayLevel[0.2]}}]
```

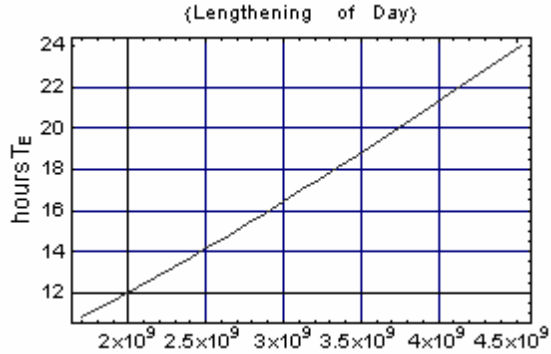

**Figure (XII.7) Lengthening of Day Curve theoretically based on canstant C.**

Show[{q13, q19}, Axes → True]

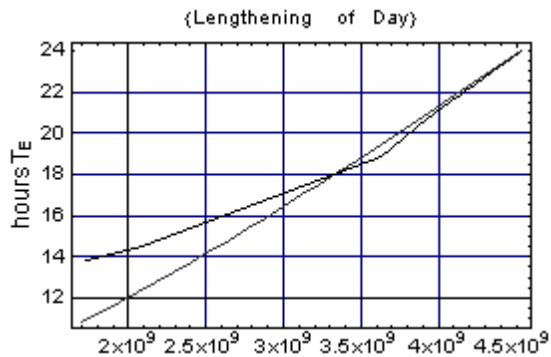

**Figure(XII.8) Superposition of the two lengthening of day curves -one based on theory with constant C and the second based on observation.**

As seen from the superposition of the two lengthening of day curves, there is remarkable match between Observation and Theory in the recent past after the Pre-Cambrian Explosion of plant and animal life but in the remote past ,particularly in early Archean Eon, Earth seems to be spinning much slower than predicted by theory .This implies that rotational inertia was much higher than what has been assumed in this analysis. In fact there are evidence to show that early Earth was much less stratified as compared to modern Earth. It was more like Venus.[Allegre et. al. 1994,Taylor et.al.1996].

In Eq.(I.24) as well as in Eq.(I.44) C, the Principal Moment of Inertia, has been assumed to be constant whereas infact it was evolving since the Giant Impact[Runcorn 1966].

In the first phase of planet formation, Earth was an undifferentiated mass of gas,rocks and metals much like Venus. At the point of Giant Impact , the impactor caused a massive heating which led to melting and magmatic formation of total Earth. The heavier metals, Iron and Nickel, settled down to the metallic core and lighter rocky materials formed the mantle. The mantle consisted of Basalt and Sodium rich Granite.

Due to Giant Impact, Earth gained extra angular momentum. This led to a very short spin period of 6 hours. It has been calculated that the oblateness at the inception must have been 1% [Kamble 1966] whereas the modern oblateness is 0.3%. Taking these two factors into account C of Earth must have been much higher than the modern value of 8.02E37 kg-m$^2$. In this paper the early C has been taken as 9.9E37 kg-m$^2$ but this subsequently led to instability in E-M system. The determination of the correct form of the evolving C has been left for the sequel paper.

After Archean Eon the general cooling of Earth over a period of 2 billion years led to slower plate-tectonic movement. The 100 continental-oceanic plates coalesced into 12 plates initially and into 13 plates subsequently. The slower plate tectonic engine led to deep recycling of the continental crust and hence to complete magmatic distillation and differentiation of the internal structure into multi-layered onion like structure. Thus at the boundary Arcean Eon and Proterozoic Eon a definite transition occurred in the internal structure.

Before this boundary, the mantle and the outer crust was less differentiated. It was composed of a mixture of Basaly and Sodium-rich granite. After this boundary a slower plate-tectonic dynamo helped create the onion-like internal structur with sharply differentiated basaltic mantle and potassium-rich granitic crust. This highly heterogeneous internal structure and less oblate geometry leads to the modern value of C equal to 8.02E37 Kg-m$^2$.

[Plate tectonic is the continual creation, motion and destruction of planet's surface-the outer crust at the oceanic plates are being ploughed inward at the sub-duction zones and magnmatic Basalt and lava are coming up and erupting through the mid-oceanic ridges.]

The form the evolving C is as follows:

**f[(t-2E9)_]:=If[(t-2E9)>0,1,0]
{9.9E37-(9.9E37-8.02E37){1-Exp[-t/16E9]}-f[(t-2E9)_](1.4E37){1-Exp[-t/(0.5E9)]}}            (XII.46)**

Here f[(t-2E9)_] is defined as a step function which is 0 before 2 billion years and is Unit Step at 2 billion years and at greater times.

By the following Plot Command the evolution of C with time is obtained:

```
ListPlot[{{0, 9.89999999999999857`*^37 }, {0.5×10^9, 9.84215848081552735`*^37 },
    {1.0×10^9, 9.78609655808933354`*^37 }, {1.5×10^9, 9.73175947939446217`*^37},
    {2.0×10^9, 9.67909417685903861`*^37 }, {2.5×10^9, 8.22099419105685669`*^37 },
    {3.0×10^9, 8.16548614251261639`*^37 }, {3.5×10^9, 8.11531421052783841`*^37},
    {4.0×10^9, 8.06802068853292908`*^37 }, {4.5×10^9, 8.02267327406780239`*^37}}, PlotJoined→True,
  GridLines→Automatic, Frame→True, FrameLabel→I_Earth, PlotLabel→{"Evolution of I_Earth"}]
```

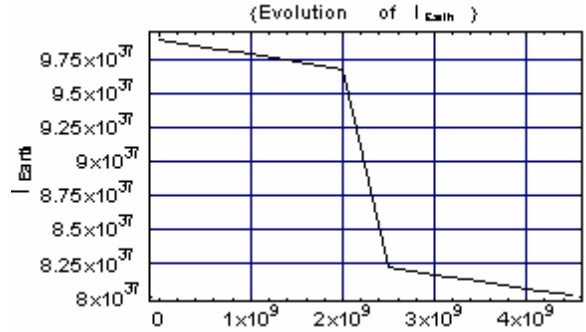

**Figure(XII.9) Profile of evolving C with time.**

**XII-E-ii. Theoretical Formulation of Lengthening of Day assuming evolving C.**

Consider Eq.(I.41) Lunar Velocity of Recession:
**$dX/dt = [(K/X^M) [(1/C)\{(MX^2)/(Y\ B) - D - E\ X^2\}-1]]/ [F[1/X^{1/2}-5AX^{3/2}+7AX^{5/2}/Y]]$**
(XII.41)

To include the effect of evolving C , Eq.(I.41) is transformed into function of time by replacing X by Eq(XII.45).This Velocity of Recession contains Moment of Inertia term C. This C is substituted by Eq.(XII.46).
This new Eq.(XII.47) is also a function of time which includes the evolving C with time.
Eq.(XII.47) cannot be directly integrated with respect to time within the limits of 0Gyrs(the time at which Giant Impact took place after the birth of Solar Nebula) and the present time of 4.52956Gyrs(the age of Moon). Therefore Eq.(XII.47) is simplified and integrated within the time limit of 0Gyrs to 4.52956Gyrs.
Integration between 0Gyrs to 4.52956Gyrs. and Integration between 0.03044Gyrs to 4.56Gyrs.have been considered to be equivalent.
The structure factor M is adjusted to achieve the integrand as (3.844E8-0.157E8)m=3.6873E8.
M=2.0928 achieves the requisite spatial span of 3.687E8m.

By the following Replace Command , Eq.(XII.41) is transformed into a function of time:

$$\left(1.81557 \times 10^{33} \times \left(\left((1 \div (Z)) \left((3.440488884 \times 10^{34} \times x^2) \div \left(2.008433307 \times 10^7 \times \right.\right.\right.\right.\right.$$
$$\left.\left(\sqrt{x} - 9.142570518 \times 10^{-21} \times x^{5/2} + 9.142570518 \times 10^{-21} \times x^{7/2} \div (5.52887891 \times 10^8)\right)\right) -$$
$$\left.8.878598241 \times 10^{34} - 7.258980539 \times 10^{22} \times x^2\right) -$$
$$\left.1\right) \div$$
$$\left.\left.x^{2.1}\right)\right) \div$$
$$\left((2.309982 \times 10^{22}) \left(1 \div \sqrt{x} - 5 \times 9.142570518 \times 10^{-21} \times x^{3/2} + 7 \times 9.142570518 \times 10^{-21} \times x^{5/2} \div (5.52887891 \times 10^8)\right)\right) /.$$
$$x \to 5.52887891 \times 10^8 -$$
$$(5.52887891 \times 10^8 - 0.157 \times 10^8) \text{Exp}[-t/(2 \times 10^9)] - (1.6 \times 10^8) \text{Exp}[-t/(14 \times 10^9)] + (1.6 \times 10^8) \text{Exp}[-t/(1.15 \times 10^9)]$$

$$\left(7.85967163380493794\,\grave{}\,\text{*}\,\text{^}\,10 \left(-1 + \frac{1}{Z} \left(-8.87859824100000238\,\grave{}\,\text{*}\,\text{^}\,34 - 7.25898053900000128\,\grave{}\,\text{*}\,\text{^}\,22 \right.\right.\right.$$
$$\left(5.52887891000000131\,\grave{}\,\text{*}\,\text{^}\,8 + 1.60000000000000008\,\grave{}\,\text{*}\,\text{^}\,8\,E^{-8.69565217391304301\,\grave{}\,\text{*}\,\text{^}\,-10\,t} - \right.$$
$$\left.5.37187891000000039\,\grave{}\,\text{*}\,\text{^}\,8\,E^{-t/2000000000} - 1.60000000000000008\,\grave{}\,\text{*}\,\text{^}\,8\,E^{-t/14000000000}\right)\text{^}\,2 +$$
$$\left(1.71302122505579497\,\grave{}\,\text{*}\,\text{^}\,27\right.$$
$$\left(5.52887891000000131\,\grave{}\,\text{*}\,\text{^}\,8 + 1.60000000000000008\,\grave{}\,\text{*}\,\text{^}\,8\,E^{-8.69565217391304301\,\grave{}\,\text{*}\,\text{^}\,-10\,t} - \right.$$
$$\left.5.37187891000000039\,\grave{}\,\text{*}\,\text{^}\,8\,E^{-t/2000000000} - 1.60000000000000008\,\grave{}\,\text{*}\,\text{^}\,8\,E^{-t/14000000000}\right)\text{^}\,2 \bigg/$$
$$\left(\sqrt{(5.52887891000000131\,\grave{}\,\text{*}\,\text{^}\,8 + 1.60000000000000008\,\grave{}\,\text{*}\,\text{^}\,8\,E^{-8.69565217391304301\,\grave{}\,\text{*}\,\text{^}\,-10\,t}} - \right.$$
$$\left.5.37187891000000039\,\grave{}\,\text{*}\,\text{^}\,8\,E^{-t/2000000000} - 1.60000000000000008\,\grave{}\,\text{*}\,\text{^}\,8\,E^{-t/14000000000}\right) -$$
$$9.14257051799999942\,\grave{}\,\text{*}\,\text{^}\,-21$$
$$\left(5.52887891000000131\,\grave{}\,\text{*}\,\text{^}\,8 + 1.60000000000000008\,\grave{}\,\text{*}\,\text{^}\,8\,E^{-8.69565217391304301\,\grave{}\,\text{*}\,\text{^}\,-10\,t} - \right.$$
$$\left.5.37187891000000039\,\grave{}\,\text{*}\,\text{^}\,8\,E^{-t/2000000000} - 1.60000000000000008\,\grave{}\,\text{*}\,\text{^}\,8\,E^{-t/14000000000}\right)\text{^}\,(5/2) +$$
$$1.65360295763829618\,\grave{}\,\text{*}\,\text{^}\,-29$$
$$\left(5.52887891000000131\,\grave{}\,\text{*}\,\text{^}\,8 + 1.60000000000000008\,\grave{}\,\text{*}\,\text{^}\,8\,E^{-8.69565217391304301\,\grave{}\,\text{*}\,\text{^}\,-10\,t} - \right.$$
$$\left.\left.\left.\left.5.37187891000000039\,\grave{}\,\text{*}\,\text{^}\,8\,E^{-t/2000000000} - 1.60000000000000008\,\grave{}\,\text{*}\,\text{^}\,8\,E^{-t/14000000000}\right)\text{^}\,(7/2)\right)\right)\right)\bigg/$$
$$\left(\left(5.52887891000000131\,\grave{}\,\text{*}\,\text{^}\,8 + 1.60000000000000008\,\grave{}\,\text{*}\,\text{^}\,8\,E^{-8.69565217391304301\,\grave{}\,\text{*}\,\text{^}\,-10\,t} - \right.\right.$$
$$\left.5.37187891000000039\,\grave{}\,\text{*}\,\text{^}\,8\,E^{-t/2000000000} - 1.60000000000000008\,\grave{}\,\text{*}\,\text{^}\,8\,E^{-t/14000000000}\right)\text{^}\,2.10000000000000008\,\grave{}$$
$$\left(1/\left(\sqrt{(5.52887891000000131\,\grave{}\,\text{*}\,\text{^}\,8 + 1.60000000000000008\,\grave{}\,\text{*}\,\text{^}\,8\,E^{-8.69565217391304301\,\grave{}\,\text{*}\,\text{^}\,-10\,t}} - \right.\right.$$
$$\left.\left.5.37187891000000039\,\grave{}\,\text{*}\,\text{^}\,8\,E^{-t/2000000000} - 1.60000000000000008\,\grave{}\,\text{*}\,\text{^}\,8\,E^{-t/14000000000}\right)\right) -$$
$$4.5712852590000006\,\grave{}\,\text{*}\,\text{^}\,-20\,(5.52887891000000131\,\grave{}\,\text{*}\,\text{^}\,8 + 1.60000000000000008\,\grave{}\,\text{*}\,\text{^}\,8\,E^{-8.69565217391304301\,\grave{}\,\text{*}\,\text{^}\,-10\,t} -$$
$$\left.5.37187891000000039\,\grave{}\,\text{*}\,\text{^}\,8\,E^{-t/2000000000} - 1.60000000000000008\,\grave{}\,\text{*}\,\text{^}\,8\,E^{-t/14000000000}\right)\text{^}\,(3/2) +$$
$$1.15752207034680743\,\grave{}\,\text{*}\,\text{^}\,-28\,(5.52887891000000131\,\grave{}\,\text{*}\,\text{^}\,8 + 1.60000000000000008\,\grave{}\,\text{*}\,\text{^}\,8\,E^{-8.69565217391304301\,\grave{}\,\text{*}\,\text{^}\,-10\,t} -$$
$$5.37187891000000039\,\grave{}\,\text{*}\,\text{^}\,8\,E^{-t/2000000000} - 1.60000000000000008\,\grave{}\,\text{*}\,\text{^}\,8\,E^{-t/14000000000})\,\text{^}$$
$$\left.\left.(5/2)\right)\right)$$

In the transformed Equation there is a marker Z. This marker is substituted by Eq.(XII.46) and the resulting Equation is as follows:

$$f[(t - 2 \times 10^9)\_] := If[(t - 2 \times 10^9) > 0, 1, 0]$$

$$\text{ReplaceAll}[(7.85967163380493794`\text{*}^\wedge 10\,(-1 + 1/(\{9.9 \times 10^{37} - (9.9 \times 10^{37} - 8.02 \times 10^{37})\{1 - \exp[-t \div (16 \times 10^9)]\} -$$
$$f[(t - 2 \times 10^9)\_] \times (1.4 \times 10^{37}) \times \{1 - \exp[-t \div (0.5 \times 10^9)]\}\}))$$

$$(-8.87859824100000238`\text{*}^\wedge 34 - 7.25898053900000128`\text{*}^\wedge 22$$
$$(5.52887891000000131`\text{*}^\wedge 8 + 1.60000000000000008`\text{*}^\wedge 8\, E^{-8.69565217391304301`\text{*}^\wedge -10\, t} \_$$
$$5.37187891000000039`\text{*}^\wedge 8\, E^{-t/2000000000} - 1.60000000000000008`\text{*}^\wedge 8\, E^{-t/14000000000})^{\wedge} 2 +$$
$$(1.71302122505579497`\text{*}^\wedge 27$$
$$(5.52887891000000131`\text{*}^\wedge 8 + 1.60000000000000008`\text{*}^\wedge 8\, E^{-8.69565217391304301`\text{*}^\wedge -10\, t} \_$$
$$5.37187891000000039`\text{*}^\wedge 8\, E^{-t/2000000000} - 1.60000000000000008`\text{*}^\wedge 8\, E^{-t/14000000000})^{\wedge} 2)/$$
$$(\sqrt{(5.52887891000000131`\text{*}^\wedge 8 + 1.60000000000000008`\text{*}^\wedge 8\, E^{-8.69565217391304301`\text{*}^\wedge -10\, t} \_}$$
$$5.37187891000000039`\text{*}^\wedge 8\, E^{-t/2000000000} - 1.60000000000000008`\text{*}^\wedge 8\, E^{-t/14000000000}) -$$
$$9.14257051799999942`\text{*}^\wedge -21$$
$$(5.52887891000000131`\text{*}^\wedge 8 + 1.60000000000000008`\text{*}^\wedge 8\, E^{-8.69565217391304301`\text{*}^\wedge -10\, t} \_$$
$$5.37187891000000039`\text{*}^\wedge 8\, E^{-t/2000000000} - 1.60000000000000008`\text{*}^\wedge 8\, E^{-t/14000000000})^{\wedge}(5/2) +$$
$$1.65360295763829618`\text{*}^\wedge -29$$
$$(5.52887891000000131`\text{*}^\wedge 8 + 1.60000000000000008`\text{*}^\wedge 8\, E^{-8.69565217391304301`\text{*}^\wedge -10\, t} \_$$
$$5.37187891000000039`\text{*}^\wedge 8\, E^{-t/2000000000} - 1.60000000000000008`\text{*}^\wedge 8\, E^{-t/14000000000})^{\wedge}(7/2)))$$
$$((5.52887891000000131`\text{*}^\wedge 8 + 1.60000000000000008`\text{*}^\wedge 8\, E^{-8.69565217391304301`\text{*}^\wedge -10\, t} \_$$
$$5.37187891000000039`\text{*}^\wedge 8\, E^{-t/2000000000} - 1.60000000000000008`\text{*}^\wedge 8\, E^{-t/14000000000})^{\wedge} 2.10000000000000008`$$
$$(1/(\sqrt{(5.52887891000000131`\text{*}^\wedge 8 + 1.60000000000000008`\text{*}^\wedge 8\, E^{-8.69565217391304301`\text{*}^\wedge -10\, t} \_}$$
$$5.37187891000000039`\text{*}^\wedge 8\, E^{-t/2000000000} - 1.60000000000000008`\text{*}^\wedge 8\, E^{-t/14000000000})) -$$
$$4.5712852590000006`\text{*}^\wedge -20\,(5.52887891000000131`\text{*}^\wedge 8 + 1.60000000000000008`\text{*}^\wedge 8\, E^{-8.69565217391304301`\text{*}^\wedge -10\, t} \_$$
$$5.37187891000000039`\text{*}^\wedge 8\, E^{-t/2000000000} - 1.60000000000000008`\text{*}^\wedge 8\, E^{-t/14000000000})^{\wedge}(3/2) +$$
$$1.15752207034680743`\text{*}^\wedge -28\,(5.52887891000000131`\text{*}^\wedge 8 + 1.60000000000000008`\text{*}^\wedge 8\, E^{-8.69565217391304301`\text{*}^\wedge -10\, t} \_$$
$$5.37187891000000039`\text{*}^\wedge 8\, E^{-t/2000000000} - 1.60000000000000008`\text{*}^\wedge 8\, E^{-t/14000000000})^{\wedge}(5/2))),$$
$$t \to 4.52956 \times 10^9]$$

**Eq.(XII.47)**

By the above Command, Eq.(I.47) is simultaneously calculated at the present time =4.52956Gyrs.
The answer is:
{{0.0379077443525112078`}}**m/yr.=3.8cm/yr=Moon's Velocity of Recession.**

This implies that Eq(XII.47) has been correctly transformed to include the evolving C.
By the following command Eq(XII.47) is simplified:

```mathematica
f[(t - 2×10^9) _] := If[(t - 2×10^9) > 0, 1, 0]
Simplify[(7.85967163380493794`*^10 (-1 + 1 / ({9.9×10^37 - (9.9×10^37 - 8.02×10^37) {1 - Exp[-t ÷ (16×10^9)]} -
            f[(t - 2×10^9) _] × (1.4×10^37) × {1 - Exp[-t ÷ (0.5×10^9)]}}))
    (-8.87859824100000238`*^34 - 7.25898053900000128`*^22
        (5.52887891000000131`*^8 + 1.60000000000000008`*^8 E^-8.69565217391304301`*^-10 t _
            5.37187891000000039`*^8 E^-t/2000000000 - 1.60000000000000008`*^8 E^-t/14000000000)^2 +
        (1.71302122505579497`*^27
            (5.52887891000000131`*^8 + 1.60000000000000008`*^8 E^-8.69565217391304301`*^-10 t _
                5.37187891000000039`*^8 E^-t/2000000000 - 1.60000000000000008`*^8 E^-t/14000000000)^2) /
        (√(5.52887891000000131`*^8 + 1.60000000000000008`*^8 E^-8.69565217391304301`*^-10 t _
                5.37187891000000039`*^8 E^-t/2000000000 - 1.60000000000000008`*^8 E^-t/14000000000) -
            9.14257051799999942`*^-21
            (5.52887891000000131`*^8 + 1.60000000000000008`*^8 E^-8.69565217391304301`*^-10 t _
                5.37187891000000039`*^8 E^-t/2000000000 - 1.60000000000000008`*^8 E^-t/14000000000)^(5/2) +
            1.65360295763829618`*^-29
            (5.52887891000000131`*^8 + 1.60000000000000008`*^8 E^-8.69565217391304301`*^-10 t _
                5.37187891000000039`*^8 E^-t/2000000000 - 1.60000000000000008`*^8 E^-t/14000000000)^(7/2)))
    ((5.52887891000000131`*^8 + 1.60000000000000008`*^8 E^-8.69565217391304301`*^-10 t _
        5.37187891000000039`*^8 E^-t/2000000000 - 1.60000000000000008`*^8 E^-t/14000000000)^2.10000000000000008`
        (1/(√(5.52887891000000131`*^8 + 1.60000000000000008`*^8 E^-8.69565217391304301`*^-10 t _
            5.37187891000000039`*^8 E^-t/2000000000 - 1.60000000000000008`*^8 E^-t/14000000000)) -
        4.5712852590000006`*^-20 (5.52887891000000131`*^8 + 1.60000000000000008`*^8 E^-8.69565217391304301`*^-10 t _
            5.37187891000000039`*^8 E^-t/2000000000 - 1.60000000000000008`*^8 E^-t/14000000000)^(3/2) +
        1.15752207034680743`*^-28 (5.52887891000000131`*^8 + 1.60000000000000008`*^8 E^-8.69565217391304301`*^-10 t _
            5.37187891000000039`*^8 E^-t/2000000000 - 1.60000000000000008`*^8 E^-t/14000000000)^(5/2)))]
```

$$\{\{(7.85967163380493794{}^{\wedge}10\,(-1+(-8.87859824100000238{}^{\wedge}34-7.25898053900000128{}^{\wedge}22$$
$$(5.52887891000000131{}^{\wedge}8+1.60000000000000008{}^{\wedge}8\,E^{-8.69565217391304301{}^{\wedge}-10\,t}\,-$$
$$5.37187891000000039{}^{\wedge}8\,E^{-t/2000000000}-1.60000000000000008{}^{\wedge}8\,E^{-t/14000000000})^{\wedge}2+$$
$$(1.713021225055579523{}^{\wedge}27$$
$$(5.52887891000000131{}^{\wedge}8+1.60000000000000008{}^{\wedge}8\,E^{-8.69565217391304301{}^{\wedge}-10\,t}\,-$$
$$5.37187891000000039{}^{\wedge}8\,E^{-t/2000000000}-$$
$$1.60000000000000008{}^{\wedge}8\,E^{-t/14000000000})^{\wedge}(3/2))/$$
$$(1-9.1425705180000012{}^{\wedge}-21$$
$$(5.52887891000000131{}^{\wedge}8+1.60000000000000008{}^{\wedge}8\,E^{-8.69565217391304301{}^{\wedge}-10\,t}\,-$$
$$5.37187891000000039{}^{\wedge}8\,E^{-t/2000000000}-1.60000000000000008{}^{\wedge}8\,E^{-t/14000000000})^{\wedge}2+$$
$$1.65360295763829645{}^{\wedge}-29$$
$$(5.52887891000000131{}^{\wedge}8+1.60000000000000008{}^{\wedge}8\,E^{-8.69565217391304301{}^{\wedge}-10\,t}\,-$$
$$5.37187891000000039{}^{\wedge}8\,E^{-t/2000000000}-1.60000000000000008{}^{\wedge}8\,E^{-t/14000000000})^{\wedge}3))/$$
$$(8.01999999999999957{}^{\wedge}37+1.88000000000000042{}^{\wedge}37\,E^{-t/16000000000}\,+$$
$$(-1.39999999999999968{}^{\wedge}37+1.39999999999999968{}^{\wedge}37\,E^{-2.{}^{\wedge}-9\,t})\,\mathrm{If}[-2000000000+t>0,1,0])))/$$
$$((5.52887891000000131{}^{\wedge}8+1.60000000000000008{}^{\wedge}8\,E^{-8.69565217391304301{}^{\wedge}-10\,t}\,-$$
$$5.37187891000000039{}^{\wedge}8\,E^{-t/2000000000}-1.60000000000000008{}^{\wedge}8\,E^{-t/14000000000})^{\wedge}1.60000000000000008$$
$$(1-4.5712852590000006{}^{\wedge}-20\,(5.52887891000000131{}^{\wedge}8+1.60000000000000008{}^{\wedge}8\,E^{-8.69565217391304301{}^{\wedge}-10\,t}\,-$$
$$5.37187891000000039{}^{\wedge}8\,E^{-t/2000000000}-1.60000000000000008{}^{\wedge}8\,E^{-t/14000000000})^{\wedge}2+$$
$$1.15752207034680765{}^{\wedge}-28\,(5.52887891000000131{}^{\wedge}8+1.60000000000000008{}^{\wedge}8\,E^{-8.69565217391304301{}^{\wedge}-10\,t}\,-$$
$$5.37187891000000039{}^{\wedge}8\,E^{-t/2000000000}-1.60000000000000008{}^{\wedge}8\,E^{-t/14000000000})^{\wedge}$$
$$3))\}\}$$

.

The above Eq(XII.47) is integrated within the limits 0.03044Gyrs and 4.56Gyrs.

Oscillatory in NIntegrate[{{(7.85967163380493794`*^10
    (-1 + (-8.87859824100000238`*^34 - 7.25898053900000128`*^22 (5.52887891000000131`*^8 +
              1.60000000000000008`*^8 E^-8.69565217391304301`*^-10 t - 5.37187891000000039`*^8
              E^-t/2000000000 - 1.60000000000000008`*^8 E^-t/14000000000)^2 +
           (1.71302122505579523`*^27
              (5.52887891000000131`*^8 + 1.60000000000000008`*^8 E^-8.69565217391304301`*^-10 t -
                 5.37187891000000039`*^8 E^-t/2000000000 -
                 1.60000000000000008`*^8 E^-t/14000000000)^(3/2)) /
            (1 - 9.1425705180000012`*^-21 (5.52887891000000131`*^8 +
                    1.60000000000000008`*^8 E^-8.69565217391304301`*^-10 t - 5.37187891000000039`*^8
                    E^-t/2000000000 - 1.60000000000000008`*^8 E^-t/14000000000)^2 +
                 1.65360295763829645`*^-29 (5.52887891000000131`*^8 +
                    1.60000000000000008`*^8 E^-8.69565217391304301`*^-10 t - 5.37187891000000039`*^8
                    E^-t/2000000000 - 1.60000000000000008`*^8 E^-t/14000000000)^3)) /
        (8.01999999999999957`*^37 + 1.88000000000000042`*^37 E^-t/16000000000 +
           (-1.39999999999999968`*^37 + 1.39999999999999968`*^37 E^-2.`*^-9 t) If[-2000000000 + t > 0, 1, 0])
    ((5.52887891000000131`*^8 + 1.60000000000000008`*^8 E^-8.69565217391304301`*^-10 t -
        5.37187891000000039`*^8 E^-t/2000000000 - 1.60000000000000008`*^8 E^-t/14000000000)^
     1.60000000000000008` (1 - 4.5712852590000006`*^-20
          (5.52887891000000131`*^8 + 1.60000000000000008`*^8 E^-8.69565217391304301`*^-10 t -
             5.37187891000000039`*^8 E^-t/2000000000 - 1.60000000000000008`*^8 E^-t/14000000000)^2 +
        1.15752207034680765`*^-28
          (5.52887891000000131`*^8 + 1.60000000000000008`*^8 E^-8.69565217391304301`*^-10 t -
             5.37187891000000039`*^8 E^-t/2000000000 - 1.60000000000000008`*^8 E^-t/14000000000)^3))}},
{t, 0.03044×10^9, 4.56×10^9}]

The integrand is:
{{3.22550496604630287`*^8 in Oscillatory}} m

Our integrand should be 3.687E8m. Therefore exponent M=2.1 which is visible in the Eq.(XII.47) before Simplification, is adjusted to 2.0928 at the pre-Simplification stage in the following manner:

```
f[(t - 2×10^9)_] := If[(t - 2×10^9) > 0, 1, 0]
Simplify[(7.85967163380493794`*^10 (-1 + 1 / ({9.9×10^37 - (9.9×10^37 - 8.02×10^37) {1 - Exp[-t ÷ (16×10^9)]} -
            f[(t - 2×10^9)_] × (1.4×10^37) × {1 - Exp[-t ÷ (0.5×10^9)]}}))
        (-8.87859824100000238`*^34 - 7.25898053900000128`*^22
            (5.52887891000000131`*^8 + 1.60000000000000008`*^8 E^-8.69565217391304301`*^-10 t _
                5.37187891000000039`*^8 E^-t/2000000000 - 1.60000000000000008`*^8 E^-t/14000000000)^2 +
            (1.71302122505579497`*^27
                (5.52887891000000131`*^8 + 1.60000000000000008`*^8 E^-8.69565217391304301`*^-10 t _
                    5.37187891000000039`*^8 E^-t/2000000000 - 1.60000000000000008`*^8 E^-t/14000000000)^2) /
            (√(5.52887891000000131`*^8 + 1.60000000000000008`*^8 E^-8.69565217391304301`*^-10 t _
                5.37187891000000039`*^8 E^-t/2000000000 - 1.60000000000000008`*^8 E^-t/14000000000) -
            9.14257051799999942`*^-21
                (5.52887891000000131`*^8 + 1.60000000000000008`*^8 E^-8.69565217391304301`*^-10 t _
                    5.37187891000000039`*^8 E^-t/2000000000 - 1.60000000000000008`*^8 E^-t/14000000000)^(5/2) +
            1.65360295763829618`*^-29
                (5.52887891000000131`*^8 + 1.60000000000000008`*^8 E^-8.69565217391304301`*^-10 t _
                    5.37187891000000039`*^8 E^-t/2000000000 - 1.60000000000000008`*^8 E^-t/14000000000)^(7/2)))
        ((5.52887891000000131`*^8 + 1.60000000000000008`*^8 E^-8.69565217391304301`*^-10 t _
            5.37187891000000039`*^8 E^-t/2000000000 - 1.60000000000000008`*^8 E^-t/14000000000)^2.0928
        (1 / (√(5.52887891000000131`*^8 + 1.60000000000000008`*^8 E^-8.69565217391304301`*^-10 t _
                5.37187891000000039`*^8 E^-t/2000000000 - 1.60000000000000008`*^8 E^-t/14000000000)) -
            4.5712852590000006`*^-20 (5.52887891000000131`*^8 + 1.60000000000000008`*^8 E^-8.69565217391304301`*^-10 t _
                5.37187891000000039`*^8 E^-t/2000000000 - 1.60000000000000008`*^8 E^-t/14000000000)^(3/2) +
            1.15752207034680743`*^-28 (5.52887891000000131`*^8 + 1.60000000000000008`*^8 E^-8.69565217391304301`*^-10 t _
                5.37187891000000039`*^8 E^-t/2000000000 - 1.60000000000000008`*^8 E^-t/14000000000)^(5/2)))]
```

As can be seen, the exponent has been changed from 2.1 to 2.0928.

$$\{\{(7.85967163380493794\text{`}*\text{^}10\,(-1 + (-8.87859824100000238\text{`}*\text{^}34 - 7.25898053900000128\text{`}*\text{^}22$$
$$(5.52887891000000131\text{`}*\text{^}8 + 1.60000000000000008\text{`}*\text{^}8\,E^{-8.69565217391304301\text{`}*\text{^}-10\,t} -$$
$$5.37187891000000039\text{`}*\text{^}8\,E^{-t/2000000000} - 1.60000000000000008\text{`}*\text{^}8\,E^{-t/14000000000})\text{^}2 -$$
$$(1.71302122505579523\text{`}*\text{^}27$$
$$(5.52887891000000131\text{`}*\text{^}8 + 1.60000000000000008\text{`}*\text{^}8\,E^{-8.69565217391304301\text{`}*\text{^}-10\,t} -$$
$$5.37187891000000039\text{`}*\text{^}8\,E^{-t/2000000000} -$$
$$1.60000000000000008\text{`}*\text{^}8\,E^{-t/14000000000})\text{^}\,(3/2))\,/$$
$$(1 - 9.1425705180000012\text{`}*\text{^}-21$$
$$(5.52887891000000131\text{`}*\text{^}8 + 1.60000000000000008\text{`}*\text{^}8\,E^{-8.69565217391304301\text{`}*\text{^}-10\,t} -$$
$$5.37187891000000039\text{`}*\text{^}8\,E^{-t/2000000000} - 1.60000000000000008\text{`}*\text{^}8\,E^{-t/14000000000}$$
$$1.65360295763829645\text{`}*\text{^}-29$$
$$(5.52887891000000131\text{`}*\text{^}8 + 1.60000000000000008\text{`}*\text{^}8\,E^{-8.69565217391304301\text{`}*\text{^}-10\,t} -$$
$$5.37187891000000039\text{`}*\text{^}8\,E^{-t/2000000000} - 1.60000000000000008\text{`}*\text{^}8\,E^{-t/14000000000}$$
$$(8.01999999999999957\text{`}*\text{^}37 + 1.88000000000000042\text{`}*\text{^}37\,E^{-t/16000000000} +$$
$$(-1.39999999999999968\text{`}*\text{^}37 + 1.39999999999999968\text{`}*\text{^}37\,E^{-2.\text{`}*\text{^}-9\,t})\,\text{If}[-2000000000 + t > 0,$$
$$((5.52887891000000131\text{`}*\text{^}8 + 1.60000000000000008\text{`}*\text{^}8\,E^{-8.69565217391304301\text{`}*\text{^}-10\,t} -$$
$$5.37187891000000039\text{`}*\text{^}8\,E^{-t/2000000000} - 1.60000000000000008\text{`}*\text{^}8\,E^{-t/14000000000})\text{^}1.5927999999$$
$$(1 - 4.5712852590000006\text{`}*\text{^}-20\,(5.52887891000000131\text{`}*\text{^}8 + 1.60000000000000008\text{`}*\text{^}8\,E^{-8.6956521739130}$$
$$5.37187891000000039\text{`}*\text{^}8\,E^{-t/2000000000} - 1.60000000000000008\text{`}*\text{^}8\,E^{-t/14000000000})\text{^}2 +$$
$$1.15752207034680765\text{`}*\text{^}-28\,(5.52887891000000131\text{`}*\text{^}8 + 1.60000000000000008\text{`}*\text{^}8\,E^{-8.6956521739130}$$
$$5.37187891000000039\text{`}*\text{^}8\,E^{-t/2000000000} - 1.60000000000000008\text{`}*\text{^}8\,E^{-t/14000000000})\text{^}\,$$
$$3))\}\}$$

Next Integration is performed:

```
Oscillatory in NIntegrate[{{(7.85967163380493794`*^10
    (-1 + (-8.87859824100000238`*^34 - 7.25898053900000128`*^22 (5.52887891000000131`*^8 +
        1.60000000000000008`*^8 E^-8.69565217391304301`*^-10 t - 5.37187891000000039`*^8
        E^-t/2000000000 - 1.60000000000000008`*^8 E^-t/14000000000)^2 +
      (1.71302122505579523`*^27
        (5.52887891000000131`*^8 + 1.60000000000000008`*^8 E^-8.69565217391304301`*^-10 t -
         5.37187891000000039`*^8 E^-t/2000000000 -
         1.60000000000000008`*^8 E^-t/14000000000)^(3/2))/
      (1 - 9.1425705180000012`*^-21 (5.52887891000000131`*^8 +
         1.60000000000000008`*^8 E^-8.69565217391304301`*^-10 t - 5.37187891000000039`
         E^-t/2000000000 - 1.60000000000000008`*^8 E^-t/14000000000)^2 +
       1.65360295763829645`*^-29 (5.52887891000000131`*^8 +
         1.60000000000000008`*^8 E^-8.69565217391304301`*^-10 t - 5.37187891000000039`
         E^-t/2000000000 - 1.60000000000000008`*^8 E^-t/14000000000)^3))/
     (8.01999999999999957`*^37 + 1.88000000000000042`*^37 E^-t/16000000000 +
      (-1.39999999999999968`*^37 + 1.39999999999999968`*^37 E^-2.`*^-9 t) If[-2000000000 + t >
    ((5.52887891000000131`*^8 + 1.60000000000000008`*^8 E^-8.69565217391304301`*^-10 t -
      5.37187891000000039`*^8 E^-t/2000000000 - 1.60000000000000008`*^8 E^-t/14000000000)^
      1.59279999999999999` (1 - 4.5712852590000006`*^-20
      (5.52887891000000131`*^8 + 1.60000000000000008`*^8 E^-8.69565217391304301`*^-10 t -
       5.37187891000000039`*^8 E^-t/2000000000 - 1.60000000000000008`*^8 E^-t/14000000000)^2 +
      1.15752207034680765`*^-28
      (5.52887891000000131`*^8 + 1.60000000000000008`*^8 E^-8.69565217391304301`*^-10 t -
       5.37187891000000039`*^8 E^-t/2000000000 - 1.60000000000000008`*^8 E^-t/14000000000)^3))}},
  {t, 0.03004×10^9, 4.56×10^9}]
```

The Integrand is:
{{3.69790563609877143`*^8 in Oscillatory}} m

This is very near the answer 3.687E8m.

Next the Integration is performed for all the given epochs in which the observational data exists and the new profile of Lunar Orbital Expansion is determined. Table(I.7) gives the new set of Lunar Orbital Radii.

**Table XII. 8. The Lunar Orbital Radius Expansion including the evolving C.**

| T(B.P.)Gyrs | T*(Gyrs after the Giant Impact) | The definite Integrand=[$r_N$-0.157E8m] | Theoretically cal. $r_N$ (*$10^8$m) |
|---|---|---|---|
| 0 | 4.52956G | 3.68503 | 3.84203 |
| 65Ma | 4.46456G | 3.65639 | 3.81339 |
| 135Ma | 4.39456G | 3.62474 | 3.78174 |
| 180Ma | 4.34956G | 3.60439 | 3.76139 |
| 230Ma | 4.29956 | 3.58169 | 3.73869 |
| 280Ma | 4.24956 | 3.55888 | 3.71588 |
| 300Ma | 4.22956 | 3.54944 | 3.70644 |
| 345Ma | 4.18456 | 3.52817 | 3.68517 |
| 380Ma | 4.14956 | 3.5118 | 3.6688 |
| 405Ma | 4.12456 | 3.49978 | 3.65678 |
| 500Ma | 4.02956 | 3.45388 | 3.61088 |
| 600Ma | 3.92956 | 3.40412 | 3.56112 |
| 900Ma | 3.62956 | 3.24826 | 3.4052 |
| 2.45Ga | 2.07956 | 2.19804 | 2.35504 |
| 2.8Ga | 1.72956 | 1.9167 | 2.0737 |

From the new set of Lunar Orbital Radii we formulate a new function of time to predict Lunar Orbital Radius in a given epoch and which includes the evolving C .

$$5.52887891 \times 10^8 - (5.52887891 \times 10^8 - 0.157 \times 10^8) \mathrm{Exp}[-t/(2.2 \times 10^9)] -$$
$$(1.6 \times 10^8) \mathrm{Exp}[-t/(10.3 \times 10^9)] + (1.6 \times 10^8) \mathrm{Exp}[-t/(1.15 \times 10^9)] /.$$
$$t \to 4.52956 \times 10^9$$

(XII.48)

By the above Mathematica Command we have defined new form of Lunar Orbital Expansion which includes the evolving C as well as it calculates Lunar Orbital Radius in the present epoch.

The result is:

$3.84390555846364492`*{}^{\wedge}8$ m.

By the following command , in Eq.(XII.24) X is replaced by Eq.(I.48)

$$((2\pi \times (C)) \div 3600) \div (3.440488884 \times 10^{34} - (2.008433303 \times 10^7 \div x^2)(8.878598241 \times 10^{34} + 7.258980539 \times 10^{22} \times x^2)$$
$$(x^{0.5} - 9.142570518 \times 10^{-21} \times x^{2.5} + 9.142570518 \times 10^{-21} \times x^{3.5} \div (5.52887891 \times 10^8))) /.$$
$$x \to 5.52887891 \times 10^{\wedge}8 - (5.52887891 \times 10^{\wedge}8 - 0.157 \times 10^{\wedge}8) \text{Exp}[-t/(2.2 \times 10^{\wedge}9)] - (1.6 \times 10^{\wedge}8) \text{Exp}[-t/(10.3 \times 10^{\wedge}9)] +$$
$$(1.6 \times 10^{\wedge}8) \text{Exp}[-t/(1.15 \times 10^{\wedge}9)]$$

As can be seen ,in the above equation a marker has been put in place of Moment of Inertia for the convenience of substituting the evolving form of C.
Following is the transformed form of Eq.(XII.24) which gives the length of Solar Day:

$$(C\pi)/(1800$$
$$(3.44048888400000052`*{}^{\wedge}34 - (2.00843330299999944`*{}^{\wedge}7 (8.87859824100000238`*{}^{\wedge}34 + 7.25898053900000128`*{}^{\wedge}2$$
$$(5.52887891000000131`*{}^{\wedge}8 + 1.60000000000000008`*{}^{\wedge}8 E^{-8.69565217391304301`*{}^{\wedge}-10\,t} -$$
$$5.37187891000000039`*{}^{\wedge}8 E^{-4.54545454545454585`*{}^{\wedge}-10\,t} -$$
$$1.60000000000000008`*{}^{\wedge}8 E^{-9.70873786407766914`*{}^{\wedge}-11\,t})^{\wedge}2)$$
$$((5.52887891000000131`*{}^{\wedge}8 + 1.60000000000000008`*{}^{\wedge}8 E^{-8.69565217391304301`*{}^{\wedge}-10\,t} -$$
$$5.37187891000000039`*{}^{\wedge}8 E^{-4.54545454545454585`*{}^{\wedge}-10\,t} -$$
$$1.60000000000000008`*{}^{\wedge}8 E^{-9.70873786407766914`*{}^{\wedge}-11\,t})^{\wedge}0.5` -$$
$$9.14257051799999942`*{}^{\wedge}-21$$
$$(5.52887891000000131`*{}^{\wedge}8 + 1.60000000000000008`*{}^{\wedge}8 E^{-8.69565217391304301`*{}^{\wedge}-10\,t} -$$
$$5.37187891000000039`*{}^{\wedge}8 E^{-4.54545454545454585`*{}^{\wedge}-10\,t} -$$
$$1.60000000000000008`*{}^{\wedge}8 E^{-9.70873786407766914`*{}^{\wedge}-11\,t})^{\wedge}2.5` +$$
$$1.65360295763829618`*{}^{\wedge}-29 (5.52887891000000131`*{}^{\wedge}8 + 1.60000000000000008`*{}^{\wedge}8$$
$$E^{-8.69565217391304301`*{}^{\wedge}-10\,t} - 5.37187891000000039`*{}^{\wedge}8 E^{-4.54545454545454585`*{}^{\wedge}-10\,t} -$$
$$1.60000000000000008`*{}^{\wedge}8 E^{-9.70873786407766914`*{}^{\wedge}-11\,t})^{\wedge}3.5`))/$$
$$(5.52887891000000131`*{}^{\wedge}8 + 1.60000000000000008`*{}^{\wedge}8 E^{-8.69565217391304301`*{}^{\wedge}-10\,t} -$$
$$5.37187891000000039`*{}^{\wedge}8 E^{-4.54545454545454585`*{}^{\wedge}-10\,t} - 1.60000000000000008`*{}^{\wedge}8 E^{-9.70873786407766914`*{}^{\wedge}-11\,t}$$
$$2))$$

In the above Equation marker C is replaced by Eq.(XII.46) and following is the result:

$$((f[(t-2\times 10^9)\_] := If[(t - 2\times 10^9) > 0, 1, 0]$$
$$\{9.9\times 10^{37} - (9.9\times 10^{37} - 8.02\times 10^{37})\{1 - \operatorname{Exp}[-t \div (16\times 10^9)]\} -$$
$$f[(t-2\times 10^9)\_]\times (1.4166\times 10^{37})\times \{1 - \operatorname{Exp}[-t \div (0.5\times 10^9)]\}\})\pi)/$$
$$(1800\,(3.44048888400000052`{*}\wedge 34 -$$
$$(2.00843330299999944`{*}\wedge 7\,(8.87859824100000238`{*}\wedge 34 + 7.25898053900000128`{*}\wedge 22$$
$$(5.52887891000000131`{*}\wedge 8 + 1.60000000000000008`{*}\wedge 8\,E^{-8.69565217391304301`{*}\wedge{-10}\,t} -$$
$$5.37187891000000039`{*}\wedge 8\,E^{-4.54545454545454585`{*}\wedge{-10}\,t} -$$
$$1.60000000000000008`{*}\wedge 8\,E^{-9.70873786407766914`{*}\wedge{-11}\,t})\wedge 2)$$
$$((5.52887891000000131`{*}\wedge 8 + 1.60000000000000008`{*}\wedge 8\,E^{-8.69565217391304301`{*}\wedge{-10}\,t} -$$
$$5.37187891000000039`{*}\wedge 8\,E^{-4.54545454545454585`{*}\wedge{-10}\,t} -$$
$$1.60000000000000008`{*}\wedge 8\,E^{-9.70873786407766914`{*}\wedge{-11}\,t})\wedge 0.5` -$$
$$9.14257051799999942`{*}\wedge{-21}$$
$$(5.52887891000000131`{*}\wedge 8 + 1.60000000000000008`{*}\wedge 8\,E^{-8.69565217391304301`{*}\wedge{-10}\,t} -$$
$$5.37187891000000039`{*}\wedge 8\,E^{-4.54545454545454585`{*}\wedge{-10}\,t} -$$
$$1.60000000000000008`{*}\wedge 8\,E^{-9.70873786407766914`{*}\wedge{-11}\,t})\wedge 2.5` +$$
$$1.65360295763829618`{*}\wedge{-29}\,(5.52887891000000131`{*}\wedge 8 + 1.60000000000000008`{*}\wedge 8$$
$$E^{-8.69565217391304301`{*}\wedge{-10}\,t} - 5.37187891000000039`{*}\wedge 8\,E^{-4.54545454545454585`{*}\wedge{-10}\,t} -$$
$$1.60000000000000008`{*}\wedge 8\,E^{-9.70873786407766914`{*}\wedge{-11}\,t})\wedge 3.5`))/$$
$$(5.52887891000000131`{*}\wedge 8 + 1.60000000000000008`{*}\wedge 8\,E^{-8.69565217391304301`{*}\wedge{-10}\,t} - 5.37187891000000039`{*}\wedge$$
$$E^{-4.54545454545454585`{*}\wedge{-10}\,t} - 1.60000000000000008`{*}\wedge 8\,E^{-9.70873786407766914`{*}\wedge{-11}\,t})\wedge 2))\,/.$$
$$t \to 4.52956\times 10^9$$

In the above Command, length of Solar Day is simultaneously calculated for the present epoch as a verification of the correctness of the transformation. The result is:

{{23.9986973569117889` Null}} **hours.**

Once the correct form of Equation for determining lod is achieved it is used to determine the lod for different epoch. By a Table Command a whole set of lod for the respective epoch is determined and plotted by the following ListPlot Command:

```
q26 = ListPlot[{{1.72956000000000003`*^9, 12.6808511439204485`}, {1.82956000000000003`*^9,
       13.1188349128043357`}, {1.92956000000000003`*^9, 13.5656303223507879`},
    {2.52956000000000003`*^9, 14.0316977578274215`},
    {2.62956000000000011`*^9, 14.4616341776643331`},
    {2.72956000000000002`*^9, 14.8994799816791996`},
    {2.82955999999999985`*^9, 15.3451361265950119`},
    {2.92956000000000003`*^9, 15.7985001961228395`},
    {3.02956000000000003`*^9, 16.2594657464214975`},
    {3.12956000000000011`*^9, 16.7279218035649224`},
    {3.22956000000000002`*^9, 17.2037524987958212`},
    {3.32955999999999985`*^9, 17.6868368280474053`},
    {3.42955999999999994`*^9, 18.1770485230451726`},
    {3.52956000000000003`*^9, 18.6742560221718818`},
    {3.62956000000000011`*^9, 19.1783225301368602`},
    {3.72956000000000002`*^9, 19.6891061563050158`},
    {3.97956000000000002`*^9, 20.9944780902675498`},
    {4.22955999999999932`*^9, 22.3384891748382008`},
    {4.47955999999999932`*^9, 23.7185818254087799`}}, PlotJoined → True, GridLines → Automatic,
   Frame → True, FrameLabel → T_E, PlotLabel → {"LOD including evolutionary C"}]
```

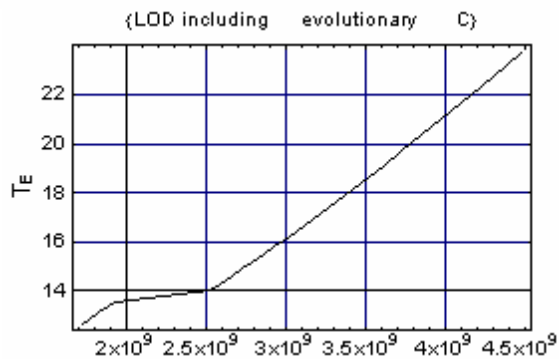

**Figure I.10. Lengthening of Day curve based on evolving C.**

Show[{q27, q28}, Axes → True]

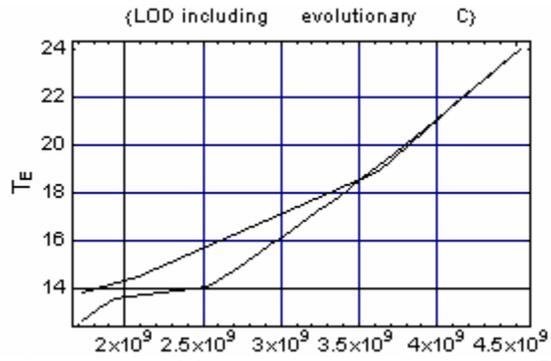

**Figure I.11..Superposition of Lengthening of Day Curves based on Observation and based on theory including evolving C.**

# APPENDIX(XIII)

## CONVERSION OF SYNODIC MONTH OF A GIVEN EPOCH INTO SOLAR DAY LENGTH(IN HOURS) OF THE SAME EPOCH.

From geometric consideration of Moon's orbit around Earth and that of Earth around Sun a geometric series is arrived at, which after infinite series summation gives the following equation:
Time interval between two full Moon phases=

$$(T_m)_{synodic} = (T_m)_{sid}[1 + \Sigma((T_m)_{sid}/Z)^n]$$
(XIII.1)

Eq.(XIII.1) is a geometric progression series upto n=∞ therefore
$(T_m)_{synodic} = a/(1-r)$;

$$(T_m)_{synodic} = (T_m)_{sidereal}/[1-(T_m)_{sidereal}/Z]$$
(XIII.2)

where $(T_m)_{synodic}$ = apparent lunar orbital time period,
$(T_m)_{sidereal}$ = actual lunar orbital time period,
Z days/solar year =
(8765.8 hours/solar year)/($T_E$ hour/solar day)
$T_E$ = solar day length on Earth in the given epoch under consideration.

Rewriting Eq.(XII.29) of App.(XII),
$\Omega_L X^2 = B[X^{1/2} - AX^{5/2} + AX^{7/2}/Y]$

$$2\pi/(T_m)_{sidereal} = (B/X^2)[X^{1/2} - AX^{5/2} + X^{7/2}/Y]$$
(XIII.3)

where X = lunar orbital radius = $r_L$,
Y = terminal point of lunar trajectory
= 5.52887891E8m
A = 9.142570518E-21E-21m$^{-2}$,
B = 2.008433303E7m$^{3/2}$/sec.

Here the sidreal lunar period is obtained in seconds. This sidereal period should be divided by 3600 second to get the same in hours. It should be further divided by 24 hours to obtain the sidereal period in modern solar days of 24 hours.
Substituting Eq.(XIII.3) in Eq.(XIII.2) and rearranging the terms,

$$1/(T_m)_{synodic} = [(86400BY')/(2\pi X^2) - (2\pi C)/\{31.55688E6\{M-(B/X^2)(D+EX^2)(Y')\}\}]$$
(XIII.4)

where
A = 9.142570518E-21m$^{-2}$, D = 8.878598241E34 Kg-m$^2$,
B = 2.008433303E7m$^{3/2}$/sec, C = 8.02E37 Kg-m$^2$ (in the recent past) and C = 10E37 Kg-m$^2$ (in the remote past before and around 2.5 b.yrs.B.P.
E = 7.258980539E22 Kg,

**M = 3.440488884E34 Kg-m²/sec.**
**Y′ = (X^{1/2} - AX^{5/2} + AX^{7/2}/Y), Y = 5.52887891E8 m.**

Further simplification leads to the expression:

$$1/(T_m)_{synodic} = [BY' \, 86400/(2\pi X^2) - (2\pi C)/(31.55688E6(M - (B/X^2)(D + EX^2)Y'))] \quad (XIII.5)$$

    Here the synodic month has been calculated in terms of modern day of 24 hours because the observed synodic month length has been calculated by dividing one mean solar year of 365.242 modern solar days by the number of lunar bands occurring within one annular band of the marine creature.

    Synodic month of 17 days of given epoch gives a lunar orbital radius X=2.75E8m by Eq.(XIII.4) assuming the modern value of C. By substituting this value of X in Eq.(XII.25), we obtain mean solar day of 13.67hours.

    Synodic month of 17 days of given epoch gives a lunar orbital radius X=2.74E8m assuming the ancient value 9.99E37Kg-m² for C. This gives the mean solar day as 16.96hours again assuming the ancient value of C.

# APPENDIX(XIV)

Modelling of Earth-Moon System as a Non-Linear Dynamic System.

> Assumptions:
> (1) E-M System is regarded as 2-body rotating system from its birth to its death in this analysis. Whereas in fact it is a 3-body system including the tidal drag of Sun for Lunar orbital radius greater than ten times Earth's radius.
> (2) Total angular momentum of E-M System has been assumed to be the scalar sum of the orbital angular momentum of E-M system, spin angular momentum of Earth and spin angular momentum of Moon. The obliquity angle of Earth's spin axis with respect to the Ecliptic plane, which is $\alpha=23.45$ degree, and the inclination angle of Moon's orbital plane with respect to the ecliptic plane, which presently is $i=5.15$ degree, make the total angular momentum the vector sum of the individual angular momentums.

By making the above two assumptions a certain loss of accuracy is involved but the calculation becomes tractable. The calculation will be further refined in a sequel paper.

**Section(A) The calculation of the centre of mass of E-M System:**

$Ed = m(r_L - d)$
Therefore $d = (m/(m+E))r_L$
    (XIV.1)

Substituting the numerical values of the present times:
$(d)_p = 4.67067E6$ m
p suffix means present.

**Section(B) First Equation of motion:**

> Centripetal force on Moon = Centrifugal force due to orbitting Moon + $\varepsilon'$.
>
> Therefore
> Centripetal acceleration = centrifugal acceleration + $\varepsilon$.
>     (XIV.2)
> Where $\varepsilon$ is residual inward acceleration because of which the outward radial velocity is being slowed until it is zero.

At the present times,
Centripetal acceleration = $GE/(r_L)^2 = 2.696738488E-3 \, m/sec^2$ .(XIV.3)

Centrifugal acceleration = $v^2/(r_L-d)$
= $[\Omega_L(r_L-d)]^2/(r_L-d) = (\Omega_L)^2 r_L \times 1/(1+m/E)$
= $2.694518066E-3 \, m/sec^2$. (XIV.4)
where $G = 6.67E-11 \, Newton-m^2/Kg$. $E = 5.9742E24 \, Kg$
$r_L = 3.844E8 \, m$.
$\Omega_L = (2\pi)/(27.3 \, solar \, days/revolution \times 86400 \, sec/solar \, day)$
$\varepsilon = 2.220422E-6 \, m/sec^2$.

This implies an inward residual acceleration which is retarding the outward radial velocity of Moon. As Moon spirals out, Earth's spin period (24 hours presently) and Lunar orbital period(27.3solar days) differential decreases hence the retarding tidal drag on Earth as well as the residual inward acceleration of Moon, both decrease. Therefore the residual inward acceleration is assumed to be:
$\varepsilon = k_3[1 - r_L/(r_L)_f]$
(XIV.5)

The tidal braking torque on Earth is assumed to be: $\tau = k_1/(r_L)^M - k_2 r_L$ (XIV.6)

**Section(C) CALCULATION OF THE TERMINAL POINT**

Sometime in remote future
at Orbital radius = $(r_L)_f$, the terminal point of the outward lunar path,
 Moon will become geosynchronous,
i.e. Moon's orbital period = Earth's spin period = 47.6 present solar days.
[This calculation will come later].
Moon will be in a geostationary orbit. At this point Sun's tidal drag will try to further slow down Earth to a synchronous or captured rotation orbit whereby Earth's spin period = Earth's orbital period = 365.25 solar days and Earth will present the same face to Sun as Moon presents the same face to Earth today. But simultaneously Moon will try to speed up Earth's spin. Because of the proximity of Moon the net effect will be the speeding up of Earth's spin and a net transfer of angular momentum to Earth. To conserve the angular momentum Moon's orbital momentum will be forced to decrease and Moon will start spiraling inward until it spirals down to Earth and crashes. But much before this celestial crash, Sun will have exhausted its nuclear fuel , expanded into a Red Giant and devoured the inner terrestrial planets including Earth. This will happen 5 billion years hence.

**Section(C-i) Calculation of total angular momentum.**

$J_T = (J_{spin})_E + (J_{spin})_m + (J_{orbit})$ (XIV.7)

$(J_{orbit}) = Ed^2\Omega_L + m(r_L-d)^2\Omega_L$
(XIV.8)

Substituting Eq.(XIV.1) in Eq.(XIV.8),
$J_{orbit} = (m/(1+m/E))(r_L)^2\Omega_L$  (XIV.9)

Substituting the present day parameters,
$\Omega_L, r_L$ and E as above and $m/E = 0.0123$,

$[J_{orbit}]_p = 2.857234392E34$ Kg-m$^2$/sec.
(XIV.10)

$[J_{spin}]_E)_p = C[\omega_E]_p = 0.5832308584E34$ Kg-m$^2$/sec.  (XIV.11)
where $C = 8.02E37$ Kg-m$^2$, $[\omega_E]_p = 2\pi/(1$ solar day $\times 86400$ secs/solar day$)$

$[J_{spin}]_m)_p = (2/5)m(R_m)^2[\omega_m]_p = 2.363359753E-5E34$ Kg-m$^2$/sec  (XIV.12)

Where $R_m = 1.738E6$ m,
$[\omega_m]_p = (2\pi)/(27.32$ solar days/rotation $\times 86400$ secs/solar day$)$

Substituting Eq.(XIV.10), (XIV.11) and (XIV.12) in (XIV.7),

$J_T = 3.440488884E34$ Kg-m$^2$/sec  (XIV.13)

Since lunar spin angular momentum constitutes an insignificant component of the total angular component therefore $(J_{spin})_m$ will be neglected throughout the analysis and $J_T$ will be expressed as follows:

$J_T = C\omega_E + (m/(1+m/E))(r_L)^2\Omega_L$
(XIV.14)

**Section(C-ii) Calculation of the terminal point $(r_L)_f$:**

At the terminal point Moon is geosynchronous and centripetal acceleration = centrifugal acceleration.

Therefore from Eq.(XIV.3) and Eq.(XIV.4),

$GE/(r_L)^2_f = (\Omega_L)^2_f (r_L)_f (1/(1+m/E))$  (XIV.15)

Solving Eq.(XIV.15),

$$(\Omega_L)_f = \sqrt{[(GE/(r_L)_f^3)((E+m)/E)]} \tag{XIV.16}$$

At the terminal point,

$$\Omega_L = \omega_E = (\Omega_L)_f. \tag{XIV.17}$$

Substituting Eq.( XIV.17) in Eq.(XIV.14),

$$J_T = [C + (mE/(m+E))(r_L)_f^2](\Omega_L)_f \tag{XIV.18}$$

Substituting Eq.( XIV.16) in Eq(XIV.18),

$$J_T = [C + (mE/(m+E))(r_L)_f^2]\sqrt{[(GE/(r_L)_f^3)((E+m)/E)]} \tag{XIV.19}$$

In the present epoch itself,
$(mE/(m+E))(r_L)_p^2 = 1.0726E40 \, Kg\text{-}m^2$

$C = 8.02E37 \, Kg\text{-}m^2$.

Therefore in all future times,

$C \ll (mE/(m+E))(r_L)^2$ and C can be neglected in Eq.( XIV.19).

Hence Eq.( XIV.19) can be rewritten as,
$$J_T = (mE/(m+E))(r_L)_f^2 (1/(r_L)_f^{3/2}) \sqrt{[G(E+m)]} \tag{XIV.20}$$

Simplifying Eq.( XIV.20),
$$J_T = mE/\sqrt{E+m} \times \sqrt{[(r_L)_f G]} \tag{XIV.21}$$

Squaring both sides of Eq.( XIV.21),
$$(J_T)^2 = (mE)^2 \times (r_L)_f \times G/(E+m) \tag{XIV.22}$$

Manipulating Eq.( XIV.22),
$$(r_L)_f = [J_T/(mE)]^2 \times (E+m)/G \tag{XIV.23}$$

Substituting the numerical values in Eq.( XIV.23)
$(r_L)_f = 5.568961629E8 \, m$
Where $J_T = 3.440488884E34 \, Kg\text{-}m^2/sec$ (XIV.24)
Rewriting Eq.( XIV.16),
$(\Omega_L)_f = 2\pi/[86400 \, sec/day \times (T_L)_f \, solar \, days]$
$= \sqrt{[(GE/(r_L)_f^3) \times ((E+m)/E)]}$ (XIV.25)

Solving Eq.( XIV.25)

$$(T_L)_f = (2\pi/86400) * (1/\sqrt{(GE/(r_L)_f^3) * ((E+m)/E)}) \qquad (XIV.26)$$

Substituting the numerical values in Eq.( XIV.26)
$(T_L)_f = 47.585$ solar days.

Therefore at the terminal point,
One solar day of Earth will be equal to 47.6 present solar days and this will be Moon's orbital period around Earth also hence Moon will come in geo-stationary orbit.

**Section(D) MAIN SOLUTION OF THE EQUATIONS OF MOTION.**

Substituting the present epoch numerical values of $\varepsilon, r_L$ and Eq.( XIV.23) in Eq.( XIV.4)

$$k_3 = 7.168533323E\text{-}6 \text{ m/sec}^2 \qquad (XIV.27)$$

Rewriting Eq.( XIV.2),

$GE/(r_L)^2 = (\Omega_L)^2 r_L E/(E+m) + k_3(1 - r_L/(r_L)_f)$ and rearranging the terms,

$$\Omega_L(r_L)^2 = \sqrt{GE(1+m/E)}[(r_L)^{1/2} - k_3(r_L)^{5/2}/(2GE) + k_3(r_L)^{7/2}/(2GE(r_L)_f)]$$
$$(XIV.28)$$

Eq.( XIV.28) is checked in the present epoch. An exact balance is obtained by finely tuning $k_3$ to $7.17E\text{-}6$ m/sec$^2$.

Eq.( XIV.28) is expressed in terms of constants and variables of Fx-570W calculator:

$k_3/(2GE) = A = 8.996706829E\text{-}21 \text{ m}^{-2}$ where $GE = 3.9847914E14 \text{ m}^3/\text{sec}^2$;
$(r_L)_f = Y = 5.568961629E8$ m;
$\sqrt{GE(1+m/E)} = B = 2.008433303E7 \text{ m}^{3/2}/\text{sec}$;
$(r_L) = X$;

Substituting these new constants and variables in Eq.( XIV.28),

$$\Omega_L X^2 = B[X^{1/2} - AX^{5/2} + AX^{7/2}/Y] \qquad (XIV.29)$$

**Section(D-i) CALCULATION OF THE SOLAR DAY ON EARTH.**

Rewriting Eq.( XIV.9),( XIV.11) and (XIV.12), the general expression of the total angular momentum is:

$J_T = 3.440488884E34$ Kg-m$^2$/sec $= (J_{spin})_E + (J_{spin})_m + J_{orb}$
$= C\omega_E + 0.4*0.0123E*R_m^2\omega_L + 0.0123E(r_L)^2(\Omega_L)/1.0123$   (XIV.30)

Since Moon is in captured rotation or synchronous rotation therefore $\omega_L = \Omega_L$.

Inserting the condition of captured rotation in Eq.( XIV.30),

$J_T = C*2\pi/(3600T_E) + [0.4*0.0123E*R_m^2 + 0.0123Er_L^2/1.0123]\Omega_L$   (XIV.31)

Substituting Eq.( XIV.29) in Eq.( XIV.31),

$J_T = C*2\pi/(3600T_E) + [0.4*0.0123ER_m^2 + 0.0123Er_L^2/1.0123]B/X^2[X^{1/2} - AX^{5/2} + AX^{7/2}/Y]$

(XIV.32)

Expressing the constants and variables of Eq.( XIV.32) in terms of those of Fx-570W,
$J_T = M = 3.440488884E34$ Kg-m$^2$,
$E = 5.9742E24$ Kg,
$R_m = D = 1.738E6$ m,
$T_E =$ Earth's solar day in hours,
$A = 8.996706829E-21$ m$^{-2}$,
$B = 2.008433303E7$ m$^{3/2}$/sec.

$M = 2\pi C/(3600T_E) + [0.4*0.0123ED^2 + 0.0123EX^2/1.0123][B/X^2][X^{1/2} - AX^{5/2} + AX^{7/2}/Y]$

(XIV.33)

The Eq.( XIV.33) is exactly balanced therefore all the constants are correct.

Rearranging Eq.( XIV.33),

$T_E = ((2\pi C)/3600)/[M - (B/X^2)(0.4*0.0123ED^2 + 0.0123EX^2/1.0123)(X^{1/2} - AX^{5/2} + AX^{7/2}/Y)]$
 (XIV.34)

$T_E = 24$ hours at $X = 3.844E8$ m and $Y = 5.568961629E8$ m.
Eq.( XIV.34) is further simplified.
$0.4*0.0123*ED^2 = 8.878598241E34$ Kg-m$^2 = D'$,
$0.0123E/1.0123 = 7.258980539E22$ Kg $= E'$.

Therefore,
$$T_E = ((2\pi C)/3600)/[M - (B/X^2)(D' + E' X^2)Y'] \quad (XIV.34a)$$

Where $M = 3.440488884E34 \, Kg\text{-}m^2/sec$,

$\quad C =$ moment of inertia of Earth around spin axis $= 8.02E37 \, Kg\text{-}m^2$,

$\quad B = 2.008433303E7 \, m^{3/2}/sec$,

$\quad X =$ Lunar orbital radus of the given epoch,

$\quad D' = 8.878598241E34 \, Kg\text{-}m^2$,

$\quad E' = 7.258980539E22 \, Kg$,

$\quad Y' = (X^{1/2} - AX^{5/2} + AX^{7/2}/Y)$,

$\quad A = 8.996706829E\text{-}21 \, m^{-2}$,

$\quad Y = 5.568961629E8 \, m$.

C=moment of inertia of Earth at the beginning before magmatic differentation was completed.At that time Earth was a homogeneous spherical mass of axial moment of inertia of $(2/5)ER_E^2 = 9.723880431E37 \, Kg\text{-}m^2$.After the Giant Impact,when Earth was heated into a molten mass , the first phase of magmatic differentiation took place and heavier elements precipitated to form the heaviest Iron-Nickel core wheras basaltic-granitic outer mantle was formed.Upto the boundry of Archean Eon (ancient)and Proterozoic Eon(early life)magmatic differentiation and internal stratification was incomplete.Before this boundry which occurs at 2.5b.yrs.B.P.there was metallic core and outer mantle composed of basalt and sodium rich granite.In the remote past that is before this boundry because of high amount of impact heating as well as because of higher level of radio-activity,intense heat was fuelling a faster plate-tectonic engine.As a result the primary continental crust was broken into hundred crustal plates.After Archean Eon the plate tectonic engine became slower,continental plates coalasced together to form twelve plates and deep recycling of the outer crustal plates led to sharply differentiated basaltic mantle and potassium-rich granitic crust.Primary crust gave way to secondary crust and secondary crust gave way to tertiary crust.Because of rapid spin of Earth the oblateness was 1%as compared to 0.3%in the modern era.In the past after the Giant Impact,Earth was homogeneous,much less stratified and much more oblate  very much like Venus. Today in the modern era Earth has an onion like internal stucture with almost one order of magnitude less oblateness.If all these factors are taken into account the axial moment of inertia was almost $10E37 \, Kg\text{-}m^2$ in the remote past.Whereas in the modern era it is only $8E37 \, Kg\text{-}m^2$.So to obtain realistic values of solar day length we must use the moment of inertia value which prevailed in the given era of interest.In our calculations we should assume a moment of inertia of Earth as $10E37 \, Kg\text{-}m^2$ before 2.5b.yrs B.P.and $8.02E37 \, Kg\text{-}m^2$ after 2.5b.yrs.B.P.In actual calculation C has been taken as $8.02E37 \, Kg\text{-}m^2$.Subsequentlty the percentage change in C is calculated to obtain the observed value of Solar Day.

### Section(D-ii) CALCULATION OF THE TIDAL TORQUE.

Substituting Eq.( XIV.28) in Eq.( XIV.9),
$J_{orb} = (mE/(m+E))\sqrt{GE(1+m/E)}[X^{1/2} - AX^{5/2} + AX^{7/2}/Y]$ (XIV.35)

X and Y are as defined in Eq.( XIV.29).

Redefinition:
$(mE/(m+E))\sqrt{GE(1+m/E)} = B' = 1.457917826E30 \text{ Kg-m}^{3/2}/\text{sec}.$
Therefore
$J_{orb} = B'[X^{1/2} - AX^{5/2} + AX^{7/2}/Y]$ (XIV.36)

Where $B' = 1.457917826E30 \text{ Kg-m}^{3/2}/\text{sec}$,
$A = 8.996706829E-21 \text{ m}^{-2}$.

Rewriting Eq.( XIV.7),

$J_T = J_{orb} + (J_{spin})_E + (J_{spin})_m$.

For all practical purposes, $(J_{spin})_m$ is negligible.
Therefore:
$J_T = J_{orb} + (J_{spin})_E$ (XIV.37)
Differentiating with respect to time(t),
$dJ_{orb}/dt + d(J_{spin})_E/dt = 0$ (XIV.38)

Therefore,
$dJ_{orb}/dt = -d(J_{spin})_E/dt$ (XIV.39)

Eq(XIV.39) implies that to conserve total angular momentum, as Earth's angular momentum is slowing down Moon's orbital angular momentum is increasing. This also implies that there is tidal braking torque which is slowing down the spinning Earth and there is a accelerating torque in reaction which is increasing the orbital momentum. The two torques are equal and opposite and the magnitude of each is given by:

$\tau = dJ_{orb}/dt$ (XIV.40)

Differentiting Eq(XIV.36) with respect to time we get the magnitude of the torque:

$\tau = B'[1/(2X^{1/2}) - 5AX^{3/2}/2 + 7AX^{5/2}/(2Y)] dX/dt$ (XIV.41)

Kitt's Peak Observatory and Macdonald Observatory Laser Bouncing Data gives the following rate of recession of Moon:

$(dX/dt)_p = 3.8 \text{cm/year} = 3.8\text{E-2m}/31.556909\text{E6s} = 1.204173704\text{E-9 m/sec}$. (XIV.42)

Substituting Eq.(XIV.42) in Eq.(XIV.41),
Braking torque in the present epoch is,

$(\tau)_p = 4.476136982\text{E16 N-m}$. (XIV.43)

**Section(D-iii) TIME INTEGRAL EQUATION OF MOON'S SPIRAL TRAJECTORY.**

In Eq(XIV.6) an empirical form of braking torque was assumed:
$\tau = k_1/(X)^M - k_2 X$ (XIV.44)
Where $X = r_L$.

Substituting the two boundary conditions two equations are obtained. Applying Cramer's rule the two unknowns $k_1$ and $k_2$ are obtained.

The two boundary conditions are: the present and the terminal point as defined in Section(I-C).

$\quad \tau_p = k_1/(X_p)^M - k_2 X_p$ (XIV.45)

$\quad 0 = k_1/(Y)^M - k_2 Y$ (XIV.46)
$\quad$ where $X_p$ = present epoch orbital radius,
$\quad$ and $Y$ = terminal point orbital radius.
Applying Cramer's rule to Eq(XIV.45) & (XIV.46),

$k_2 = (\tau_p/Y^M)/[Y/X_p^M - X_p/Y^M]$ and $k_1 = k_2 Y^{M+!}$ (XIV.47)

where $Y$ = orbital radius at the terminal point = 5.568961629E8m,
$\quad X_p$ = present epoch orbital radius = 3.844E8,
$\quad \tau_p = 4.476136982\text{E16 N-m}$.
In terms of Fx-570W constants,
$\quad\quad \tau_p = F$

Rewriting Eq.(XIV.47),

$k_2 = (F/Y^M)/[Y/X_p^M - X_p/Y^M]$ and $k_1 = k_2 Y^{M+1}$ (XIV.48)

Equating Eq.(41) and (44),

$$k_1/X^M - k_2 X = B/2[1/X^{1/2} - 5AX^{3/2} + 7AX^{5/2}/Y]dX/dt \quad (XIV.49)$$

Rearranging and integrating both sides we get,

$$\int dt(\text{limits}, t_0, t_p) = B/2 \int [(1/X^{1/2} - 5AX^{3/2} + 7AX^{5/2}/Y)/(k_1/X^M - k_2 X)] dX(\text{limits}, X_0, X_p)$$

(XIV.50)

$$(t_p - t_0)/(31.5569088E6 \text{ sec/solar yr.})$$
$$= B/(2*31.5569088E6) \int [[(1/X^{1/2} - 5AX^{3/2} + 7AX^{5/2}/Y)/(k_1/X^M - k_2 X)] dX, X_0, X_p]$$

(XIV.**51**)

where $t_0 = 0.06$ b.yrs = instant of Giant Impact after the birth of our Solar Nebula.
$(t_p) = 4.56$ b.yrs = the present age of Earth.
$X_0$ = Roche's radius = 18,000 km = $1.8E7$ m $\cong 3R_E$ = the distance from the center of Earth at which the Giant Impact generated debris accreted into Moon.
$X_p = 3.844E8$ m = the present orbital radius of Moon.

By determining the above integral between Roche's radius and the present orbital radius, the present age of Moon is determined which by observation has been determined to be 4.5 b.yrs. Hence the parameter M has to be finely tuned in order to get the definite integral equal to 4.5 b.yrs.

Expressing Eq.( XIV.51) in terms of Fx-570W constants,

$$[(t_p - t_0)/31.5569088E6] \text{ solar years} =$$
$$D \int [(X^{(M-1/2)} - 5AX^{(M+3/2)} + 7AX^{(M+5/2)}/Y)/(B - CX^{(M+1)}), E, F]$$
(XIV.**52**)

where $D = B/(2 \times 31.5569088E6) =$
$(1.457917826E30 \text{ Kg-m}^{3/2}/\text{sec})/(2 \times 31.5569088E6 \text{ sec/solar year}) = 2.309982E22$,
$A = 8.996706829E-21$,
$Y = 5.568961629E8$ m,
$E = X_0 = 1.8E7$ m, $F = X_p = 3.844E8$ m, $k_1 = B, k_2 = C$.
After several iterations the primary bench mark of 4.5 b.yrs is achieved at $M = 1.031$;

For $M = 1.031$,
$(k_1) = B = 6.002958088E25$, $k_2 = C = 1.036806853E8$,
$(t_p - t_0)/(31.5569088E6) = 4.499151941E9$ yrs.
Eq.( XIV.52) can be interpolated to any epoch in the past or to any future epoch. If the orbital radius in any epoch is known then by integrating between the limits $X_0$(Roche's Radius)and $X_N$(Orbital radius of Moon in the unknown epoch),the chronological date of the unknown epoch can be pinned down. If the epoch is known then through several iterations of Eq.( XIV.52)the correct lunar orbital radius can be arrived at.

**TABLE(I.1):The symbols and the values of the different parameters of Earth-Moon System.**

| Symbols | Definition | Values |
|---|---|---|
| E | Mass of Earth | 5.9742E24 Kg |
| m | Mass of Moon | E/81 = 7.348E22 Kg |
| m/E | | 0.0123 |
| G | Gravitational Constant | 6.67E-11 N-m$^2$/Kg$^2$ |
| S | Barycentre(centre of mass of the Earth-Moon System) | |
| $(d)_p$ | Distance between the centre of Earth and the Barycentre | 4.67067E6 m |
| GE | | 3.986E14 m$^3$/sec$^2$ |
| 1 A.U. | Astronomical Unit = mean orbital radius of Earth around Sun | 1.496E11 m |
| $(r_L)_p$ | Lunar mean orbital radius at the present | 0.00255*1A.U.= 3.844E8 m |
| $(r_L)_N$ | Lunar orbital radius in the past Nth epoch | |
| $R_{Earth}$ | Equatorial radius of Earth | 6.37814E6 m |
| $R_{Moon}$ | Equatorial radius of Moon | 1.738E6 m |
| a | Semi-major axis of Earth | 6.3788E6 m |
| b | Semi-minor axis of Earth | 6.3562E6 m |
| e | Oblateness of Earth = (a-b)/a | 0.00335 |

| C | Moment of Inertia of Earth around Polar axis | $0.3308ER^2_{Earth} = 8.02E37$ Kg-m$^2$ |
|---|---|---|
| $T_E$ | Spin period of Earth with respect to Sun i.e. mean solar day | 24 hours(pesent mean solar day)=86400 seconds |
| $(T_E)_{sidereal}$ | Spin period of Earth with respect to a fixed Star | 23hrs.56mins.4.09secs.=23.9344694hrs=86164 secons |
| 1 Solar Year=1 Tropical Year | The time between two successive vernal equinoxes | 365.242 days =365.242*86400=31.5569088E6 seconds per solar year |
| 1 Sidereal Year | The time between successive occupation by Earth of the same point in the orbit with respect to stars | 365.256 days=365.256*86164= 31.47191798E6 seconds per sidereal year |
| Z | Number of Mean Solar Days per Solar Year | 365.242 days |
| $(T_m)_{sidereal}$ | Time required for Moon to circle Earth and return to the same celestial longitude of a given star | 27.3 Solar Days per sidereal month |
| $(T_m)_{synodic}$ | Time between successive alignment of Eart, Moon and Sun | 29.50537844 Solar Days per synodic month or lunar month |
| Sidereal Period of Lunar Spin | Time required for Moon to rotate once on its polar axis and return to given celestial longitude of a given star | 27.32 Solar Days per rotation |

| $(\omega_E)_p$ | Present rate of rotation of Earth on its polar axis | $2\pi$ radians/86400seconds |
|---|---|---|
| $(\omega_L)_p$ | Present sidereal rate of rotation of Moon on its polar axis | $2\pi$ radians/[(27.32days)(86400s/d)] |
| $(\Omega_L)_p$ | Present sidereal rate of revolution of Moon around Earth | $2\pi$ radians/[(27.3days)(86400s/d) |
| $J_{orb}$ | Orbital angular momentum of Earth-Moon System around barycentre | $[E/(E+m)](mr_L^2)(\Omega_L)$ |
| $(J_{spin})_E$ | Spin angular momentum of Earth around spin axis | $C\omega_E$ |
| $(J_{spin})_m$ | Spin angular momentum of Moon around its spin axis | $(2/5)m(R_m)^2(\omega_L)$ |
| $\alpha$ | Earth's obliquity with respect to Ecliptic plane | 23.45 degrees |

| i | **Moon's Orbital Plane** **With respect to Ecliptic plane** | **5.15 degrees** |
|---|---|---|
| β | **Moon's obliquity with respect to Moon's orbital plane** | **6.6833 degrees** |
|   |   |   |